\documentclass[fleqn,usenatbib,useAMS]{mnras}

\usepackage{graphicx}	
\usepackage{amsmath}	
\usepackage{multicol}        
\usepackage{bm}		
\usepackage{pdflscape}	

\usepackage{caption}
\usepackage{tabularx}
\usepackage{booktabs}
\usepackage{txfonts}
\usepackage{multirow}
\usepackage{pbox}
\usepackage{color}
\usepackage{pdflscape}
\usepackage{enumitem}

\newcommand{\kms}{km s$^{-1}$}
\newcommand{\myr}{M$_{\odot}$ yr$^{-1}$}
\newcommand{\msun}{M$_{\odot}$}

\newcommand{\ci}{[C\,{\sc i}]}
\newcommand{\cii}{[C\,{\sc ii}]}
\newcommand{\hi}{H\,{\sc i}}

\newcommand{\bbarolo}{$^{\rm 3D}$\textsc{Barolo}}
\newcommand{\vrot}{$V_\mathrm{rot}$}

\usepackage[T1]{fontenc}
\usepackage{ae,aecompl}

\title[Resolving starbursts in a $z=2.4$ overdensity]{Revealing the nature of the starburst galaxies in the $z=2.4$ overdensity HATLAS J0849}

\author[M. Kaasinen]{Melanie Kaasinen$^{1}$%
\thanks{Contact e-mail: \href{mailto:melanie.kaasinen@anu.edu.au}{melanie.kaasinen@anu.edu.au}}, %
F.~Rizzo$^{2}$,
F.~Valentino$^{3,4}$,
C.~Bacchini$^{5}$,
J.~Chen$^{6}$,
T.~Tsukui$^{7}$,
A.~Amvrosiadis$^{8}$
\\
$^{1}$Research School of Astronomy and Astrophysics, Australian National University, Canberra, ACT 2611, Australia\\
$^{2}$Kapteyn Astronomical Institute, University of Groningen, Landleven 12, 9747 AD Groningen, The Netherlands\\
$^{3}$Cosmic Dawn Center (DAWN), Denmark\\
$^{4}$DTU Space, Technical University of Denmark, Elektrovej 327, DK-
2800 Kgs. Lyngby, Denmark\\
$^{5}$DARK, Niels Bohr Institute, University of Copenhagen, Jagtvej 155, 2200 Copenhagen, Denmark\\
$^{6}$Max-Planck-Institut f\"{u}r Extraterrestrische Physik (MPE), Gie{\ss}en-
bachstr. 1, D-85748 Garching, Germany\\
$^{7}$Kavli Institute for the Physics and Mathematics of the Universe (WPI), The University of Tokyo Institutes for Advanced Study,\\~~~The
University of Tokyo, Kashiwa, Chiba 277-8583, Japan\\
$^{8}$Sub-department of Astrophysics, Department of Physics, University of Oxford, Keble Road, Oxford, OX1 3RH, UK
}

\date{Preprint. Submitted to MNRAS on March 18; under peer review.}

\pubyear{{\the\year{}}}

\begin{document}
\label{firstpage}
\pagerange{\pageref{firstpage}--\pageref{lastpage}}
\maketitle

\begin{abstract}
    Today's most massive ellipticals are proposed to originate from starbursting galaxies in $z\gtrsim2$ overdensities. To discern what triggers these starbursts, and their $z=0$ descendants, we performed a detailed case study of five gas-rich galaxies in the $z=2.41$ overdensity, HATLAS J084933.4+021443. Using $0\farcs15$ resolution CO(4--3), \ci 1--0, and dust-continuum observations, we characterised their cold gas morphology and kinematics. We find two rotating discs, W and C, both exhibiting non-axisymmetric radial gas motions (consistent with bars). Of the two extreme starbursts, W is a lopsided, rotation-dominated disc with a rotation velocity of $\sim520$\,\kms, whereas T is most likely a late-stage merger. Combined with recent studies, we find that $\gtrsim42\%$ of gas-rich, massive starbursts in overdensities are rotation-dominated discs, a fraction not yet systematically reproduced by galaxy evolution models. Beyond $z=1$, disc galaxies with rotation velocities of $>400$\,\kms\ reside almost exclusively in overdensities, consistent with early mass assembly in dense environments. By comparing to local early-type galaxies with cold gas discs, we confirm that these systems already reside in halos comparable to the most massive $z\sim0$ ellipticals at the centres of groups and clusters. Despite their extreme star-formation rates, these discs lie on the same $\sigma-$SFR locus as lower-SFR field galaxies, implying that stellar feedback remains the dominant turbulence driver. We postulate that this is because inflowing gas is effectively transported through ordered streaming, such that only a small fraction of kinetic energy feeds disc-wide turbulence.  \
\end{abstract}


    \section{Introduction}
        \label{sec:intro}

            Over the last decade, sub-/millimetre interferometers have transformed our view of the most intensely star-forming galaxies at $z\gtrsim2$. Tens of what were originally identified as bright, single-dish sources have each been resolved into multiple dusty star-forming galaxies (DSFGs), clustered in redshift and on sky -- signposting galaxy overdensities \citep[][and references therein]{alberts_noble_2022}. Collectively, these DSFG groupings reach star formation rates (SFRs) of $\gtrsim 5000$\,\myr, motivating their interpretation as brief, rapid-growth phases that build the stellar mass of today's most massive ellipticals. Yet, higher resolution has not eliminated the extremes; even after deblending, individual galaxies with $\mathrm{SFR}\gtrsim 10^3$\myr\ have been revealed in these overdense systems \citep[e.g.][]{ivison_2013, dannerbauer_2014,oteo_2018,kamieneski_2024}. Such extreme, clustered starbursts remain hard to explain, leaving open which modes of gas assembly and star formation drive the stellar mass assembly of today’s most massive ellipticals.  
        
            In cosmological simulations, massive galaxies in overdense regions are mainly fed by gas accretion along filaments and mergers with gas-rich satellites, enabling them to reach SFRs of a few 100\,\myr\ \citep{dekel_2009l,dave_2010,narayanan_2015,lovell_2021}. Thus, massive $z=1-3$ starbursts may represent the extreme extension of normal accretion-fed discs. But most simulations and empirical models struggle to reproduce the SFRs of $\gtrsim 10^3$\myr measured in DSFG-rich overdensities \citep{bassini_2020,lim_2021,remus_2023}. Instead, these top-percentile starbursts are generated in high-resolution zoom-in simulations through gas-rich major mergers \citep{hopkins_2013, sparre_2016,lower_2023}. Bridging these two scenarios, semi-analytic models find that the brightest, most starbursting DSFGs -- which reside in rare overdensities -- are half merger and half disc-instability-induced \citep{lagos_2020,araya_2025}, although the trigger of these disc instabilities is not modelled directly. Several simulation-based studies also find environmental phenomena like tidal effects and ram pressure to play a major role in enhancing SFRs in dense environments \citep{kapferer_2008,bekki_2014,lee_2020}, in part by making these discs less stable. Without high-resolution studies to compare against, theory cannot converge on what fraction of these starbursting systems represent (i) the extreme extension of normal, accretion-fed discs, (ii) brief, merger-driven starbursts, or (iii) a more environment-specific mode of star formation.

            To test which star-formation modes dominate in $z=1-3$ clustered starbursts, it is critical to resolve their cold-gas kinematics and morphology. Starbursts triggered through major mergers typically exhibit highly disturbed velocity fields and irregular morphologies; starbursts fed by smooth, symmetric accretion (aligned with the disc) tend to show rotation-dominated kinematics and more symmetric structures; starbursts shaped by environmental effects (e.g. tidal interactions, ram-pressure stripping, harassment) often produce skewed/lopsided morphologies and velocity fields \citep{bournaud_2005,shapiro_2008,dekel_2009l,bellocchi_2012,yozin_2014}. Recent ALMA studies resolving cold gas on 1.4--2.5\, kpc scales have provided strong evidence of rotating discs in several $z=2.1-4.3$ proto-/clusters \citep{lee_2019, rizzo_2023, venkateshwaran_2024, umehata_2025_kin,wang_2025,pensabene_2025}, suggesting that at least some of the embedded starbursts represent the extreme, high-mass extension of ``normal'', accretion-fed discs. However, many of these appear significantly disturbed, hinting at the prevalence of interactions/minor mergers. Moreover, several proto-/clusters -- including some of those hosting rotating discs -- have been found to host multiple ongoing major mergers \citep{tadaki_2014,hine_2016,coogan_2018,umehata_2021}. Together, these observations paint a mixed picture of what processes drive such rapid stellar mass build up. 
            
            Beyond improved statistics, one of the main challenges in determining what drives the starbursting cores of $z=1-3$ overdensities has been in resolving several members of the same system, up to the individual galaxy outskirts. Only a handful of such galaxies have sufficiently deep and high-resolution observations to kinematically classify them and, if they are discs, constrain whether they are rotation- or dispersion-dominated \citep[as in][]{umehata_2025_kin,pensabene_2025}. This has made it hard to systematically test what drives their high SFRs. To address this challenge, we analyse deep and high-resolution observations of cold gas and dust within the starbursts at the heart of the $z=2.41$ overdensity HATLAS J084933.4+021443 \citep[studied in][]{ivison_2013,Gomez_2018,ivison_2019,gomez_2025}, suggested to be a protocluster core. The individual SFRs of the five known member galaxies span 600--3500 \myr, with a combined total SFR of $\sim7400$ \myr \citep{ivison_2013}, making HATLAS J0849 one of the most highly star-forming protocluster cores detected so far \citep{alberts_noble_2022}. Rotation-dominated gas kinematics were reported for W and T in \cite{Gomez_2018,gomez_2025} on the basis of $0\farcs3$ resolution CO(7--6) and \ci\,2–1 observations. Yet, these observations did not achieve sufficient depth (or resolution) to probe the extended, potentially flat part of the rotation curve. \cite{rizzo_2023} also find W (their ID18) to be a rotation-dominated disc based on resolved \ci\,1--0 emission (albeit with the same limitations for the rotation velocity), whereas the classification for T was uncertain. We hereby push these efforts into a new regime, analysing deep, $0\farcs15$-resolution observations of CO(4--3), \ci\,1–0, and the underlying dust-continuum emission of all five member galaxies to unambiguously classify and constrain their kinematic and morphological properties. 
           
            In Section~\ref{sec:sample_obs}, we discuss the reduction and imaging of these new data. In Section~\ref{sec:kin_analysis}, we describe the kinematic analysis. We discuss the results and implications of our analysis in Section~\ref{sec:discussion} and summarise the findings in Section~\ref{sec:summary}. Throughout this work, we adopt a flat $\Lambda$CDM cosmology, consistent with \cite{planck_2020}, and a Chabrier IMF \citep{chabrier_2003}.

    \begin{table*}
          \centering
             \caption{Properties of HATLAS J084933.4+021443 member galaxies \label{tab:source_prop}} 
             \begin{tabularx}{\textwidth}{@{}>{\raggedright\arraybackslash}X *{5}{>{\centering\arraybackslash}X}@{}}
             \toprule
             Property    &   HyLIRG-W   &   HyLIRG-T   &   ULIRG-C   &  ULIRG-M   & ULIRG-E \\
             \midrule 
             RA (J2000)         &  \phantom{+}08:49:33.59    &   \phantom{+}08:49:32.96    &   \phantom{+}08:49:33.91    &  \phantom{+}08:49:33.80    &  \phantom{+}08:49:32.87 \\
             DEC  (J2000)        &  +02:14:44.6\phantom{0}    &  +02:14:39.7\phantom{0}    &  +02:14:45.0\phantom{0}   & +02:14:45.6\phantom{0}    &  +02:14:53.1\phantom{0}\\
             \midrule
              \textbf{Literature$^{a}$}  &      &      &      &    &  \\
             \midrule
             $\log(L_\mathrm{IR}/L_\odot)$         &  $13.52\pm0.04$    &   $13.16\pm0.05$   & $12.8\pm0.2$    &  $12.9\pm0.2$   &  $12.8\pm0.2$ \\
             $\log(M_\star/\mathrm{M}_\odot)$          &  $11.38\pm0.12$    &  $11.01\pm0.12$    &  $10.36\pm0.18$ &    ---   & $10.3\pm0.2$ \\
              $\log(M_\mathrm{mol}/\mathrm{M}_\odot)$$^b$         &  $11.04\pm0.05$    &  $10.92\pm0.06$    &  $10.25\pm0.07$   &   $10.11\pm0.09$  & $10.8\pm0.2$ \\
             \midrule
             SFR$_\mathrm{IR}$$^c$ (\myr)         &  $3600\pm300$    & $1600\pm200$     &   \phantom{0}$680\pm320$  &  \phantom{0}$860\pm400$    & \phantom{0}$680\pm320$ \\
             \midrule
              \textbf{This work}  &      &      &      &    &  \\
             \midrule
              $z_\mathrm{CO(4-3)}$      &  $2.4074\pm0.0002$  &   $2.4090\pm0.0002$     &   $2.4145\pm0.0003$     &   $2.4177\pm0.0003$     &  $2.4121\pm0.0004$   \\
             \bottomrule
             \end{tabularx}
             \vspace{0.5cm}
                \begin{minipage}{\textwidth}
                \small
                \textbf{Notes:} \\
                $^{a}$ Literature values taken from \cite{ivison_2013} for galaxies W, T, C and M and from \cite{ivison_2019} for galaxy E. Values for HyLIRG-T are corrected for a small magnification of $\mu=1.5\pm0.2$. Stellar masses were derived by fitting the available photometry with \texttt{magphys}. \\
                $^{b}$ determined from the CO(1--0) luminosities assuming $\alpha_\mathrm{CO}=0.8$ (K km s$^{-1}$ pc$^2$)$^{-1}$ \citep{ivison_2013,ivison_2019}: we investigate the assumed $\alpha_\mathrm{CO}$ and hence $M_\mathrm{mol}$ in a follow-up paper. \\
                $^{c}$ SFR derived from the IR luminosity by applying the relation from \cite{kennicutt_1998} and correcting to a Chabrier IMF. \\
                \end{minipage}
             \label{tab:line}
         \end{table*}

    \section{Observations and data products}
        \label{sec:sample_obs}

         \begin{figure*}
            \centering
            \includegraphics[width=0.85\textwidth,trim={0.2cm 2.5cm 0.05cm 2.2cm},clip]{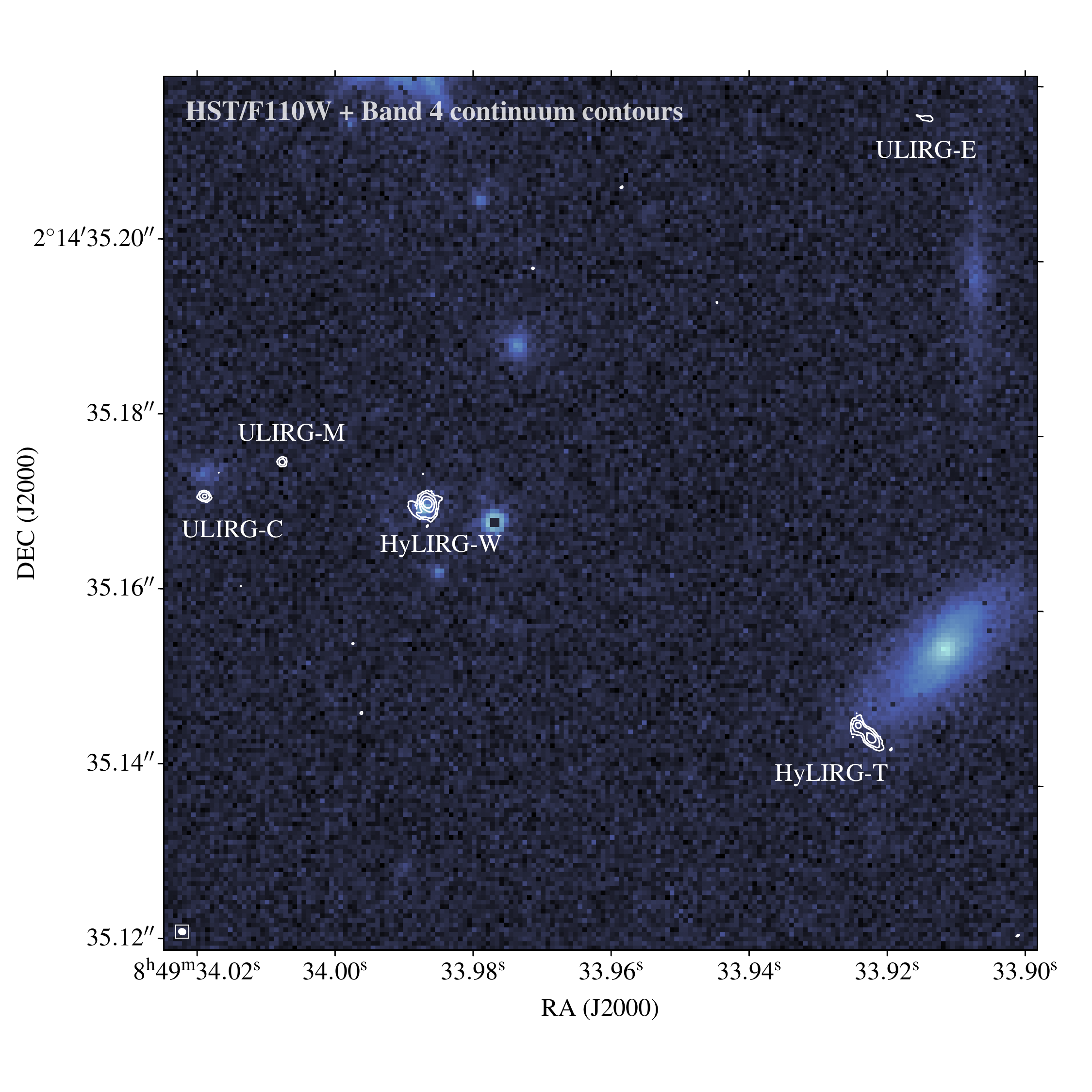}
            \caption{$9\farcs5 \times 9\farcs5$ cutout of HATLAS J084933.4+021443 comparing the HST/110W stellar emission (background colour map) vs ALMA/Band 4 dust-continuum emission (white contours, in steps of $2^n\sigma$, where $n=2,3,4,5,...$). The five known member galaxies are labelled in white. The ALMA beam is shown in the bottom left of the image. 
            }
            \label{fig:hst_cont}
        \end{figure*}

        \subsection{Targets: the galaxies in HATLAS J0849} 

             Originally detected in the Herschel-ATLAS extragalactic survey \citep{Eales_2010}, HATLAS J084933.4+021443 was initially flagged as an exceptionally IR-bright lens candidate. Follow-up CO(1–0) observations with the Green Bank Telescope instead revealed it to be an intrinsically luminous system at $z=2.41$ \citep{Harris_2012}, motivating extensive follow-up observations of the CO(1--0), CO(3--2), CO(4--3), \ci 1--0, CO(7--6), \ci 2--1 and underlying dust-continuum emission of this system \citep{ivison_2013,Gomez_2018,ivison_2019,gomez_2025}. These extensive submillimetre observations have unveiled the presence of five gas-rich, starbursting galaxies across a 15~\arcsec$\times$ 15~\arcsec region on sky, labelled W, T, C, M, and E \citep{ivison_2013,ivison_2019}. 
             
             The well-sampled optical-to-radio spectral energy distributions (SEDs) of the five member galaxies imply high star formation rates (SFRs) of 600--3500 \myr\ and the CO(1--0) observations imply molecular gas masses of 1.8--11$\times 10^{11}$ \msun.\footnote{We test these values, including deriving $\alpha_\mathrm{CO}$ in a companion paper.} The two brightest galaxies, W and T, are both Hyper Luminous Infrared Galaxies (HyLIRGs, $L_\mathrm{IR} \geq 10^{13}~\mathrm{L}_\odot$), whereas M, C, and E are Ultra Luminous Infrared Galaxies ($L_\mathrm{IR} \geq 10^{12}~\mathrm{L}_\odot$). Of the five, only HyLIRG-T is reported as being lensed, albeit marginally \citep[$\mu\sim1.5$, $\mu\sim2.82$:][]{ivison_2013,bussmann_2013}.\footnote{We discuss the potential impacts on the kinematic analysis in Sec.~\ref{sub:kin_classification} and account for the factor of $\mu=1.5\pm0.2$ when quoting the IR luminosity and SFR in Table~\ref{tab:source_prop}.} Only HyLIRG-W is confirmed as hosting an AGN, based on both X-ray observations and rest-frame optical observations, exhibiting an extremely broad H$\alpha$ line \citep{ivison_2019}.

         \subsection{ALMA observations}

            In this work, we mainly analysed the ALMA Band 4 observations from program 2023.1.00714.S (PI: F.~Valentino), although we also imaged the data from program 2018.1.01146.S (PI: N.~Nagar) used in \cite{rizzo_2023}. Both programs cover the full HATLAS J0849 field in a single pointing, including the five known member galaxies. Observations for 2023.1.00714.S were conducted in the C43-8 configuration in six execution blocks between October 21 and 28, 2023, with a minimum (maximum) antenna spacing of 85\,m (8283\,m). The amplitude and phase calibrators were J0750+1231 and J0839+0104, respectively. Observations for 2018.1.01146.S were conducted in the C43-7 configuration in two separate execution blocks on August 15, 2019, with a minimum (maximum) antenna spacing of 41\,m (3638\,m). The amplitude and phase calibrators were J0750+1231 and J0845+0439, respectively.  %
                
            The total on-source integration times for programs 2023.1.00714.S and 2018.1.01146.S were 3.56 and 1.28 hours, respectively. For program 2023.1.00714.S, the correlator was set up to cover the CO(4--3) lines of all member galaxies in the lower sideband, covering \ci\,1--0 for all but ULIRG-M in the upper sideband (although ULIRG-C and E are only barely covered). The observations were centred at RA=08:49:33.336, DEC=+02:14:44.520, and  $\nu$=139.95\,GHz. For the 2023.1.00714.S data, both sidebands were taken in FDM mode, with a spectral resolution of 1.953125 MHz ($\sim4.3$\,\kms\ for CO(4--3) at $2.41$). In contrast, program 2018.1.01146.S fully covered the \ci\,1--0 transitions of all five galaxies, but only fully covered CO(4--3) for ULIRGs-M (with ULIRG-C partially covered). These observations were centred at RA=08:49:33.400, DEC=+02:14:43.000, and $\nu$=139.18\,GHz, (i.e. at a spatial offset of $1\farcs8$ and frequency offset of 0.77\,GHz from the observations of 2023.1.00714.S). The lower sideband was taken in Frequency Division Mode, with a channel resolution of 1.953125 MHz, whereas the upper sideband (covering \ci\,1--0) was taken in Time Division Multiplexing mode, with a native channel resolution of 15.625 MHz. 

        \subsection{Data reduction and imaging}
              
            Both sets of ALMA data were consistently reduced and calibrated by the European ALMA Regional Centre's Calibrated Measurement Set service using the Common Astronomy Software Applications (CASA; McMullin et al. 2007) with the standard pipeline script. CASA versions  6.5.4-9 and 5.6.1-8 were used for 2023.1.00714.S and 2018.1.01146.S, respectively. We then imaged the two sets of data in several stages using CASA v6.6.5, concatenating the execution blocks from each program first. We imaged the 2023 data over each sideband at $8\times$ the native resolution to determine the approximate line widths from the higher S/N line: CO(4--3). Based on these, we performed continuum subtraction with \texttt{uvcontsub}, excluding the identified lines (i.e. omitting 700--1200\,\kms\ around the line centres). To create the continuum image, we cleaning down to $1\sigma$ to ensure our model flux was consistent with the cleaned image. We made one continuum image for all sources, omitting channels with line emission from all five targets (rather than making a continuum image for each source and omitting the lines for that source only). Due to the large band- vs line widths, this approach gave the same rms and flux values as when cropping the line emission and imaging source-by-source; that is, the rms values were within 4\% and the flux values were fully consistent. Although we tested different robust weightings, the continuum images shown here were created with a natural (robust=2) weighting. 
            
            For the observations from 2023.1.00714.S, we generated continuum-subtracted emission-line cubes at two spectral resolutions: $4\times$ native ($\sim 17$\,\kms), and $8\times$ native ($\sim35$\,\kms). We used these CO(4--3) cubes for kinematic modelling, as described in Sec.~\ref{sec:kin_analysis}. Using the coarser resolution cube, we extracted CO(4--3) spectra from circular apertures determined via a curve-of-growth method (with radii of $0\farcs5$ to $1\farcs3$). We fitted these spectra with a single or double Gaussian (depending on which provided the best fit) to determine the redshifts and line widths used to generate moment maps and position-velocity diagrams (PVDs). These values are provided in Table~\ref{tab:source_prop}. For HyLIRGs-W and -T we fitted the \ci\,1--0 spectra generated from the 2023.1.00714.S data, forcing the redshift to be within $\delta z=0.0004$ of the best-fit redshift for CO(4--3). We found that the CO(4--3) line widths are $1.05\times$ the \ci\,1--0 line widths. We used this to inform the spectral region for the \ci\ moment maps of all five galaxies. The best fits to the CO(4--3) global profiles and spectral regions encompassing $\sim$90\% of the flux are shown in Fig.~\ref{fig:spectra}. Unless otherwise stated, we performed our analysis on cubes created with a robust weighting of 2 (natural weighting) and a cleaning threshold of $1\sigma$.

         \subsection{Creating masks and moment maps}
            \label{sub:masking}

            We first generated moment-0 maps individually for each galaxy and emission line by imaging with \texttt{tclean} over a single channel with a width covering 90\% of the line flux, a range chosen to balance S/N yet keep important features in the wings of HyLIRG-W and -T's line profiles. 
            Imaging the full line as one channel can make the S/N easier to interpret because it concentrates the line flux into a single deconvolution problem with one PSF, ensuring that the noise behaves more like a straightforward $\sqrt{\left(\mathrm{bandwidth}\right)}$ average rather than being mixed with channel-to-channel \textsc{clean} and beam variations. 
            
\begin{figure*}
                    \centering
                    \includegraphics[width=\textwidth,trim={2.4cm 1.8cm 3.5cm 1cm},clip]{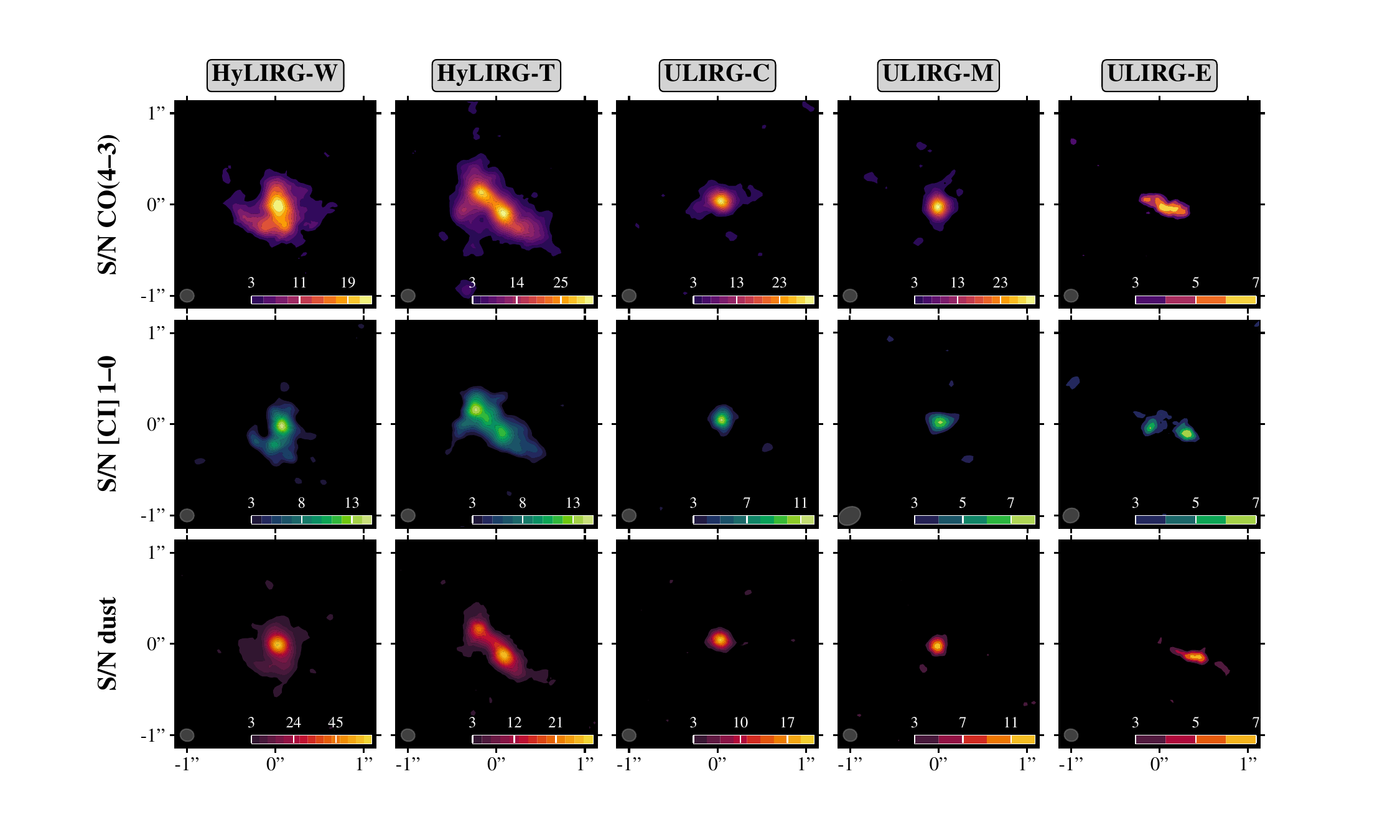}
                    \caption{Columns (from left to right): unmasked CO(4--3) moment-0, \ci\,1--0 moment-0, underlying dust-continuum emission maps, represented in terms of the signal-to-noise ratio (S/N). The moment-0 maps were made by imaging a single channel over 90\% of the line emission. 
                    }
                    \label{fig:lines_and_dust}
                \end{figure*}

            We used only the 2023.1.00714.S data to create the CO(4--3) moment-0 maps for all but ULIRG-M, for which we combined the 2018 and 2023 data, as this was the only source for which CO(4--3) was fully covered by the 2018 data. To generate the \ci\,1--0 moment-0 maps of HyLIRGs-W and -T we used only the 2023 data, thereby better matching the spatial resolution of the CO(4--3) emission. For ULIRGs C and E, for which the 2023 data barely cover 90\% of the predicted \ci\,1--0 linewidth, we created moment-0 maps using both the 2018 and 2023 data. Since \ci 1--0 was not covered for ULIRG-M in the 2023 data, we only imaged the 2018 data. The final moment-0 maps are shown in Fig.~\ref{fig:lines_and_dust}.

            To create masks for the kinematic modelling and moment-1 and 2 maps we also implemented a PYTHON-based algorithm that identifies coherent source emission, associated with a $>3.5\sigma$ peak in three dimensions \citep[see also][]{leroy_2021}. This masking process starts by identifying all $>3.5\sigma$ peaks within a user-defined region-of-interest, after which these peaks are expanded (i.e. the mask grows) in RA, DEC, and velocity until the $2\sigma$ boundaries are reached. To avoid sharp pixelated cutoffs, we smoothed the masks by convolving them with a Gaussian kernel of 1.5--2$\times$ the beam FWHM (and applied the same smoothing kernel to the frequency axis). Our approach is similar to the ``SMOOTH\&SEARCH'' option implemented in the kinematic modelling software \bbarolo\ \citep{3dbarolo}, which we use in Sec.~\ref{sec:kin_analysis}. The main difference appears to come from the smoothing step. Because convolution spreads non‑zero mask values arbitrarily far from detected emission, we applied a threshold to the smoothed mask to retain pixels with sufficient spatial support and prevent the inclusion of noise-dominated regions. We chose a threshold of 0.2--0.25 as this included pixels within $\lesssim1$ beam width of the originally identified regions, ensuring contiguous emission structures and avoiding extrapolation into regions dominated by noise. This threshold appears to be higher than in \bbarolo's smoothing function. 
            
\begin{figure*}
            \raggedright
            \includegraphics[width=0.6\textwidth,trim={1.2cm 1.4cm 3.cm 1.cm},clip]{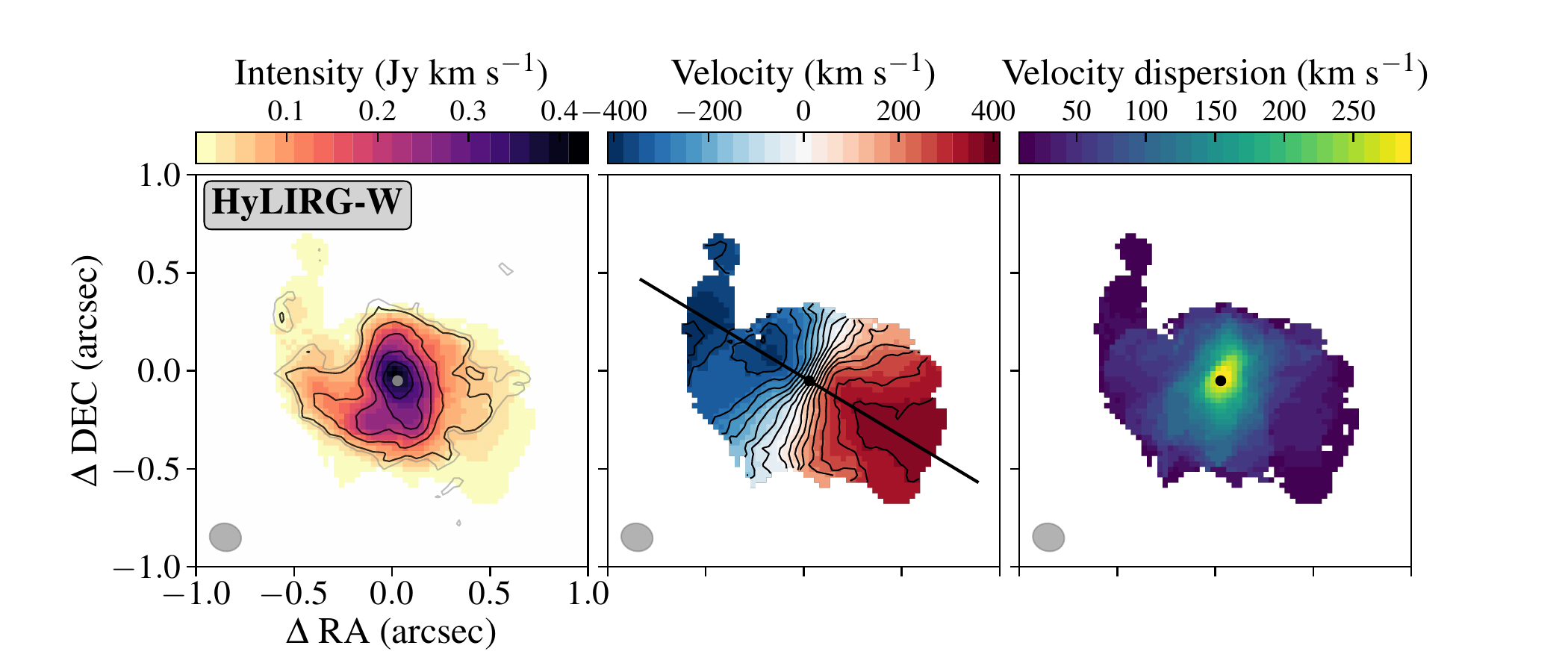}
            \raisebox{-0.2cm}{\includegraphics[width=0.39\textwidth,trim={0.2cm, 0cm 0.2cm 0cm},clip]{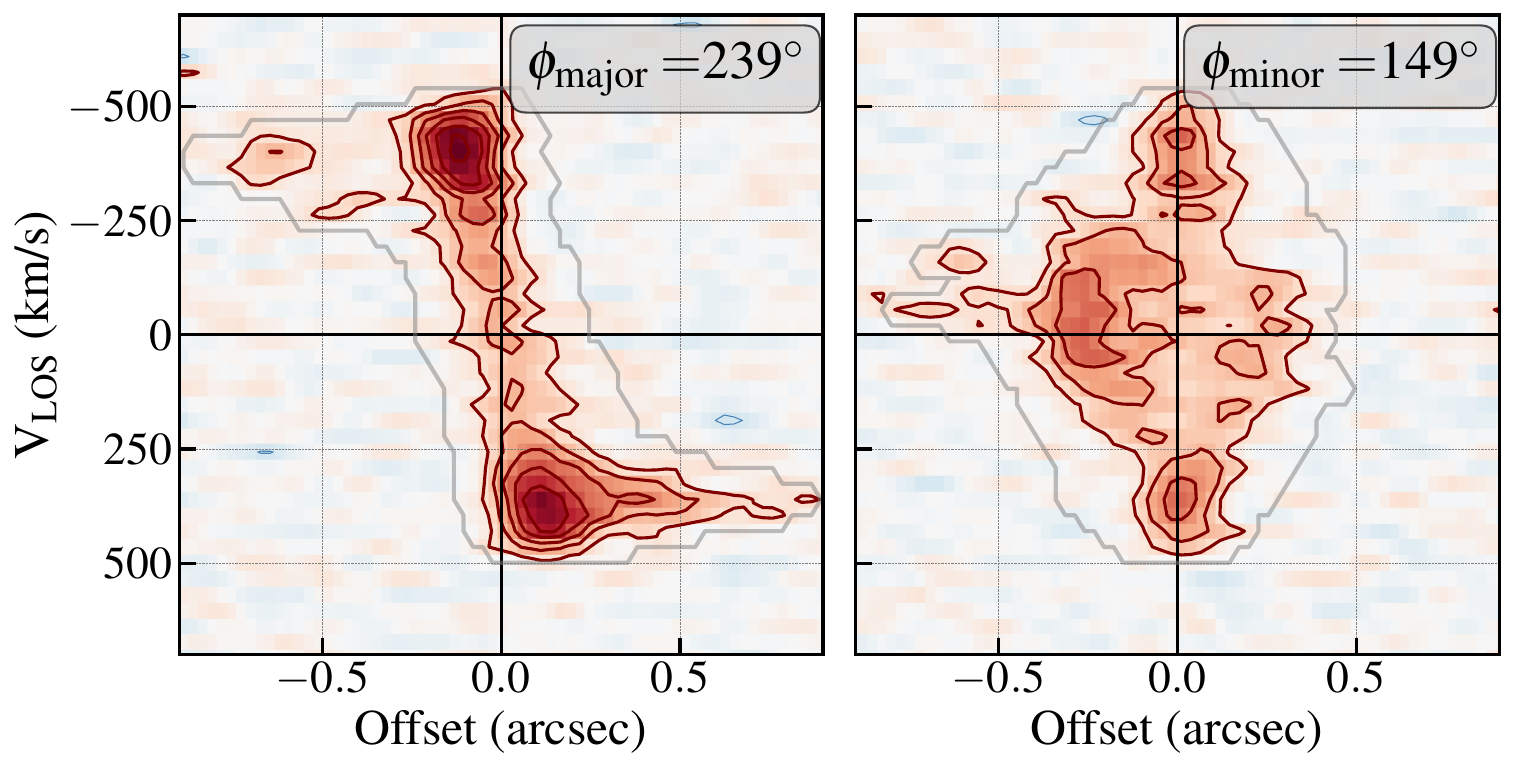}}
            \\
            \includegraphics[width=0.6\textwidth,trim={1.1cm 1.3cm 1.8cm 1.5cm},clip]{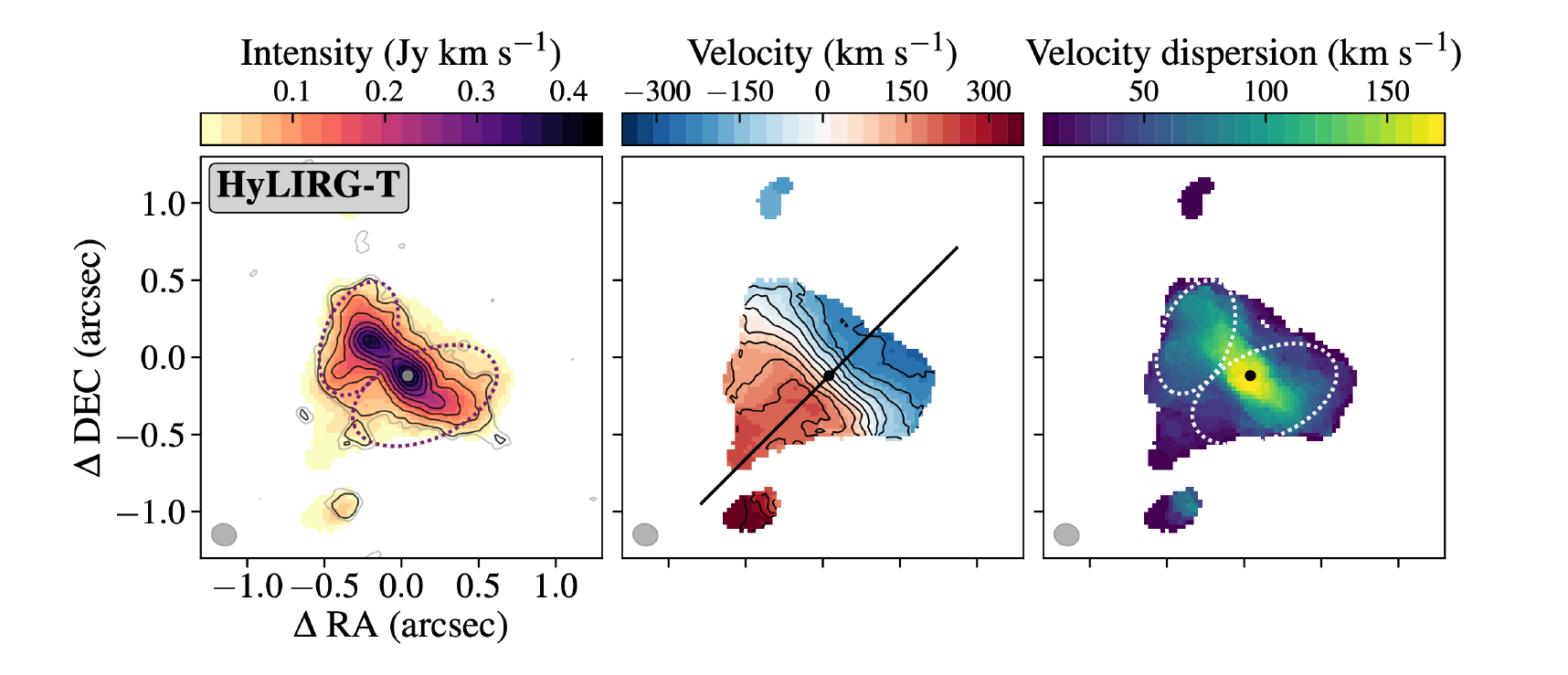}
            \raisebox{-0.1cm}{\includegraphics[width=0.39\textwidth,trim={0.2cm, 0cm 0.2cm -1.2cm},clip]{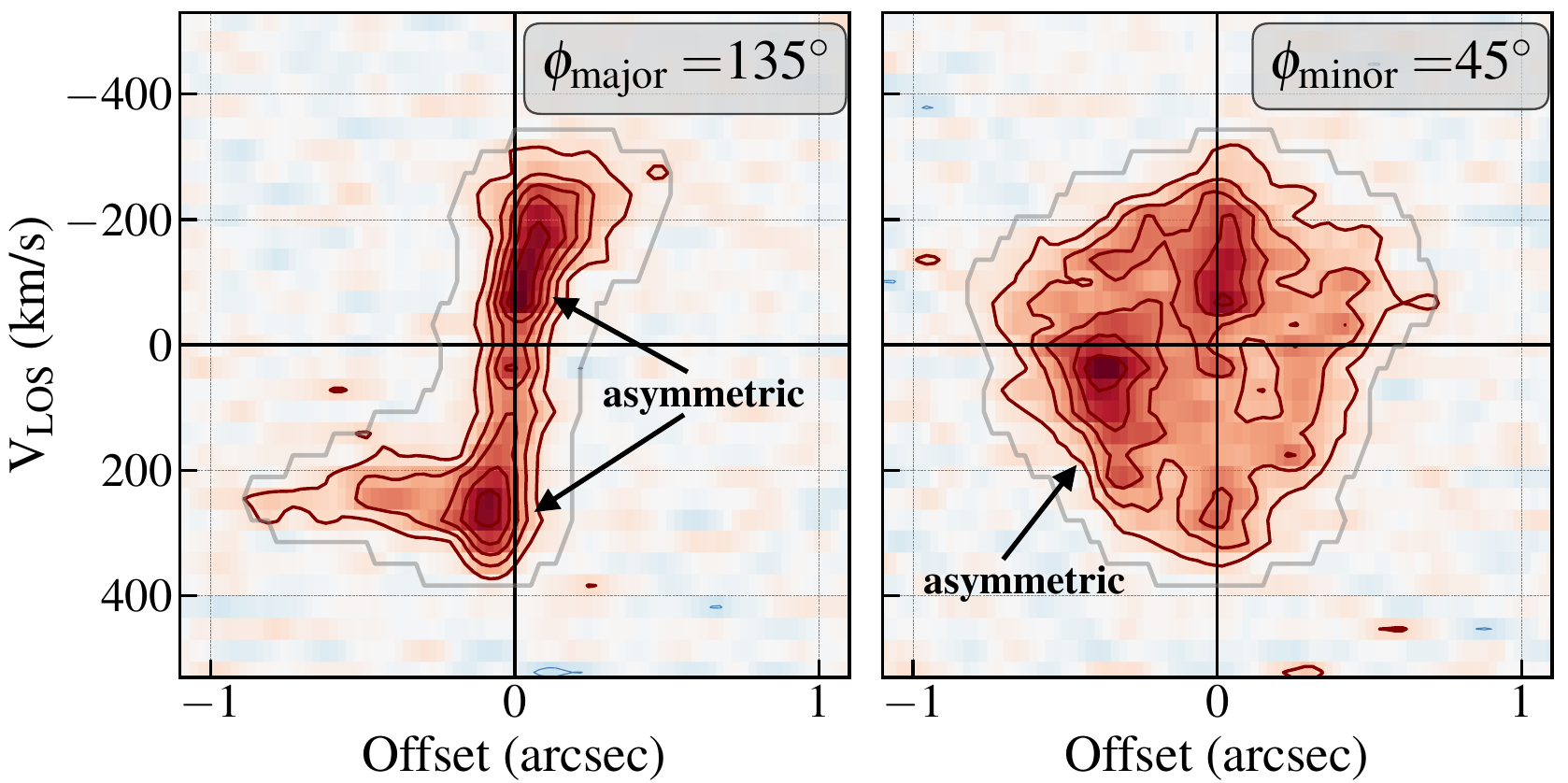}}
            \\
            \includegraphics[width=0.6\textwidth,trim={1.2cm 1.3cm 3cm 2.2cm},clip]{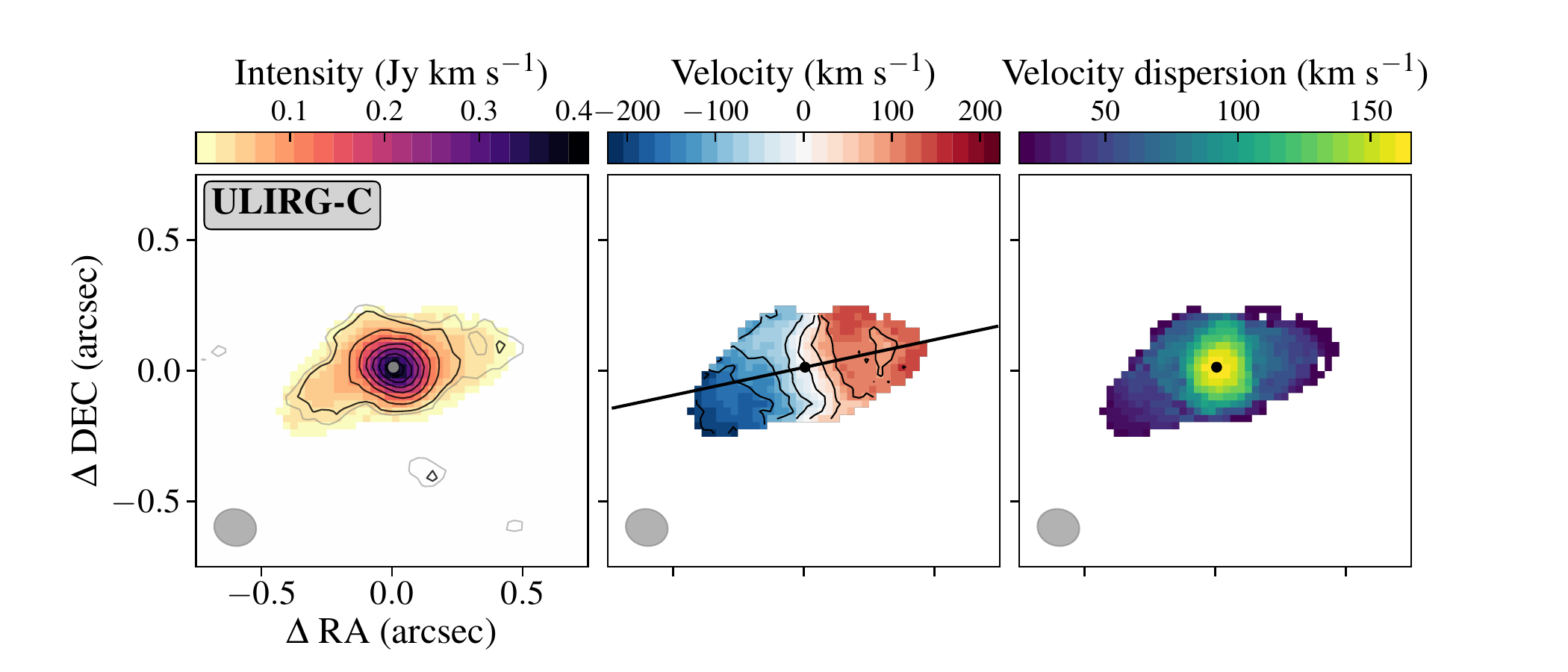}
            \raisebox{-0.1cm}{\includegraphics[width=0.39\textwidth,trim={0.2cm, 0cm 0.2cm -1.5cm},clip]{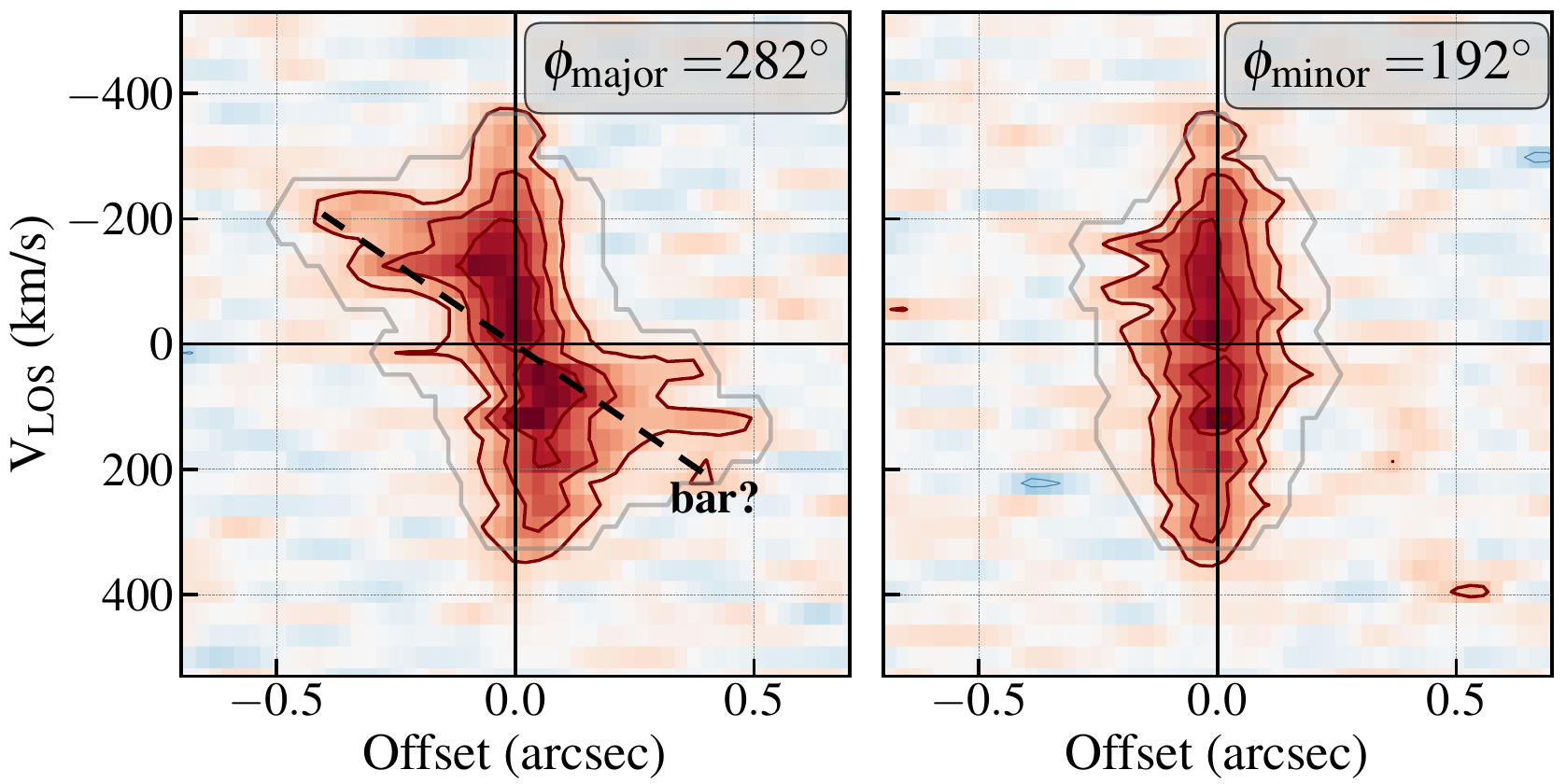}}
            \\
            \includegraphics[width=0.6\textwidth,trim={1.2cm 1.3cm 3cm 2.2cm},clip]{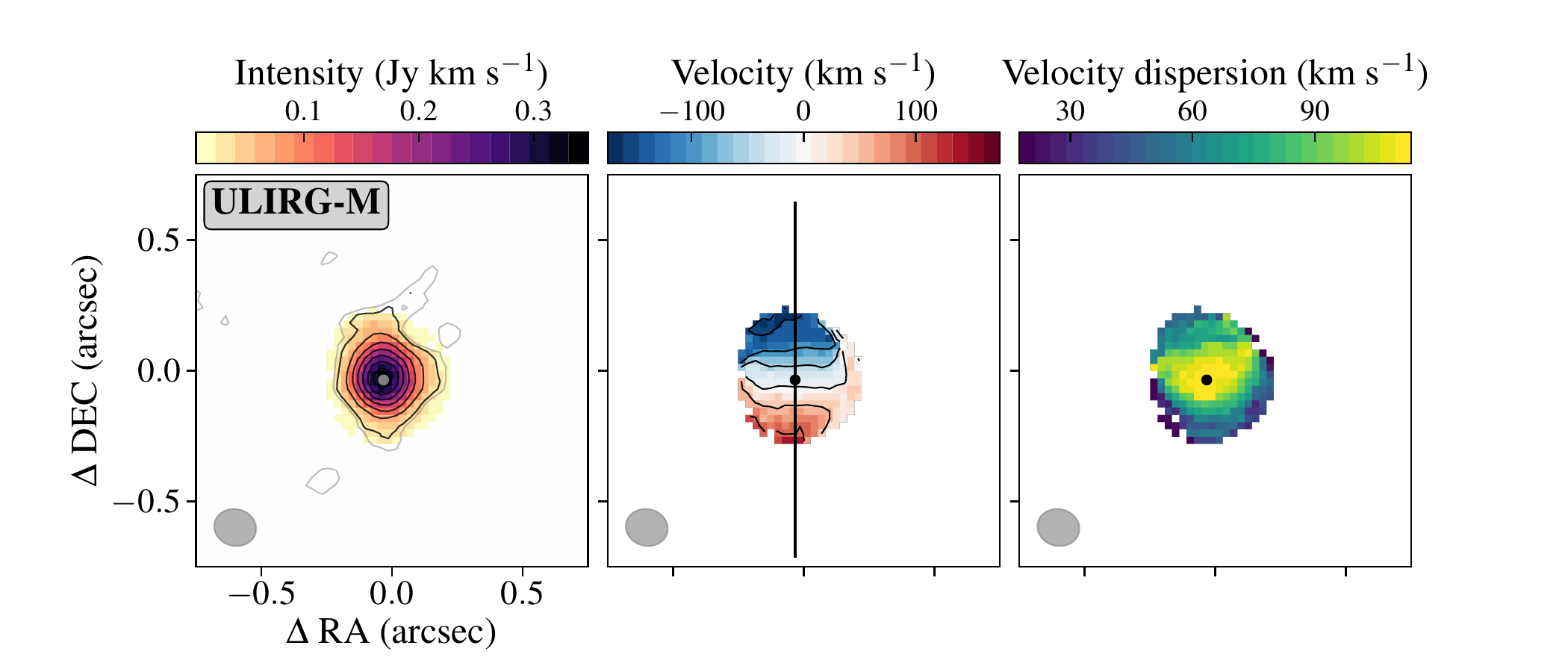}
             \raisebox{-0.1cm}{\includegraphics[width=0.39\textwidth,trim={0.2cm, 0cm 0.2cm -1.5cm},clip]{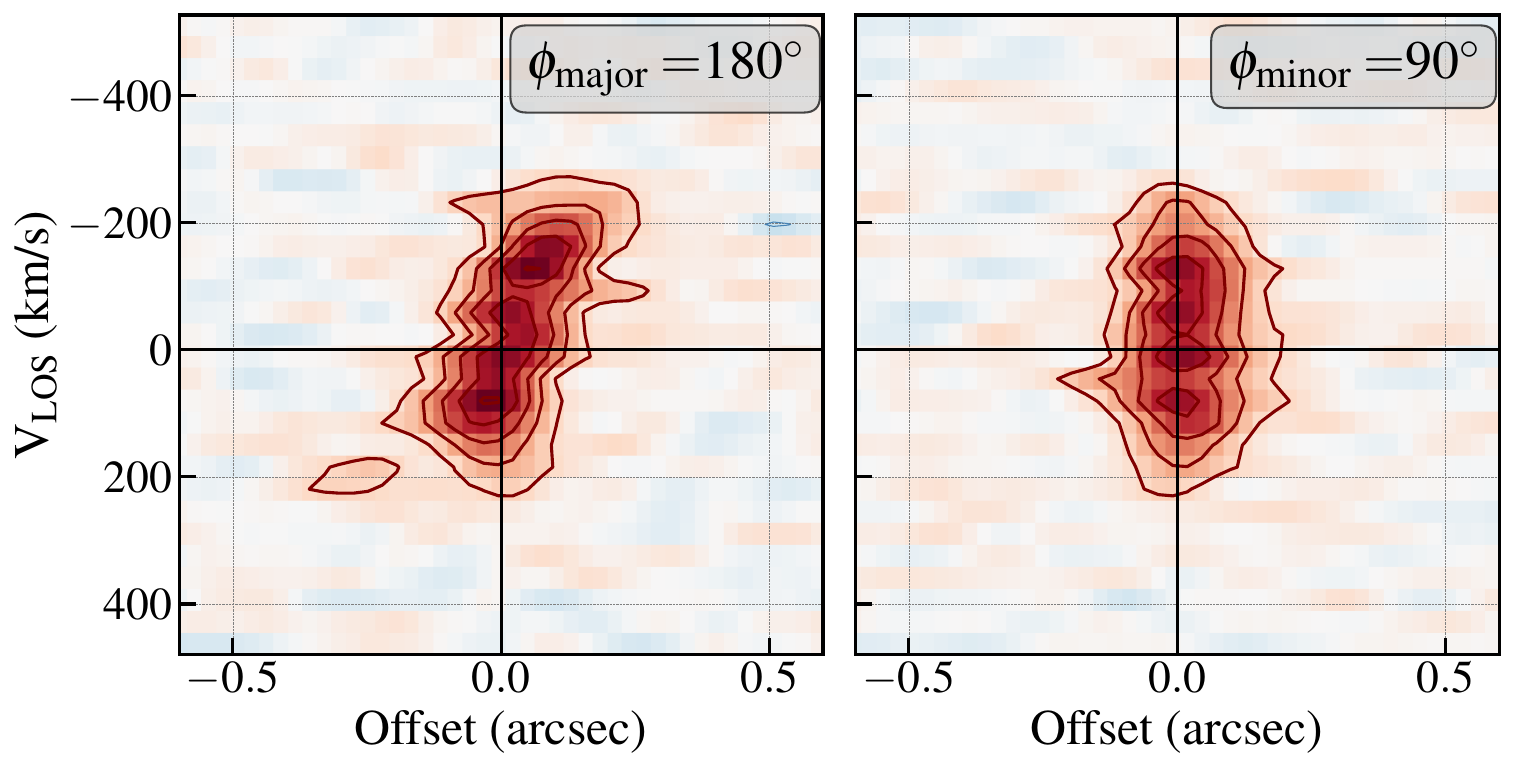}}
             \\
             \includegraphics[width=0.6\textwidth,trim={1.2cm 0.2cm 3cm 2.1cm},clip]{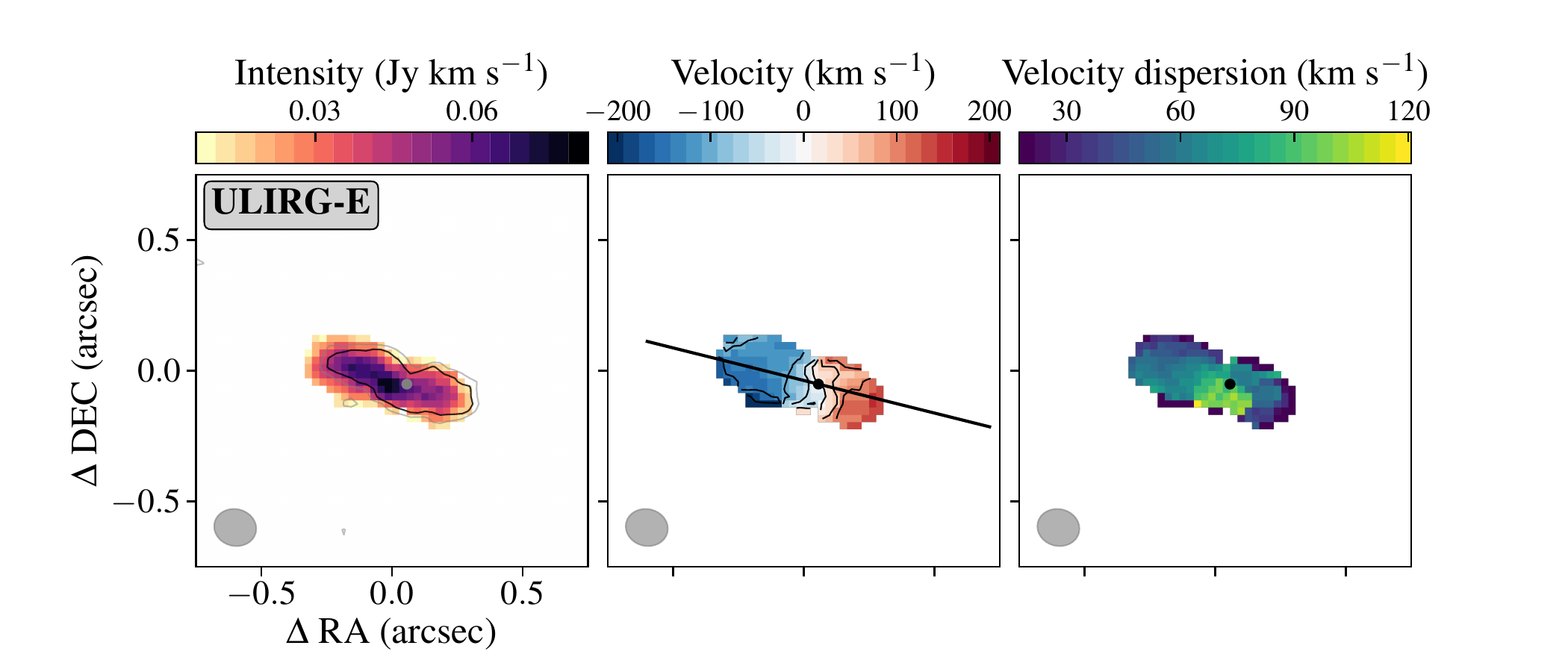}
             \raisebox{0.1cm}{\includegraphics[width=0.39\textwidth,trim={0.2cm, -0.5cm 0.2cm 0.2cm},clip]{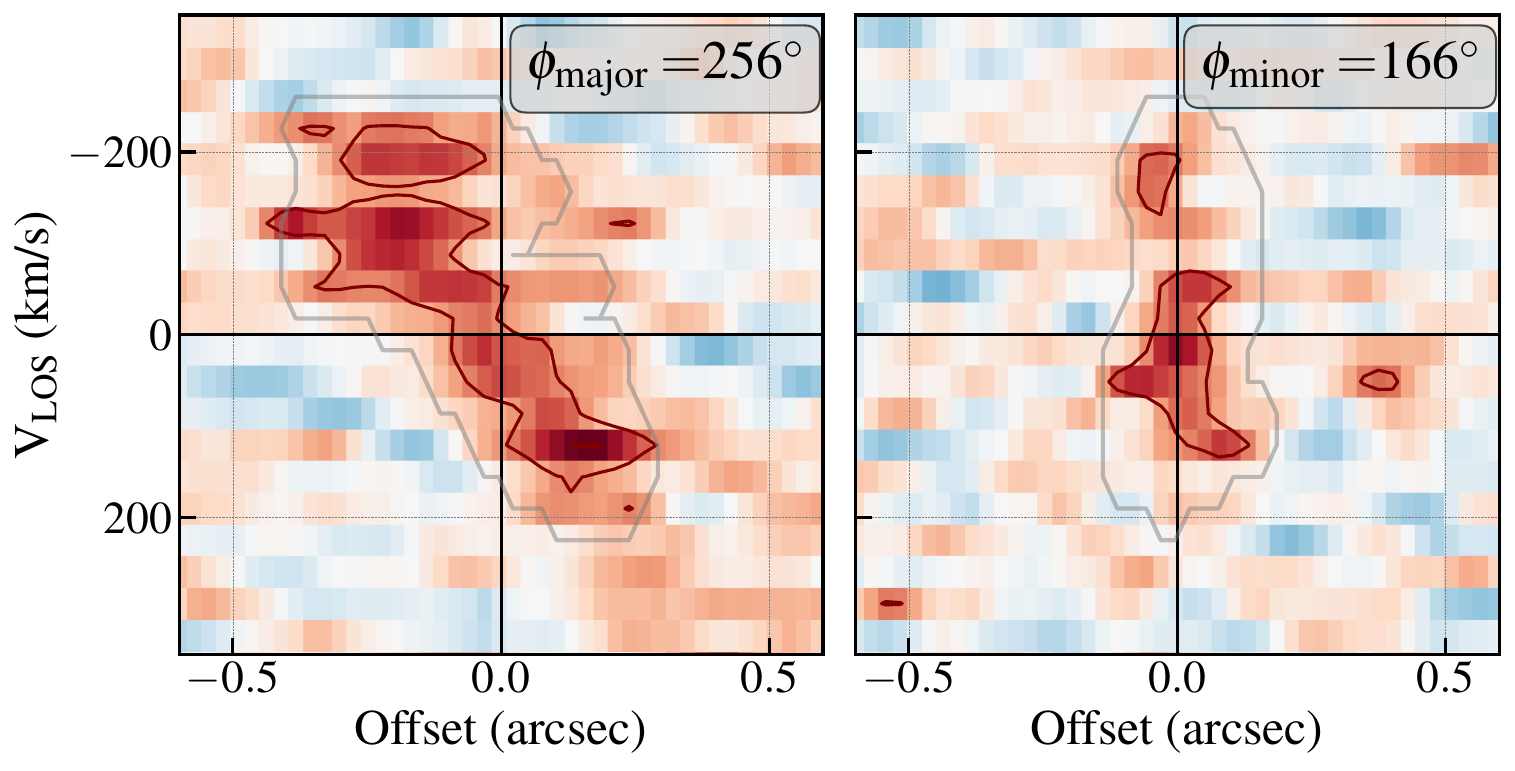}}
            \caption{Each row: CO(4--3) moment-0, 1, 2 maps, major-axis PVD, and minor-axis PVD for the galaxy labeled at left. For HyLIRGs-W and T, we show the CO(4--3) moment maps generated from the emission lines cubes with 4$\times$ the native resolution ($\sim17$\,\kms), whereas for the ULIRGs, we show the moment maps generated from the cubes with 8$\times$ the native resolution ($\sim35$\,\kms). The colormap in the left panel shows the moment-0 maps generated from the masked cubes, whereas the contours show the 1-channel-image moment-0 maps also shown in Fig.~\ref{fig:lines_and_dust}, in steps of 3$\sigma$ (grey) and $4n\sigma$ (black, where $n=1,2,3...$). For consistency, we show the PVDs extracted along the axis shown in the left panels (black line, with centre marked), from the 8$\times$native channel resolution cube, with contours at $3n\sigma$. }
            \label{fig:moments_and_pvs}
        \end{figure*}

            We chose the initial region of interest in which to search for $>3\sigma$ peaks based on the CO/dust peaks\footnote{We choose the brightest peaks for W, C, and M, a location between the two CO/dust peaks for T, and a location between the single CO and dust peak for E. Our results remained consistent when varying these centres by $0\farcs5$} and applied additional options on trimming regions with unphysical values or low S/N in the final moment maps. Our approach resulted in slightly more compact masked moment maps than when using ``SMOOTH\&SEARCH'' with fewer spurious features (e.g. high-dispersion clumps below the beam size, or regions of negligible dispersion and fixed velocity outside the $2\sigma$ contours in the moment-0 map). In general, we find that adopting masks in the kinematic modelling of the imaged data cubes with \bbarolo\ yields models that better fit the data (based on the PVDs and residuals in the channel maps). We show the moment maps for the five galaxies in the left column of Fig.~\ref{fig:moments_and_pvs}. 

    \section{Kinematic analysis}
        \label{sec:kin_analysis}

        \subsection{Kinematic classification}
            \label{sub:kin_classification}

            To determine whether the member galaxies of HATLAS J0849 are rotating discs, mergers, or something else entirely, we rely on the new ALMA data. Unfortunately, the existing observations of the stellar emission are uninformative, as all member galaxies of HATLAS J0849, except HyLIRG-W, are undetected in the one available HST image (F110W, Fig.~\ref{fig:hst_cont}) and there are no JWST observations available yet. In the optical-to-NIR observations from VISTA, Spitzer, and Herschel, the galaxies are either unresolved or barely resolved, providing no clues as to their stellar distribution.  
            
            Even though the velocity fields of the HATLAS J0849 member galaxies show clear positive-to-negative gradients and are fairly smooth, this does not guarantee that the member galaxies are all rotating discs. Several studies have shown that merging systems can exhibit smooth velocity gradients at low resolution \citep{simons_2019,kohandel_2020,rizzo_2022}. Likewise, velocity gradients caused by outflows can be misinterpreted as rotation \citep{loiacono_2019}. To determine whether these systems can be classified as rotating discs, we used both the moment maps and PVDs shown in the left and right columns of Fig.~\ref{fig:moments_and_pvs}, respectively. 

            To extract the PVDs, we first determined the centre and most appropriate kinematic major axis PA. To this end, we took the best-fit straight line through the extrema of the moment-1 velocity field and took the centre as the midpoint, at $v=0$. For HyLIRG-W and C, we refined this midpoint within $<1$ pixel to ensure the major axis PVD was as symmetric as possible (about the central position and velocity), judging the symmetry both by eye and using the tool PVSplit \citep{rizzo_2022}. For ULIRG-M, we also refined the midpoint by $<1$ pixel to better match the peak flux and dispersion (shifting the centre slightly up and to the left).   

              \begin{figure}
                    \centering
                    \includegraphics[width=0.45\textwidth,trim={0.2cm 0.3cm 0.2cm 0.2cm},clip]{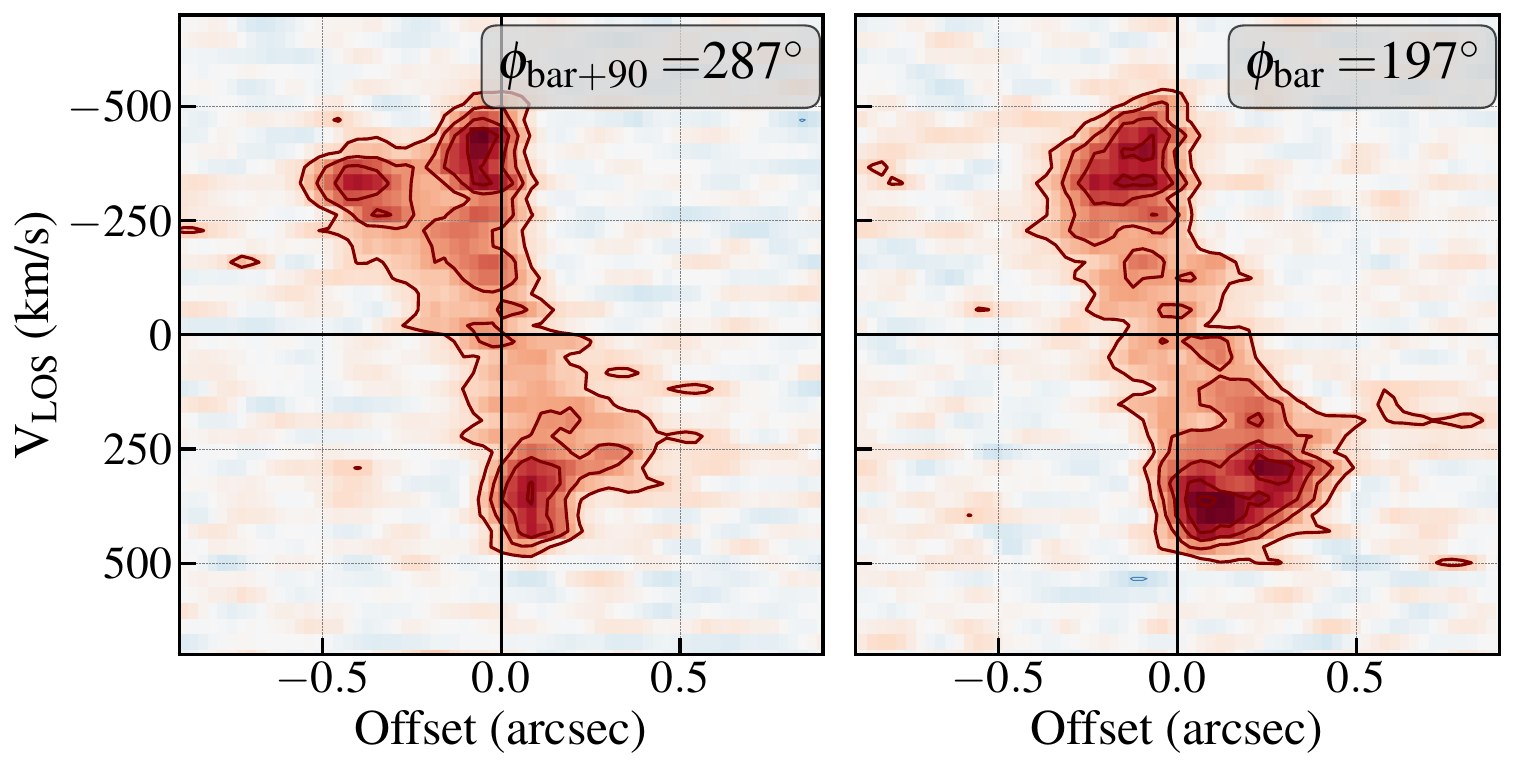}
                    \caption{PV diagrams along the potential bar axis of HyLIRG-W (right) and perpendicular to it (left). The left-hand PV exhibits the X-shape typical of a barred galaxy. \label{fig:pv_bar_w} }
                \end{figure}

            \subsubsection{The two HyLIRGs}
                \label{subsub:HyLIRG_classification}

                HyLIRG-W exhibits a clear spiral structure in both the line and continuum emission, as expected for a rotating disc (Fig.~\ref{fig:moments_and_pvs}, top). The PVDs for HyLIRG-W (top right panels, Fig.~\ref{fig:moments_and_pvs}), also exhibit the typical features of a rotating disc, with an S-shaped major-axis PVD and a diamond-shaped minor-axis PVD (albeit an asymmetric one). Moreover, the moment-2 map shows a single peak aligned with the centre of the velocity field, and no sharp discontinuity. Thus, all evidence points towards HyLIRG-W being a rotating disc, with no indication of an ongoing major merger. Although HyLIRG-W is a rotating disc, there are several kinematic asymmetries in the velocity field, dispersion map, and PVDs, which point to additional in-disc non-circular motions (rather than galaxy-scale outflows). We discuss these in Sec.~\ref{sub:deviations_axisummetric_rotator}. 
                 
                HyLIRG-T is more complicated to classify. It shows a clear velocity gradient and asymmetric S-shaped major-axis PVD (Fig.~\ref{fig:moments_and_pvs}, second row), but there are several indications that it is not a simple rotating disc. The most obvious is that the position angle of the velocity field is offset by $\sim85^\circ$ from the morphological position angle, tracing the line between the two clumps seen in the CO, \ci, and dust-continuum emission. It therefore looks to be tumbling about its major axis. In addition, the minor-axis PVD is highly asymmetric, indicating the presence of an additional low-velocity component. 

                We rule out that lensing causes the clear mismatch between the morphology and kinematics of HyLIRG-T. Although it is weakly lensed by the foreground lenticular galaxy at $z=0.34$ seen in the HST image in Fig.~\ref{fig:hst_cont} \citep{bussmann_2013,ivison_2013}, the mass, orientation, and proximity of this lenticular galaxy cannot twist the velocity field or flip the morphological axis ratio of HyLIRG-T to the extent that it would be a self-consistent rotating disc. Doing so would require an implausible amount of lensing, such that the two clumps seen in the CO and dust-continuum emission are intrinsically highly elongated along the kinematic PA. Moreover, HyLIRG-T cannot be doubly imaged: the foreground source is too far away to cause such strong lensing (and we find no counter image), there is no evidence of a closer, additional foreground source when performing an isophotal analysis of the HST data, and the moment-2 map indicates one clear peak in dispersion -- we would expect mirrored peak if the CO(4--3) emission were doubly imaged. Thus, we conclude that the strange kinematic vs morphological signatures are intrinsic to HyLIRG-T and explore three potential scenarios:
                \begin{enumerate}[label=(\roman*), leftmargin=*, itemsep=0pt, topsep=0pt]
                    \item a polar ring/disc galaxy, with a velocity field that is significantly tilted compared to the central disc, 
                    \item a highly inclined starbursting disc with a strong outflow, and 
                    \item a late-stage merger.
                \end{enumerate}

                Under scenario (i) -- the most exotic of the three -- the peak CO, \ci, and dust-continuum emission stem from a central, highly inclined disc, whereas the extended velocity field traces the kinematics of a larger, more face-on ring/disc. Kinematic misalignments approaching $\sim90^\circ$ are have been reported in ionised gas and stars in local and simulated polar-ring systems \citep{iodice_2015,smirnov_2024}. However, molecular constraints are limited, and in most well-studied cases the polar structure contains more molecular gas than the central disc/spheroid \citep{watson_1994,schinnerer_2002}, unlike for HyLIRG-T. In an exception to this, NGC 660 hosts molecular gas and star formation in both the central disc and ring \citep{combes_1992,vandriel_1995} and two \hi\ column-density peaks aligned with the central disc \citep{vandriel_1995} -- as for HyLIRG-T. However, unlike for HyLIRG-T, the PV cut along the polar ring peaks at the largest radii and the disc-axis PV is narrow and symmetric. Overall, the polar-ring/disc interpretation appears unlikely for HyLIRG-T given the combination of bright central CO and strongly asymmetric PV morphology. 
                
                If the velocity field shown in the moment-1 map and major-axis PVD traces an outflow (scenario ii), then the outflow velocity is at least $\sim300$\,\kms, a value that is not implausible given its extreme SFR. Outflow velocities of several hundred \kms\ have been measured for local starbursts with similar stellar masses but lower SFRs than HyLIRG-T \citep[e.g.][]{cicone_2014,fluetsch_2019}. However, the major-axis PVDs of such galaxies are consistently more broadened around the spine of the ``S'' than that of HyLIRG-T \citep[e.g.][]{oosterloo_2017}, which instead shows a distinct spine and sharp bend. Moreover, in local galaxies the high-dispersion component is typically co-spatial with the outflow \citep{leroy_2015,aalto_2020}, whereas for HyLIRG-T the dispersion seems highest along the line of CO clumps, indicating a spatial decoupling from the ordered velocity gradient (or the presence of multiple nuclei). Most local starbursts, including M82, also exhibit a more twisted CO velocity field, tracing both the disc and outflow \citep[e.g.][]{leroy_2015}. This twist remains evident for M82 when degrading the \hi\ and CO observations to the data quality of HyLIRG-T. Moreover, M82 reveals a declining major-axis PVD, implying that HyLIRG-T must also have a different outflow vs disc orientation for this scenario to hold. Thus, a single rotating disc plus a single outflow component would need a finely tuned orientation and/or unusually coherent outflow kinematics to reproduce both the narrow, major-axis PV spine and largely untwisted moment-1 field. 
                
                We find more consistent evidence in favour of scenario (iii). Separating the two clumps seen in emission, reveals two distinct kinematic components (marked by the ellipses in Fig.~\ref{fig:moments_and_pvs}). The brightest clump to the south-west exhibits a strong velocity gradient, with a position angle of $\sim130^\circ$, whereas the fainter clump to the north-east appears to be associated with the extended, high-dispersion (and slightly receding) component. Moreover, the moment maps and PVDs of HyLIRG-T appear very similar to those of the Medusa merger \citep{koenig_2014}, a local minor merger. HyLIRG-T exhibits two extended features, one approaching (north-east) and one receding (south-east) -- similar to the molecular tendrils feeding the Medusa. Moreover, the major- and minor-axis PVDs of HyLIRG-T exhibit similar asymmetries: the left-most peak in HyLIRG-T's minor-axis PVD (caused by the small, north-eastern clump) mimics the effects of the western arm of Medusa. Given these similarities, it seems plausible that in the case of HyLIRG-T, a smaller galaxy may be in the late stages of merging from the north-east in an almost coplanar direction, such that there is no distinct receding/approaching velocity contribution and the dominant velocity gradient is that of the larger galaxy. Although we consider this late-stage minor-merger the most likely explanation for HyLIRG-T, upcoming JWST/NIRCam data (PID 6958) will reveal which (if any) of the three scenarios is correct.

        \subsubsection{The ULIRGs}
            \label{subsub:ulirgs_kin_classification}

                ULIRG-C exhibits a clear, slightly asymmetric velocity field and its major-axis PVD is X-shaped, indicative of a disc with central non-circular streaming motions, as is typical for barred disc galaxies \citep[e.g.][]{merrifield_1999,huettemeister_2000,lundgren_2004,hernandez_2005}. The minor-axis PVD is fairly symmetric and narrow, consistent with a thin rotating disc. Such thin, minor-axis PVDs occur for barred galaxies if the bar is oriented such that radial motions do not project along the line of sight \citep{randriamampandry_2015}. This configuration appears to match the velocity field and major-axis PV of ULIRG-C, with the putative bar oriented along the kinematic minor axis. 
                ULIRG-C displays features similar to the disturbed disc ``Freesia'' in \cite{rizzo_2022}, with PVSplit placing the system just within the ``not-a-disc'' regime. 
                This is because the emission peaks are at lower velocities than those enclosed by the contours at larger radii, yielding a high $P_\mathrm{V}$ of $-0.0002$. Likewise, the X-shape leads to emission spanning the left and right sides, yielding a low $P_\mathrm{R}$ of $0.14$. Given the symmetry of the major-axis PV around zero velocity ($P_\mathrm{major}=-3.0$), and the clear X-shape, we conclude that ULIRG-C is most likely a rotating disc with additional streaming motions (we explore this further in Sec.~\ref{sub:deviations_axisummetric_rotator}). 

                Unlike ULIRG-C, ULIRG-M has a disjoint velocity field; the high receding velocity component to the south and east appears separate from the central ellipse where the dispersion is highest. Moreover, the highest S/N region in the major-axis PVD stays close to zero offset and velocity, with no high-velocity peaks offset from the centre (i.e. no S-shape typical of rotation). It is therefore unsurprising that the PVSplit parameters ($P_\mathrm{major}=-2.50$, $P_\mathrm{V}=-0.0003$, and $P_\mathrm{R}=0.13$) also place ULIRG-M well inside the ``not-a-disc'' regime. Based on the clear asymmetry of the major-axis PVD and the lack of an S-shape, we conclude that we cannot classify the component of ULIRG-M that we are seeing as a rotating disc. It may be that the system is face-on and/or that we are only picking up a central bulge-like component, in which case the warped velocity field indicates that the fainter disc must be significantly disturbed. However, based on the moment maps and major-axis PVD, ULIRG-M could instead be either a merger or outflow; it is similar to the mock observations of the simulated galaxy ``Adenia'' in \cite{rizzo_2022}.

                ULIRG-E appears most consistent with being a merger. The moment-1 map looks distinctly like poorly resolved mergers \citep[e.g.][]{rizzo_2022}. Moreover, the CO(4--3), \ci\,1--0 and continuum emission appear significantly offset from each other (with the continuum peak not evident in CO). The PVDs are fairly uninformative, thanks to the low S/N, but are certainly inconsistent with a rotating disc (by eye or using PVSplit). Even accounting for this low S/N, we find it hard to justify a scenario where deeper data would cause these disparate elongated blobs to resolve into a single rotating disc.

        \subsection{Kinematic modelling}
            \label{sub:kinematic_modelling}

            Based on the kinematic classification above, we conclude that HyLIRG-W and ULIRG-C are best described as rotating discs. To quantify the motion of their cold gas, we fitted the CO(4--3) emission-line data cubes using the kinematic modelling software, \bbarolo\ \citep{3dbarolo}, and compared these results to the values fit with \textsc{KinMS} \citep{kinms}. We chose these software because they have been extensively tested on real and mock data, spanning a range of data quality \citep{3dbarolo,rizzo_2022,davis_2017}. The models in both \bbarolo\ and \textsc{KinMS} are described by a set of geometric and kinematic parameters. The geometric parameters used in both models are the central position, $x_0, y_0$; inclination angle, $i$;  and position angle, PA. 
            The kinematic parameters are the systemic velocity, $v_{sys}$; rotation velocity, \vrot; and velocity dispersion, $\sigma$. Although \textsc{KinMS} can also be used to directly fit the dynamics. However, we only applied the kinematic fitting here, as we have poor constraints on the stellar light distribution. We leave the kinematic modelling of the other galaxies (not classified as rotating discs) to a future work, making use of upcoming JWST data to correctly characterise the different components. 

            \subsubsection{Estimating the inclination}
                
                Although both \bbarolo\ and \textsc{KinMS} can jointly fit the geometric and kinematic parameters, we separately estimated the centre and inclination, and fixed these to avoid model degeneracies. To this end, we used the centres fitted in Sec.~\ref{sub:kin_classification}. The inclination is more challenging to constrain, due to its large degeneracy with the rotation velocity. Most $z>1$ studies avoid this degeneracy by fixing the inclination to that estimated from the optical images \citep[e.g.][]{lelli_2016,wisnioski_2019,kaasinen_2020}, but as there are currently no optical or NIR images of sufficient data quality for HATLAS J0849, we instead used the CO(4--3) data directly. 

                For our first estimate of the inclinations, we applied the tool \texttt{Cannubi} \citep{roman-oliveira_2023}, which uses \bbarolo\ to fit the total-flux map or entire cube with resolution-matched 3D tilted-ring models of rotating discs. For ULIRG-C, \texttt{CANNUBI} recovered an inclination of $64^{\circ~+3}_{\phantom{\circ~}-5}$ (and a centre and PA consistent with our estimate in Sec.~\ref{sub:kin_classification}), matching our expectation from the outermost $3-4\sigma$ contours. For HyLIRG-W, conventional approaches provided a poor fit to the contours in the outer disc. When fitting the moment-0 map or whole cube with \texttt{Cannubi}, or fitting a single S\'{e}rsic/exponential profile to the moment-0 map with \texttt{GALFIT}, we recovered a low inclination of $\lesssim30^\circ$ -- biased to the bright inner regions. We obtained the same low value when fitting a parametric model of a rotating disc to the CO(4--3) data cube with \textsc{KinMS}, leaving both the kinematic properties and inclination free (Sec.~\ref{subsub:kin_fits}). 
                This low inclination provides a poor fit to the velocity field and PVD, when performing the full kinematic fits and results in an extremely high rotation velocity of $>750$\,\kms. 
                
                To avoid biasing our results to the central structure, we instead fitted the outer disc contours of HyLIRG-W directly. We measured the source geometry by fitting an ellipse to the moment-0 isophotes at $3\pm1\sigma$, using the largest contiguous contour components near the expected centre to suppress spurious noise islands. To estimate the uncertainties, we performed Monte Carlo realisations: perturbing the centre and position angles within their uncertainties ($\pm1$ pixel, $\pm5^\circ$), refitting the observed (beam-smeared) ellipse, and deconvolving the beam by subtracting the beam covariance matrix from the ellipse covariance to infer the axis ratio. We then converted the axis-ratio distribution to an inclination distribution using an assumed intrinsic axial ratio of $q_0=0.2$. We adopt the median and its difference to the 16th(84th) percentile as the quoted value and lower(upper) uncertainty intervals. This yielded an inclination of $48_{-8}^{+4}\,^{\circ}$.

            \subsubsection{Kinematic fits}
                \label{subsub:kin_fits}
             
                \bbarolo's tilted-ring model model works under the assumption of a thin disc. Assuming that there are no axisymmetric radial motions, the line-of-sight velocity at each radius, $R$, is given by 
                \begin{align}
                    V_\mathrm{los} = V_\mathrm{sys} + V_\mathrm{rot}(R) \cos(\phi) \sin(i)
                \end{align}
                where $\phi$ is the azimuthal angle relative to the plane of the disc.\footnote{We also tested fitting the radial motions, finding that the average of all rings is consistent with zero (spanning $\sim-12\to15$\,\kms), as expected given the non-axisymmetric residuals (Sec.~\ref{sub:deviations_axisummetric_rotator}).} To obtain the best-fit model \bbarolo\ employs a least-square-minimisation approach, where, for each model optimisation step, the model disc is convolved with a Gaussian kernel with the same size and position angle as the beam. We applied the azimuthal normalisation, used an exponent of weight for blank pixels of $bweight=0$, adopted the weighting function, $wfunc=1$ ($|\cos(\theta)|$), and the minimise function, $ftype=2$ (the default, $\left| mod-obs\right|$). We fixed the systemic velocity to 0\,\kms, fixed the centre to that identified from the PVDs, and fixed the inclinations to the values described above. 
                
                We set a lower bound on the velocity dispersion consistent with the instrumental velocity dispersion, $\sigma_\mathrm{inst}$ \citep[see also][]{lelli_2023}; for 8$\times$ the native channel resolution $\sigma_\mathrm{inst} = 0.3 \times \Delta v \approx 10.5$ \kms and set the $linear$ value used in \bbarolo\ to the standard deviation equating to $\mathrm{FWHM}=1$ output channel, namely $1/(2\sqrt{2 \ln(2)})$ (see Appendix~\ref{sub:spectra_response} for a discussion). For the initial values of rotation velocity and dispersion, we used 90\% of the maximum l.o.s velocity and 30\,\kms, respectively. We fixed the disc thickness to be 0.3\,kpc, consistent with the molecular gas scale height of disc galaxies at $z\sim2.4$ simulated with the cosmological simulation COLIBRE (McGregor et al.~in prep.). This value is 10--15\% of the CO(4--3) half-light radii of the two galaxies (Kaasinen et al. in prep.), and is therefore consistent with a thin disc for both galaxies. Using smaller values of disc thickness has a negligible impact on the best-fit \vrot\ and $\sigma$. We chose rings that oversample the beam, with a ring width of $0\farcs09 \approx 0.6 \sqrt(\mathrm{FWHM_{major}}\times\mathrm{FWHM_{minor}})$. For HyLIRG-W, we used seven rings sampling out to $0\farcs7$, whereas for the much smaller galaxy, ULIRG-C, we used four rings sampling to $0\farcs4$. Compared to using beam-width rings, this choice of oversampling enables more flexible best-fit models that capture small features in the rotation curve, consistent with recent z=1--3 studies, which sample the beam by a factor of 0.5--0.8 \citep{rizzo_2023, lelli_2023,roman-oliveira_2023,pensabene_2025}). 

                \bbarolo\ allows the user to choose several masking options. We applied the mask described in Sec.~\ref{sub:masking}, but derive consistent kinematic properties when masking with \bbarolo's SMOOTH\&SEARCH function (which yields more extended masks). We tested that our masking is not too aggressive, in that it effectively crops the broad wings of the line, leading to lower $\sigma$ values \citep[e.g.][]{lee_2025}. To this end, we used \bbarolo's \texttt{SPACEPAR} task. For both galaxies, the values of \vrot\ are well constrained for all rings (for both the 35 and 17\,\kms cubes). However, for HyLIRG-W we find that for the 35\,\kms cube, the $\sigma$ value of the 7th ring is ill-constrained and for the 17\,\kms cube, $\sigma$ is already ill-constrained by the 6th ring. Given these tests, we excluded these outer one (two) rings when calculating the mean and outer $\sigma$ for HyLIRG-W using the 35\,\kms (17\kms) data cubes. For ULIRG-C, we find that all rings remain well-constrained. 

                For both HyLIRG-W and ULIRG-C, we find that the dispersion declines more with radius in the fits to the 17\,\kms\ cube because, at the higher spectral resolution, the line wings are no longer detected at significant S/N (Fig.~\ref{fig:kinfit_3dbarolo_HyLIRGw_17kms}). This also leads to a poorer fit to the centre of the PVDs. Thus, we do not quote the outer dispersion for the 17\,\kms\ cubes, and use the values from the 35\,\kms cube from here onwards. For ULIRG-C, the dispersion profile decreases steeply as a result of additional non-circular motions not included in our model (see sec.~\ref{sub:deviations_axisummetric_rotator}). We therefore treat the $\sigma$ values derived for ULIRG-C as upper limits on the intrinsic velocity dispersion of the disc. 

                We initially fitted for the rotation velocity, velocity dispersion and PA per ring (with TWOSTAGE=True). However, we found that for HyLIRG-W, the best-fit value of the PA per ring varies in an unpredictable way depending on the input PA and side being fit (increasing with radius in some cases and declining in others). We therefore iterated over the input PA ($239\pm2$, Sec.~\ref{sub:kin_classification}) and choice of sides to fit (both, approaching, and receding) for both the 35\,\kms\ and 17\,\kms\ data cubes to determine the median, 16th and 84th percentile values for PA of the outer two rings (quoted in Table~\ref{tab:kin_props}). We then fixed this initial best-fit outer PA for the kinematic models shown in Figs.~\ref{fig:kinfit_3dbarolo_HyLIRGw} and \ref{fig:kinfit_3dbarolo_HyLIRGw_17kms}. Since the formal uncertainties on this fit returned by \bbarolo\ do not capture the additional uncertainty from the inclination (which we fixed due to the degeneracy with $V_\mathrm{rot}$) or the choice of side being fit, we estimated representative systematic uncertainties by repeating the fit across a grid of inclinations spanning the asymmetric uncertainty and all three side options per inclination. We took the 16th--84th percentile range of the resulting ensemble as the inclination plus side contribution and combined this with the lower/upper uncertainties on the best fit returned by \bbarolo\ by adding these in quadrature, reporting the final asymmetric error bars on the quoted fiducial values.      

                In Table~\ref{tab:kin_props}, we report the mean rotation velocity and velocity dispersion of all rings, $\overline{V}_\mathrm{rot}$ and $\overline{\sigma}$, as well as the external values, taken by averaging the outer two rings, $V_\mathrm{ext}$ and $\sigma_\mathrm{ext}$. We also determined the external circular velocities, $V_\mathrm{circ}$ by repeating the \bbarolo\ fit as described above but setting $\mathrm{ADRIFT}=True$. Since the rotation velocity in the inner regions may be impacted by central mass components, we take $V_\mathrm{ext}$ as the rotation velocity representative of the disc. To allow a fair comparison to other studies, we use $\overline{\sigma}$ to represent the disc dispersion. Thus, we use $V_\mathrm{ext}/\overline{\sigma}$ as a measure of the rotational support of these discs. HyLIRG-W is clearly rotation-dominated ($V\/sigma=10_{-2}^{+3}$), whereas we provide a lower limit of $V/\sigma>2.2$ for ULIRG-C.   

                We tested the \bbarolo\ tilted-ring fits against a parametric fitting approach with \textsc{KinMS}, for which we assumed an exponential flux profile, constant \vrot, and constant $\sigma_\mathrm{rot}$. We forward modelled the data cubes at 17 and 35 \kms resolution, using the MCMC sampler \textsc{GAStimator} \citep{gastimator}. At each step, a model cube, $M(\theta)$, was generated, and compared voxel-by-voxel to the observed data cube, $D$, with a Gaussian likelihood and per-voxel uncertainty. Applying the same mask to the 35\,\kms cube, and fixing the centre, inclination, and PA to the same values used in \bbarolo, we find constant $V_\mathrm{rot}=565\pm8$, which lies between the maximum and mean \vrot\ values found with \bbarolo. We also find a slightly higher -- but still consistent -- constant dispersion of $\sigma=64\pm7$\,\kms. These higher values seem to be the result of the forced constant $V_\mathrm{rot}$ and $\sigma$ profiles, yielding a good fit to the bright central regions (and hence better capturing the total flux) but yielding a poorer fit to the disc outskirts. When fitting the data using \textsc{KinMS} without the application of a mask, we find values of $\sigma$ that are only 4--5\,\kms higher on average than using \textsc{KinMS} without the mask.

            \subsection{Deviations from an axisymmetric rotating disc}
                \label{sub:deviations_axisummetric_rotator}

\begin{figure*}
                \raggedright
                \includegraphics[width=\textwidth,trim={0cm 0cm 0cm 0cm},clip]{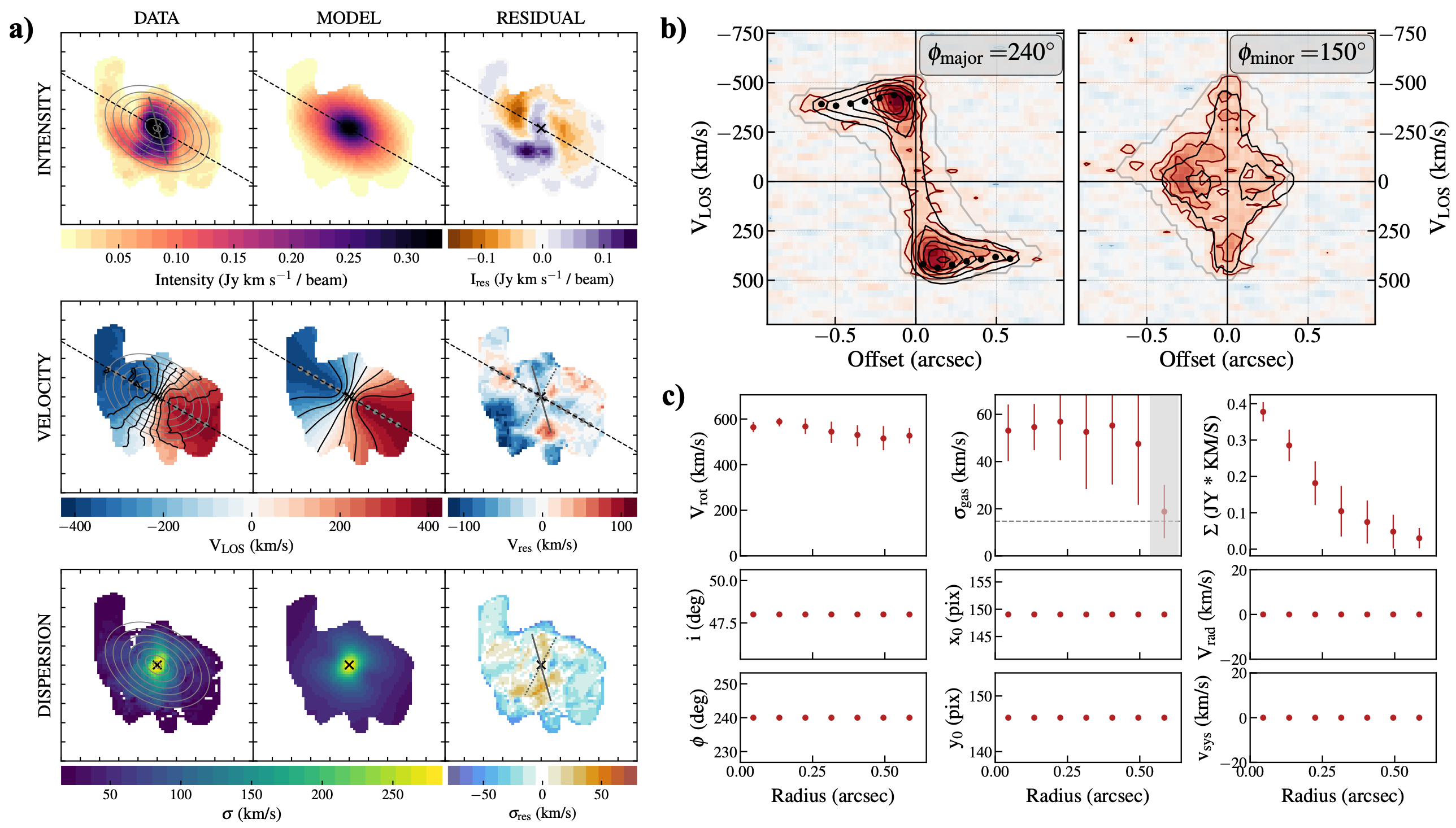}
                \\
                {\large\textbf{d)}}
                \includegraphics[width=\textwidth,trim={1.cm 0.7cm 1.1cm 0.2cm},clip]{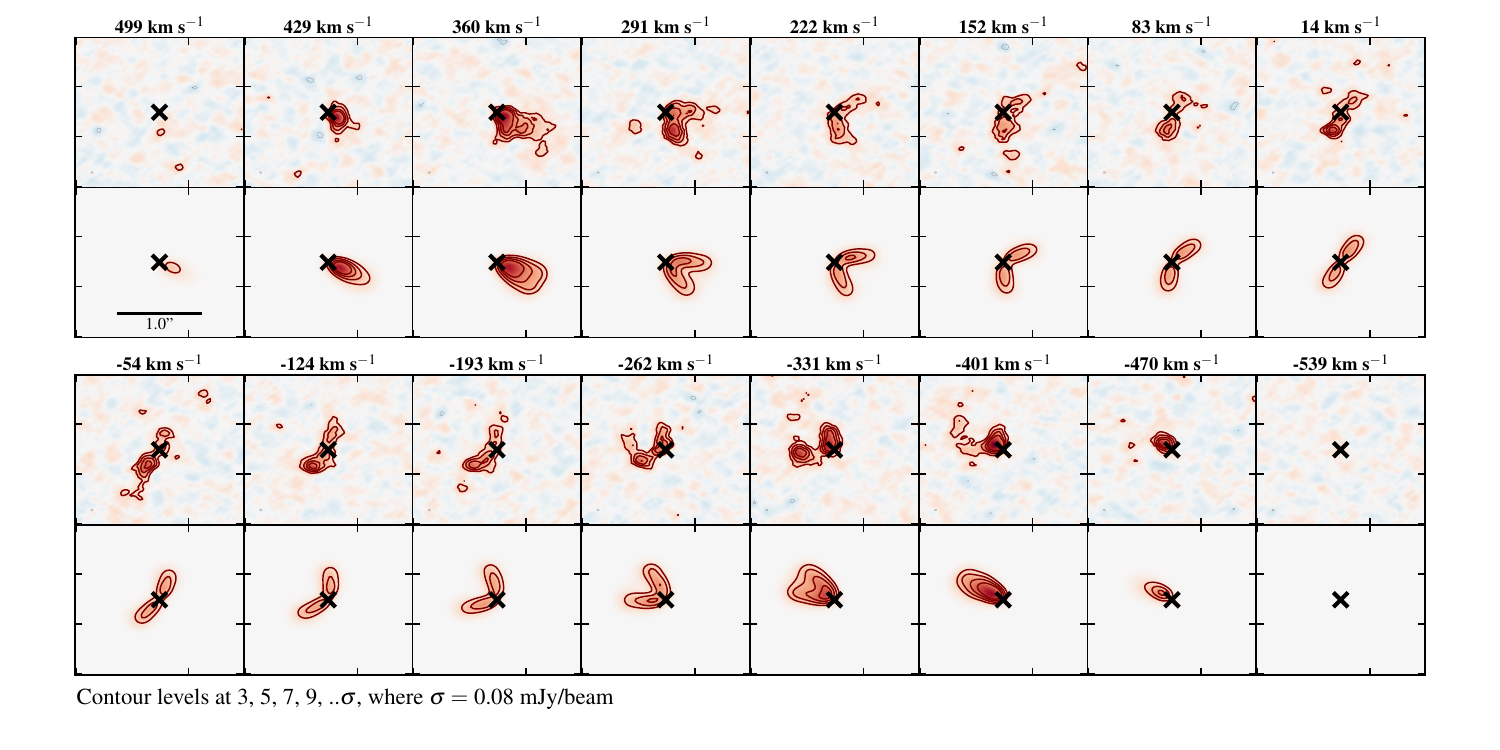}
                \caption{Best-fit kinematic model for the CO(4--3) emission of HyLIRG-W, imaged at $\sim35$ \kms. \textbf{a)} Moment-0, 1, and 2 maps (left), best-fit model (centre), and residual (right). 
                \textbf{b)} Position-velocity diagrams for the CO(4--3) emission of HyLIRG-W, extracted along the major (top) and minor (bottom) kinematic axes. The data is shown by the red-blue colour map and  the mask is outlined in transparent grey. Contours for the data and best-fit model are shown in red and black respectively, in levels of $\pm3n\sigma$, with the rotation curve depicted by the overlaid black points. \textbf{c)} Best-fit (with error bars) and fixed parameters for the example model fit showing the 2nd iteration with the PA fixed to the mean of the initial fit. 
                \textbf{d)} Comparison of the data and model for every second 35\,\kms channel. The top panels in each row depict the data (red-blue colour map), whereas the bottom panels depict the best-fit model. Contours are shown in steps of $3+2n\sigma$ (red lines, steps labelled at bottom), with the mask applied to the data outlined in the top panels (thick grey line). 
                }
                \label{fig:kinfit_3dbarolo_HyLIRGw}
            \end{figure*}

\begin{figure*}
                \raggedright
                \includegraphics[width=\textwidth,trim={0cm 0cm 0cm 0cm},clip]{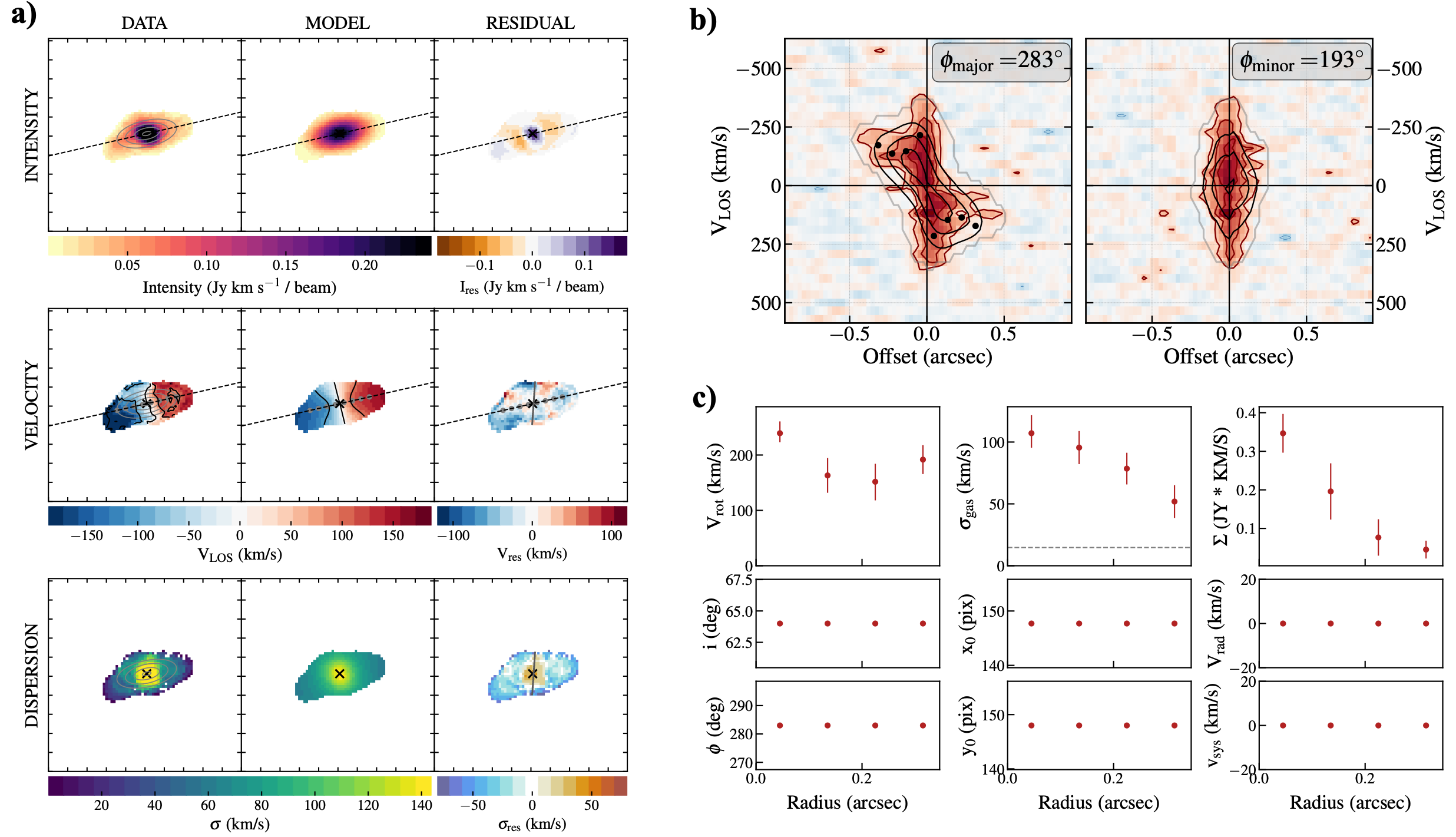}
                \\
                {\large\textbf{d)}}
                \includegraphics[width=\textwidth,trim={1.cm 0.7cm 1.1cm 0.2cm},clip]{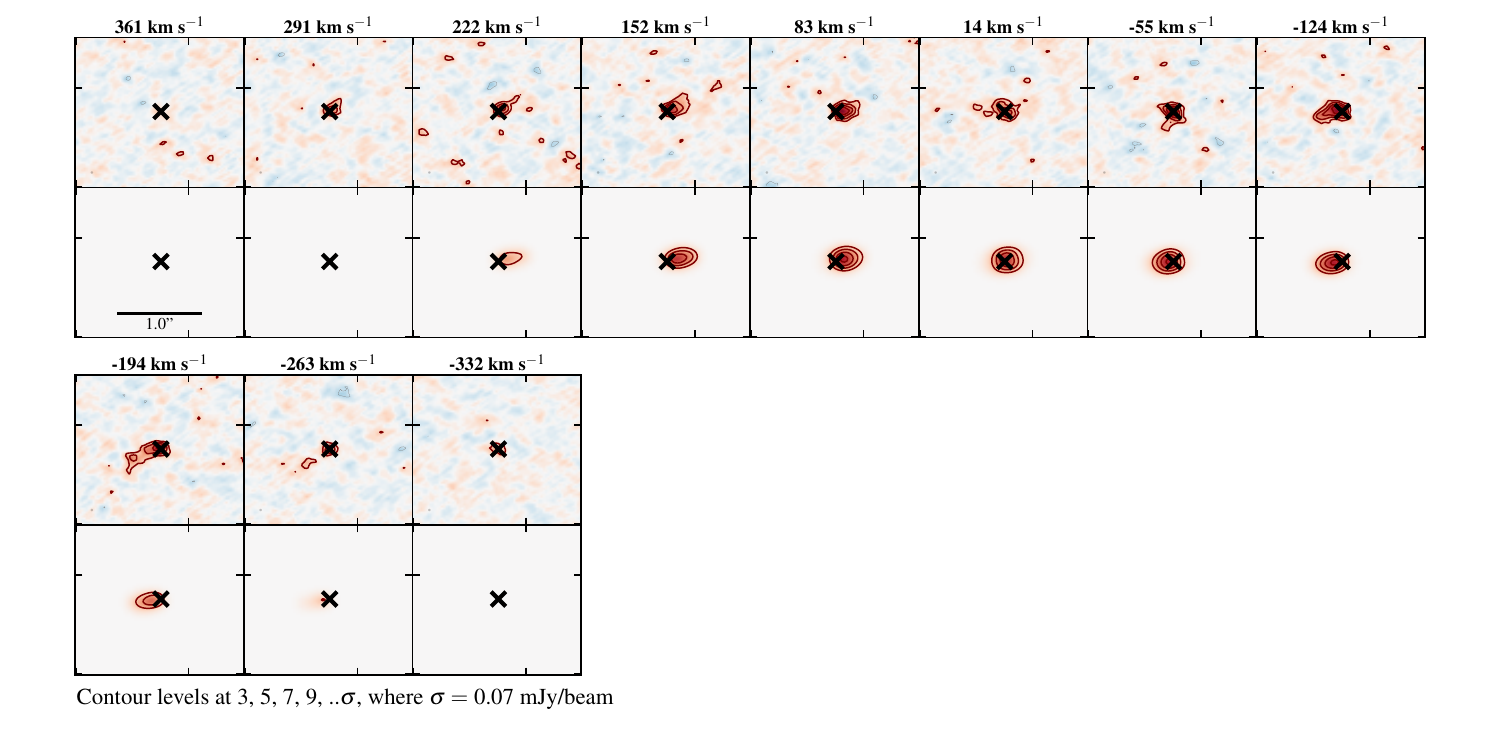}
                \caption{Best-fit kinematic model for the CO(4--3) emission of ULIRG-C, imaged at $\sim35$ \kms. Each panel is the same as in Fig.~\ref{fig:kinfit_3dbarolo_HyLIRGw}
                }
                \label{fig:kinfit_3dbarolo_ULIRGc}
            \end{figure*}

                Channel-by-channel comparisons between the data and models for HyLIRG-W and ULIRG-C show significant evidence of non-axisymmetric kinematic features, resulting in three-arm spiral $V$ residuals and $\sigma$ residuals of $20-50$\,\kms. These signatures are particularly pronounced for HyLIRG-W (Fig~\ref{fig:kinfit_3dbarolo_HyLIRGw}), for which the model cannot capture the rapidly widening and twisting butterfly pattern (and asymmetric line profiles) on the approaching side, resulting in a region of elevated $\sigma$ that appears between the axis tracing the central CO/dust emission ridge and the minor axis (solid vs dashed grey lines in Fig.~\ref{fig:kinfit_3dbarolo_HyLIRGw}). For ULIRG-C, this axis is offset $\gtrsim20^\circ$ from the minor axis (Fig.~\ref{fig:kinfit_3dbarolo_ULIRGc}). For ULIRG-C, we find an X-shaped major axis PVD (Sec.~\ref{subsub:ulirgs_kin_classification}). If we create PV diagrams along the axis tracing the CO/dust ridges of HyLIRG-W, we find a similar X-shaped PVD (Fig.~\ref{fig:pv_bar_w}). We also find asymmetric isovelocity contours for HyLIRG-W, with different opening angles on each side. ULIRG-C shows a twisting of the kinematic PA. We explore two potential causes for these non-axisymmetric signatures below.

                
                    \textbf{Outflows:} We find no evidence of large-scale molecular gas outflows in HyLIRG-W or ULIRG-C; there are no broad CO(4--3) components nor any forking in the PVDs \citep[typically used to diagnose outflows, e.g.][]{walter_2002,cicone_2014,stuber_2021}. HyLIRG-W is known to host an AGN with a strong ionised gas outflow \citep[$\mathrm{FWHM}_\mathrm{H\alpha}\sim9700$\,\kms,][]{ivison_2019}, potentially directed along the NE-SW axis. If it were impacting the disc, we might expect localised, bright CO(4--3) emission and high-velocity residuals, such as the bright CO(4--3) clump without a \ci\ counterpart to the SW (consistent with shock compression). However, it is unlikely that a disc-outflow interaction would dominate at the 1--2 kpc scales seen for HyLIRG-W; in local galaxies, such high-dispersion features are typically present at scales of a few 100\,pc \citep{zanchettin_2023,esposito_2024}. Moreover, an outflow would typically produce a bipolar velocity residual signature \citep[][]{shimizu_2019} rather than the three-armed spirals seen here for the moment-1 residuals.
    
                    \textbf{Bars:} Several of HyLIRG-W's and ULIRG-C's kinematic signatures are consistent with a bar. The three-armed spiral velocity residuals are typical of $m=2$ (bar or spiral) perturbations \citep{schoenmakers_1997,fathi_2005,vanDeVen_2010}. The X-shaped PVD of ULIRG-C's kinematic major axis (and the axis perpendicular to the putative bar in HyLIRG-W) are a typical signature of different bar orbit families and bar-driven streaming \citep{bureau_1999_diagnostics}. The high-$\sigma$ residuals aligning with the CO/dust ridges in HyLIRG-W are consistent with orbit crowding and shocks enhancing the dispersion along bars and the bar/spiral interface \citep[e.g.][]{reynaud_1998,emsellem_2003,liang_2025}. The asymmetric velocity-field opening angles are also typical of barred galaxies, with bars causing streaming motions that cause skewed isovelocity contours and non-axisymmetric opening angles  \citep{pence_1984,chemin_2003,randriamampandry_2015}. In addition, the wide, vertical streak in the minor-axis PVD of C -- not captured by the axisymmetric disc model -- is expected for overlapping non-circular bar orbits (). Overall, we find the CO kinematic signatures to be well-explained by bar-driven in-plane motions, although central spiral arms would cause similar signatures.

        \begin{table}
          \centering
             \caption{Kinematic properties fit with \bbarolo \label{tab:kin_props}} 
             \renewcommand{\arraystretch}{1.3}
            \begin{tabularx}{\columnwidth}{@{}p{0.25\columnwidth} *{4}{>{\centering\arraybackslash}X}@{}}
            \toprule
             \textbf{Galaxy}                                 & \multicolumn{2}{c}{\textbf{HyLIRG-W}}  &  \multicolumn{2}{c}{\textbf{ULIRG-C}}   \\
             \midrule
             Fitting inputs & & & & \\
             \midrule
             $\mathrm{n_{rings}}$                             & \multicolumn{2}{c}{7$^a$}  &   \multicolumn{2}{c}{4}              \\   
             incl. ($^\circ$)                    & \multicolumn{2}{c}{$48_{-8}^{+4}$}    &   \multicolumn{2}{c}{$64_{-5}^{+3}$}          \\
             \midrule
              $\Delta v_\mathrm{chan}$ (\kms)                       & 35    & 17  &  35   & 17  \\
            \midrule
            Best-fit values & & & &\\
            \midrule
             $\langle\mathrm{PA}\rangle$  ($^\circ$)    & $240_{-5}^{+3}$  &  $239_{-4}^{+15}$   & $283_{-7}^{+4}$  &      $285_{-5}^{+2}$          \\
             $V_\mathrm{max}$ (\kms)                   & $588_{-41}^{+85}$ &  $579_{-38}^{+102}$   &  $239_{-24}^{+28}$ &    $243_{-29}^{+28}$          \\
             $\overline{V}_\mathrm{rot}$ ~(\kms)    & $548_{-27}^{+84}$ &  $541_{-22}^{+94}$    &  $186_{-16}^{+23}$ &     $190_{-27}^{+27}$          \\
             $V_\mathrm{ext}$ ~(\kms)                 & $521_{-34}^{+84}$ &  $522_{-37}^{+82}$    & $172_{-40}^{+31}$  &      $182_{-88}^{+42}$          \\
             $V_\mathrm{circ}$$^b$(\kms)    & $529_{-35}^{+85}$  &  --   &  $209_{-38}^{+49}$ &    --          \\
             $\overline{\sigma}$ ~~\,(\kms)            & $\phantom{1}53_{-13}^{+8}$   &  $\phantom{1}48_{-16}^{+9}$     &  $\phantom{1}83_{-22}^{+8}$  &    $\phantom{1}76_{-37}^{+10}$             \\
             $\sigma_\mathrm{ext}$ (\kms)            & $\phantom{1}51_{-25}^{+18}$  &  --     &  $\phantom{1}65_{-16}^{+12}$  &   --              \\
             $ V_\mathrm{ext} /\overline{\sigma}$  & $\phantom{1}10_{-2}^{+3}$   &  $\phantom{1}11_{-2}^{+3}$     &  $\phantom{1}2.2_{-0.5}^{+0.5}$ &   $\phantom{1}2.2_{-0.7}^{+1.2}$             \\
             \bottomrule
             \end{tabularx}
             \vspace{0.5cm}
                \begin{minipage}{\columnwidth}
                \small
                \textbf{Notes:} 
                $^a$The $\sigma$ value of the 7th ring is poorly constrained, so we omit this ring for the mean and outer $\sigma$. \\
                $^b$We take the circular velocity of the outer two rings, as for $V_\mathrm{ext}$. \\
                \end{minipage}
         \end{table}

    \section{Discussion}
        \label{sec:discussion}

        \subsection{2 rotating discs, 2 mergers, 1 unknown}

            Based on the new, high-resolution CO(4-3) observations of HATLAS J0849, we classified the kinematics of its five known member galaxies. HyLIRG-W and ULIRG-C are consistent with rotating discs, with significant non-axisymmetric radial gas motions. HyLIRG-T is not a rotating disc; no plausible amount of lensing would allow this scenario. Instead, it is most consistent with a late-stage merger scenario. ULIRG-E is most consistent with being a merger, whereas ULIRG-M could be either a merger or a slightly disturbed galaxy, for which we are mainly viewing the central dispersion-dominated component. These results indicate that mergers must play a role in maintaining the high SFRs of $z>1$ overdensities, but that top-percentile starbursts like HyLIRG-W need not be undergoing a major merger for the full duration of the starbursting phase. We return to the implications in sec.~\ref{sub:v_evolution}.
            
            Our kinematic classifications are only partially consistent with previous studies of HATLAS J0849. \cite{ivison_2013} classified both HyLIRGs-W and T as rotating discs, based on $1\farcs0\times0\farcs5$ resolution CO(4--3) observations. \cite{rizzo_2023} also classify HyLIRG-W as a rotating disc using the $0\farcs25\times0\farcs19$ \ci\,1--0 observations from 2018.1.01146.S, but were unable to classify HyLIRG-T with these data. More recently, \cite{gomez_2025} presented $0\farcs27\times0\farcs24$ observations of the CO(7--6), and \ci\,2--1 emission in W, T, M, and C -- concluding that the kinematics of all four are rotation-dominated. Yet, our deeper CO(4--3) observations clearly rule out this scenario for HyLIRG-T, and find more plausible explanations for ULIRG-M. Moreover, our best-fit \vrot\ and $\sigma$ values for HyLIRG-W are slightly higher than those derived in \cite{rizzo_2023}, who report $V_\mathrm{rot} = 509\pm 27$\,\kms\ and $\sigma=40^{+10}_{-8}$\,\kms, based on the slightly higher inferred inclination of $\sim50^\circ$. These comparisons highlight the importance of deep, high-resolution data in classifying and modelling the cold gas kinematics.           

        \subsection{A lopsided gas spiral at the heart of HATLAS J0849}
            \label{sec:superspiral}

\begin{figure*}
                    \raggedleft
                    \includegraphics[width=0.49\textwidth,trim={0.5cm 1.98cm 0.1cm 1cm},clip]{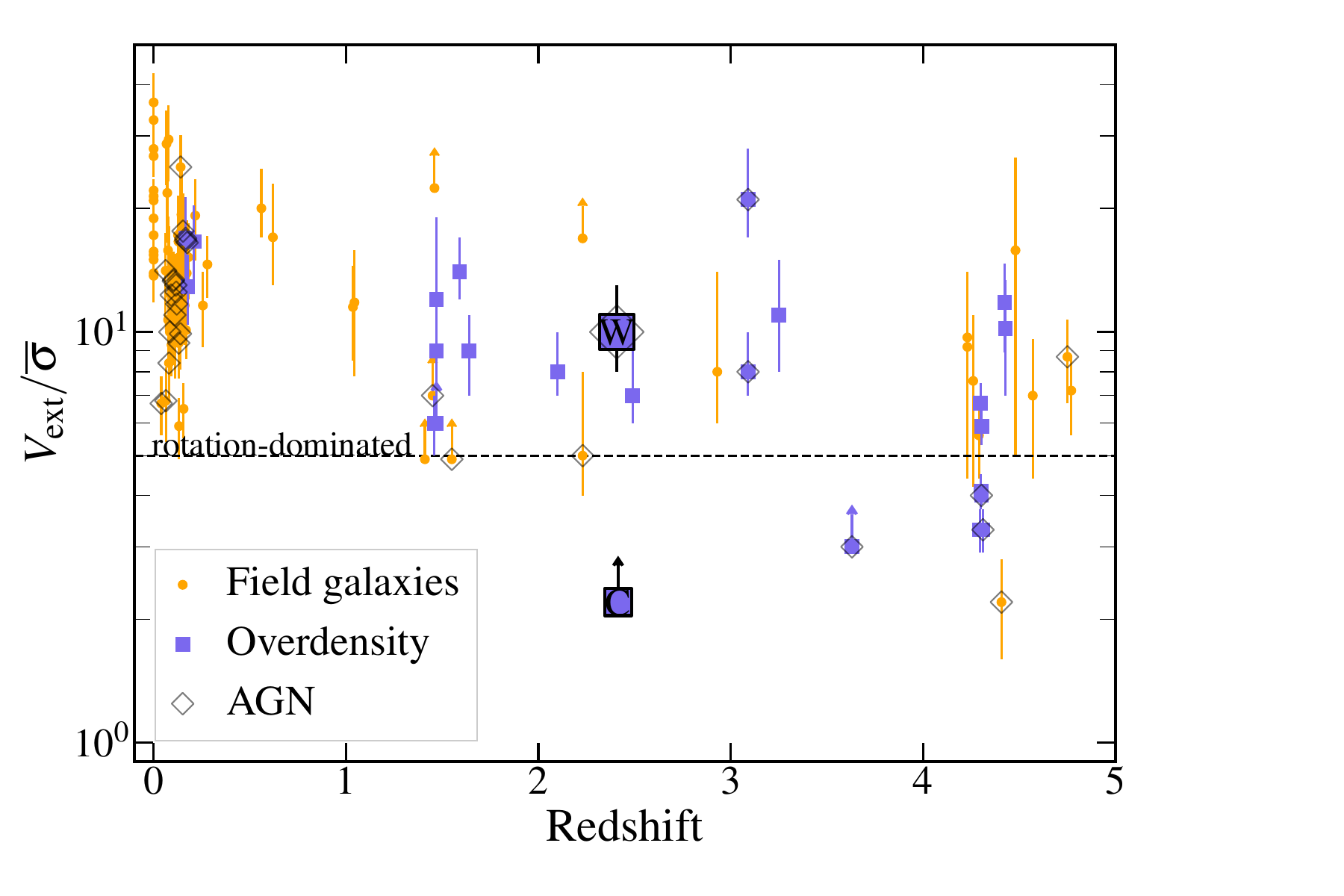}
                    ~
                    \includegraphics[width=0.49\textwidth,trim={0.5cm 1.98cm 0.1cm 1cm},clip]{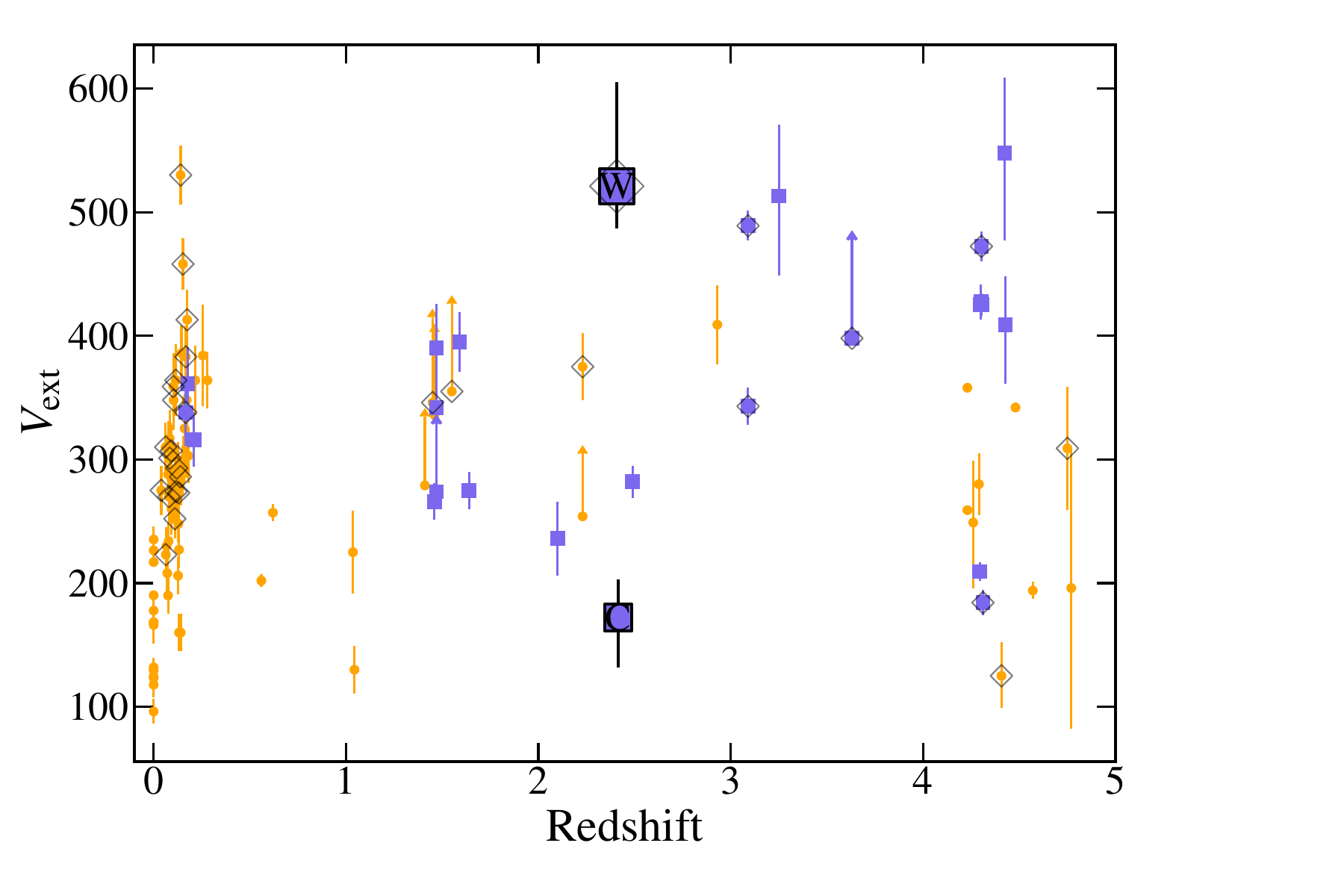}
                    \\
                    \includegraphics[width=0.49\textwidth,trim={0.5cm 0.5cm 0.1cm 1cm},clip]{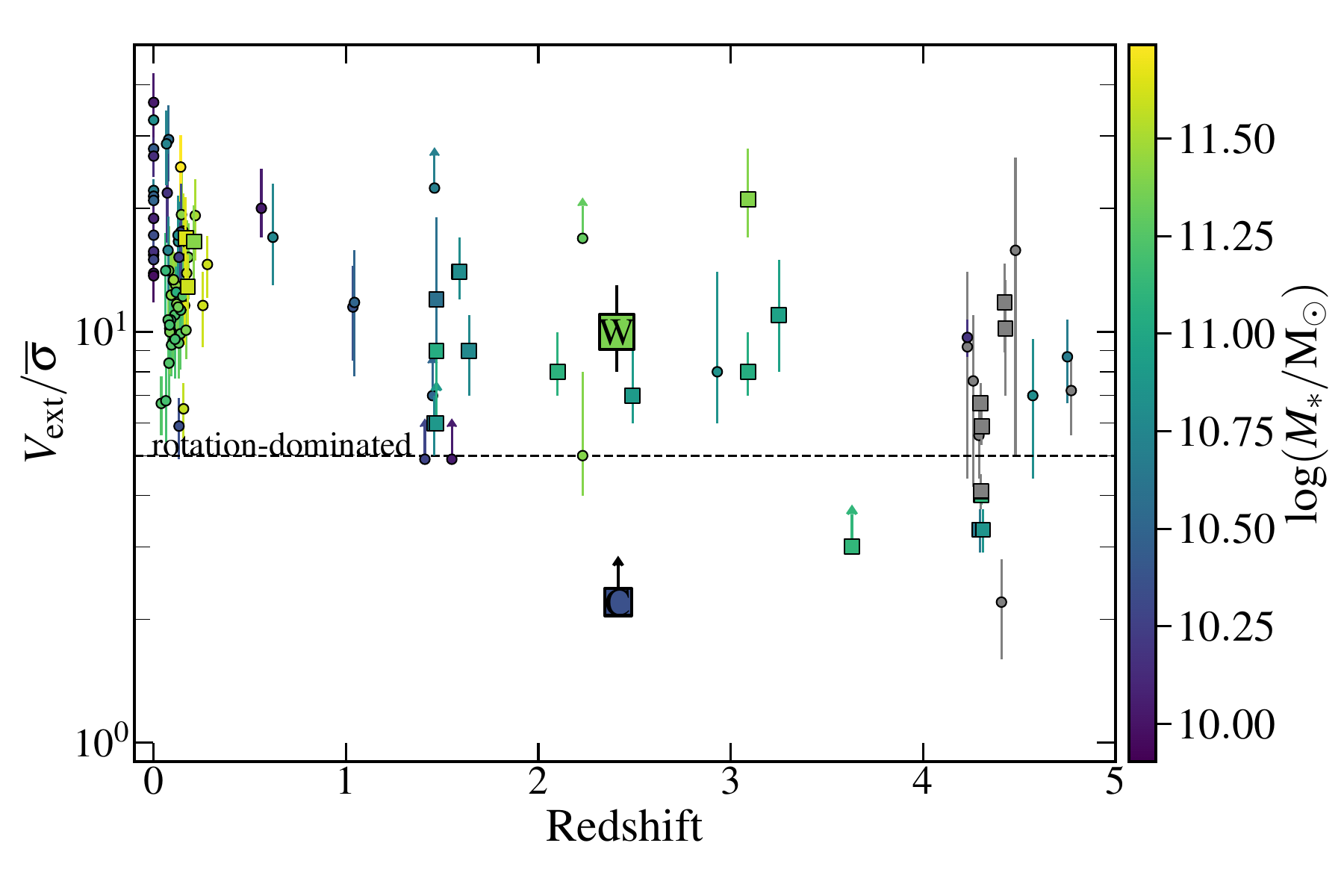}
                    ~
                    \includegraphics[width=0.49\textwidth,trim={0.5cm 0.5cm 0.1cm 1cm},clip]{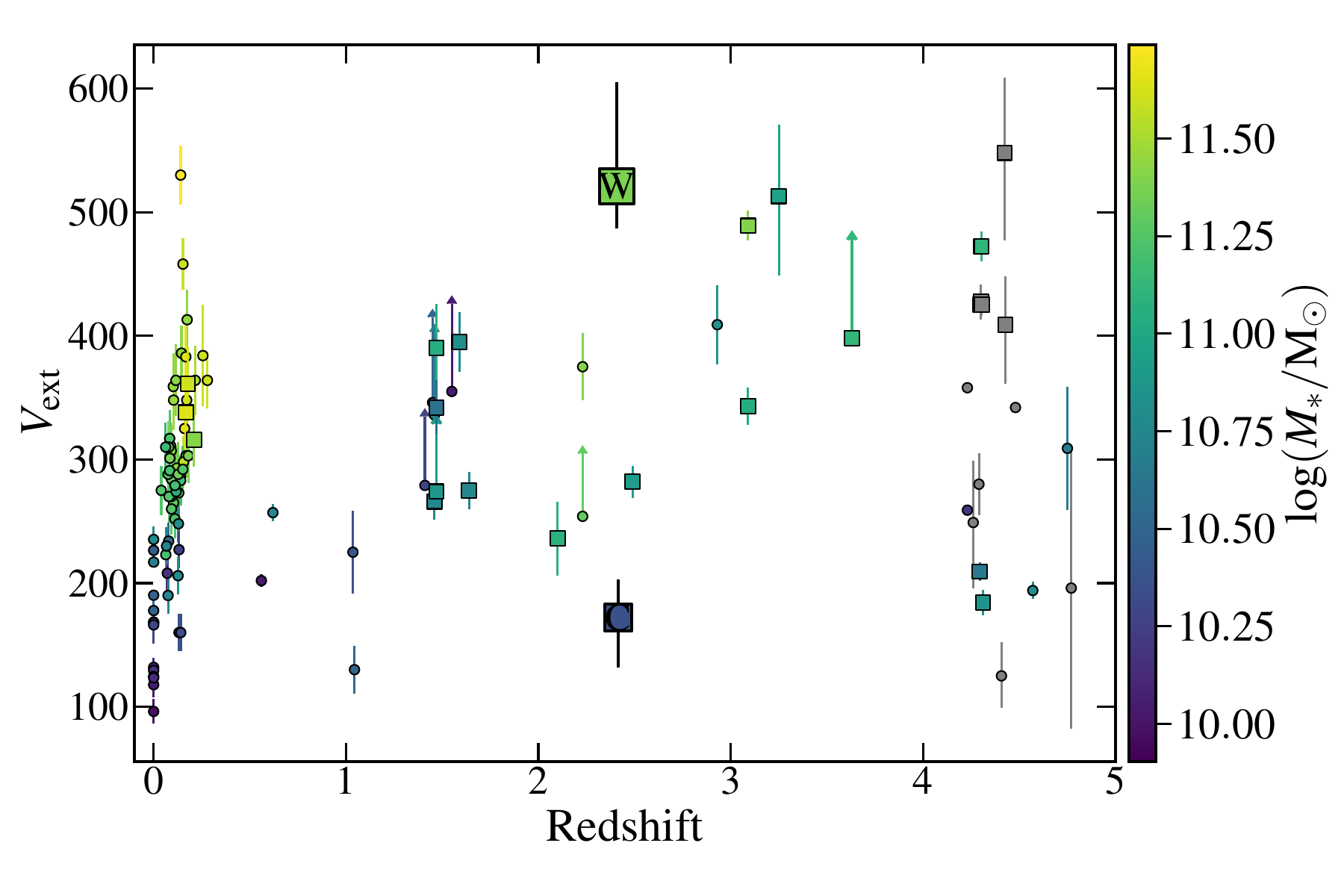}
                    \caption{Cold gas kinematics in context -- Top left: Rotation-to-dispersion ratio vs redshift as measured from observations of cold gas (literature sample in Appendix~...), distinguishing galaxies based on whether they reside in an overdensity. Top right: rotation velocity vs redshift, with the same sample and colour+symbol scheme as in the left panel. Bottom left: same as top left but colour-coded by stellar mass (where available). Bottom right: same as top right but colour-coded by stellar mass.
                    }
                    \label{fig:v_evolution}
                \end{figure*}

            The new CO(4-3) data presented here reveal that HyLIRG-W has a rotation velocity of $521_{-34}^{+84}$ \kms (Sec.~\ref{sub:kinematic_modelling}). As shown in Fig.~\ref{fig:v_evolution} and \ref{fig:ETG}, this value is consistent with the highest rotation velocities reached by local ``superspirals'' \citep{ogle_2019,diteodoro_2021} and $z>3$ starbursts in overdensities, with local superspirals exhibiting similar stellar and gas masses ($<10^{11.1}$\msun\ and $>10^{10}$\msun, respectively). The CO(4--3) and \ci\,1--0 emission of HyLIRG-W also show strong evidence of lopsidedness. Both reveal a strong $m=1$ asymmetry, with an off-centred light distribution, one dominant spiral arm and an asymmetric velocity field/line profile (Figs~\ref{fig:moments_and_pvs}, \ref{fig:kinfit_3dbarolo_HyLIRGw}). Lopsidedness is a common feature of the stellar and H{\sc i} content of disc galaxies with about 1/3 of local discs showing morphological and/or kinematic lopsidedness \citep[e.g.][]{rix_1995,jog_2009}. Strong lopsidedness can be caused by minor mergers, tidal encounters, and asymmetric gas accretion \citep{bournaud_2005,jog_2009}. In simulations, the lopsidedness caused by tidal encounters and minor mergers may persist for $\gtrsim10$ dynamical times (1--2 Gyr) -- well after the smaller companion fades or merges. However, \cite{bournaud_2005} show that the strongest lopsidedness can only be observed during the interaction/minor merger (when the companion is still close) and thus asymmetric gas accretion is critical to maintaining the strong lopsidedness in massive gas-rich discs with no evidence of any massive interacting companion. 
            
            The presence of a companion may help distinguish between the causes of lopsidedness in HyLIRG-W. Assuming the separation from ULIRG-M corresponds to the separation on sky, $\sim 27$\,kpc, the tidal field due to ULIRG-M would be $\sim0.1\%$ of HyLIRG-W's potential, making it unlikely that this would induce the dominant spiral arm. The same reasoning applies for HyLIRG-T, which is at $\sim85$\,kpc separation and has a similar tidal field of (at most) a few \% of HyLIRG-W's potential. For context, we note that the separation of M 51 and its companion (NGC 5195) is 10--13 kpc, yet M51 is only moderately lopsided due to recent passes of the companion through the disc plane \citep{colombo_2014,font_2024}. It therefore seems unlikely that the closest observed galaxies to HyLIRG-W are inducing its lopsidedness. However, we cannot rule out an earlier, stronger interaction (plus the response of a now axisymmetric halo), nor the perturbation by an undetected gas-poor companion.

           \begin{figure*}
                    \centering
                    \includegraphics[width=\textwidth,trim={0cm 0cm 0cm 0cm},clip]{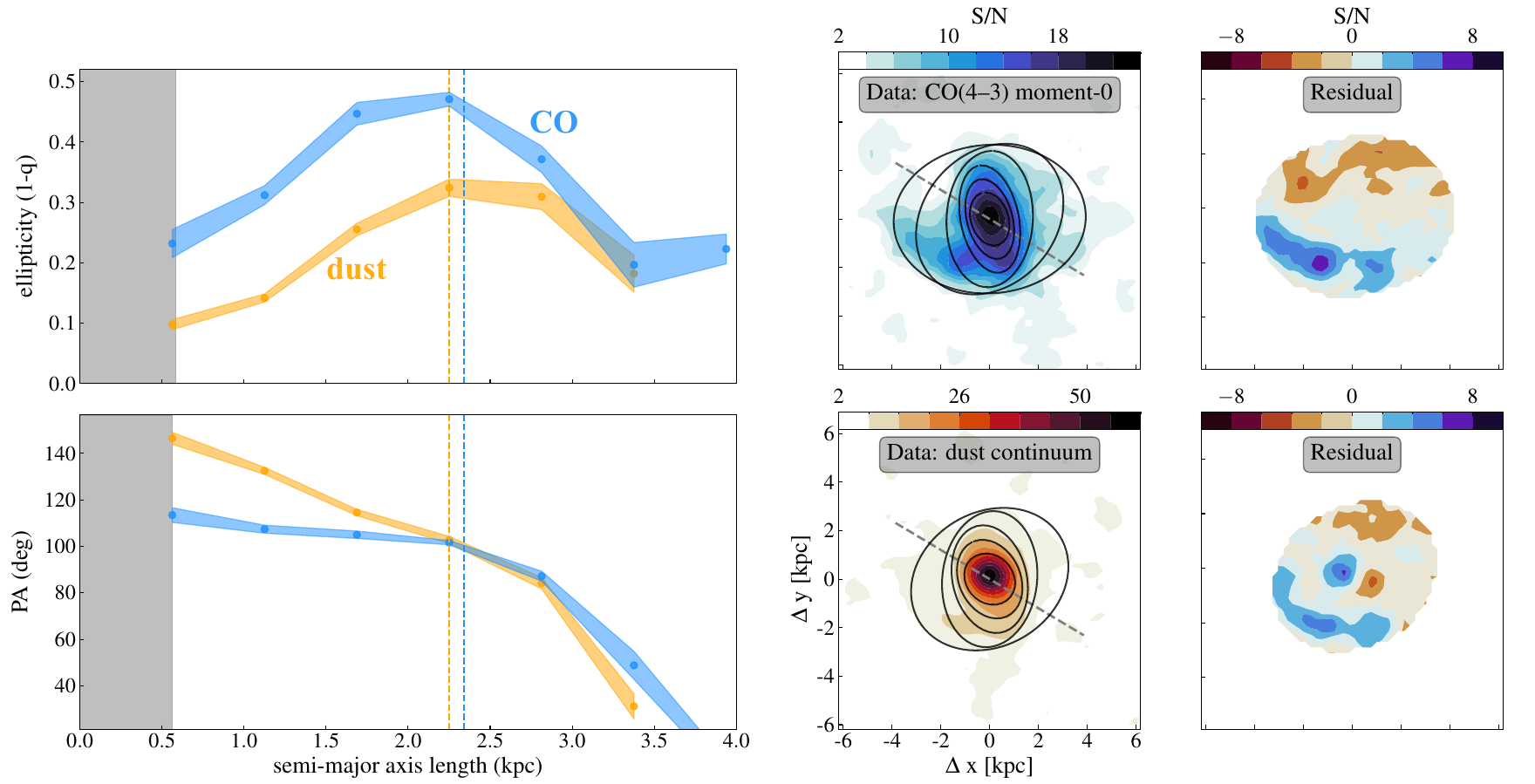}
                    \caption{Left: Ellipticity (top; where $q$ is the axis ratio) and position angle (bottom; defined east of north) of isophotes fit to the CO(4--3) moment-0 (blue) and underlying dust-continuum (orange) maps for HyLIRG-W. The minimum criteria the ellipticity peaks need to exceed to measure a bar length -- 2 FWHM -- are shown as the dashed lines (blue: CO, orange:dust). Right: Isophote fits to the CO(4--3) moment-0 map (top) and underlying continuum (bottom) for HyLIRG-W---left: observations with best-fitting isophote ellipses overlaid; right: noise-normalised residuals of the best-fit isophotes.}
                    \label{fig:isophotal_fits}
                \end{figure*}

        \subsection{Potential evidence of bars}

            Recent JWST studies have revealed that bars are more prevalent at $z>2$ than previously thought \citep[e.g.][]{leconte_2024}. Moreover, several recent $z\gtrsim2$ studies have found evidence for bars, based on cold gas and/or dust-continuum emission \citep{tsukui_2024,umehata_2025_kin,amvrosiadis_2025,pastras_2025,huang_2025}. Using the high angular resolution of our observations, we tested for kinematic signatures of bars, finding several signatures of deviations from a simple rotating disc model, which are consistent with -- although not conclusive proof of -- W and C hosting bars (Sec.~\ref{sub:deviations_axisummetric_rotator}). Moreover, the rotation curves of HyLIRG-W and ULIRG-C both peak at 10--20\% of the maximum detected radius, potentially indicating a concentrated central mass component, such as a bar \citep{randriamampandry_2015}, or non-circular motions associated with in-/outflows, which can also be induced by bars. 
             
            We also test for the presence of morphological signatures of a bar, noting that there are three common methods used to classify and quantify the properties of bars based on their 2D surface brightness distribution: 1) Fourier decomposition \citep{ee_1985,ohta_1990,laurikainen_2002}, 2) ellipse fits \citep{elmegreen_1996,gadotti_2007,guo_2025}, and 3) structural decomposition techniques \citep{laurikainen_2005,gadotti_2011}. We test the Fourier decomposition method presented in \citet{tsukui_2021}, finding that it confirms the elongated central CO(4--3) distribution and lopsidedness of HyLIRG-W, with clear $m=2$ and $m=1$ modes. However, the exact results are highly sensitive to how we mask the central region, with this approach not being adapted for such a high level of beam smearing. We also test the ellipse-fitting method employed for the $z=4.4$ galaxy BRI1335-041 \citep{tsukui_2024} and the $z=3.8$ lensed starburst SPT-2147 \citep{amvrosiadis_2025}, with the results shown in Fig.~\ref{fig:isophotal_fits}. HyLIRG-W shows a stronger peak in ellipticity than SPT-2147, more similar to that of BRI1335-041. However, as for BRI1335-041, the PA of W declines significantly up to the peak ellipticity, violating the typical bar criterion of the PA remaining fairly constant along the bar \citep[$\Delta PA <10^\circ$,][]{gadotti_2007,leconte_2024,guo_2025}. Such large PA variations imply either significant bulge/spiral contamination or the lack of a bar. We find that these types of morphological approaches yield less convincing evidence of a bar than the kinematic signatures, which is to be expected given that bars are typically stellar structures.

            Given that HyLIRG-W is clearly rotation-dominated and ULIRG-C may well be, it would be no surprise if they host bars. Bar formation is inhibited in dispersion-dominated galaxies \citep{sheth_2012,melvin_2014}, with rotation-dominated discs being a necessary but not sufficient condition for the presence of bars. Moreover, strong bars grow in discs where the inner regions are baryon-dominated (which seems to be the case for both HyLIRG-W and ULIRG-C), this being another necessary but not sufficient condition \citep{rosas_guevara_2022,bland-hawthorn_2024,fragkoudi_2025}. Both observation- and simulation-based studies find that bar fractions are elevated in dense environments \citep{thompson_1981,mendez-abreu_2012,lokas_2016,rosas_guevara_2024}. Moreover, recent studies also show that bars can form in gas-rich discs with elevated central turbulence \citep{bland-hawthorn_2024}. Thus, it should not come as a surprise if some of the galaxies in the targeted $z\sim2.4$ overdensity host bars. However, the presence of bars does little to discern between the past accretion/merger history. For galaxies as massive as W and C, simulations indicate that bars may be triggered by sustained gas accretion \citep{bournaud_2002,bland-hawthorn_2024}, tidal interactions/minor mergers \citep{lokas_2016,lokas_2025}, and even massive mergers \citep{fragkoudi_2025} -- although the latter must have been sufficiently gas-rich and non-destructive (or followed by rapid re-formation of the disc).

        \subsection{Rotation-dominated discs as a common feature of starbursts in $z\gtrsim2$ overdensities} 
            \label{sub:v_evolution}

            We place the disc galaxies in HATLAS J0849 in context with other massive, $z\lesssim5$ disc galaxies in the literature for which resolved cold gas observations are available (see Appendix~\ref{sub:lit_sample}). Based on the information available, we classify these galaxies as residing in either overdensities or the field. For the purposes of discerning the effects of environment, we count local galaxies at the outskirts of groups as being in the field. We consistently use the external rotation velocities (to avoid regions biased to bulges or bars), which are taken as $V_\mathrm{flat}$ for local studies, $V_\mathrm{ext}$ measured from the outer two rings for $z>1$  studies using \bbarolo\ and the mean $V_\mathrm{rot}$, for the handful of $z>1$ studies based on parametric models (with constant $V_\mathrm{rot}$). We use the mean dispersion, $\overline{\sigma}$, for consistency between different approaches, noting that for galaxies with additional non-circular motions, this value may be significantly higher than the actual velocity dispersion of the disc.

            We first check for any systematic difference in the kinematics of galaxies in the field vs overdensities. Based on the small number statistics currently available at $z>1$,\footnote{Two ongoing surveys are obtaining these measurements for more field galaxies -- the CONDOR-ALMA large program and the Northern Extended Millimeter Array (NOEMA) survey, NOEMA3D -- enabling more systematic studies of the role of different environmental metrics. } we find no systematic difference between the rotation-to-dispersion ratio of galaxies in the field vs in overdensities (Fig.~\ref{fig:v_evolution}, top left). However, there is a very clear difference in $V_\mathrm{ext}$ (Fig.~\ref{fig:v_evolution}, top right); at $z>1$, the highest rotation velocities ($V_\mathrm{ext} > 420$\,\kms) are almost exclusively occupied by galaxies in overdensities. This does not lead to systematically higher $V/\sigma$ values, as the increased impact of stellar feedback in these high-SFR discs appears to counter-balance the high $V_\mathrm{ext}$ by increasing $\sigma$ (see Sec.~\ref{sub:turbulence_driving}). 
            
            We also find no evidence for any dependence of $V/\sigma$ on stellar mass (Fig.~\ref{fig:v_evolution}, bottom left), albeit within the range of high stellar masses probed at $z>1$. In contrast, it is clear that below $z=0.3$, $V_\mathrm{ext}$ depends strongly on the stellar mass. At $z>1$ the trend with stellar mass is likely less pronounced because $V_\mathrm{ext}$ traces the total baryonic mass ($M_\mathrm{bar}=M_\mathrm{gas}+M_\mathrm{\star}$), which is dominated by the stellar mass at low redshift but becomes increasingly sensitive to the gas mass as gas fractions increase towards higher redshift \citep[e.g.][]{tacconi_2020}. The lack of a strong dependence of $V/\sigma$ on stellar mass likely reflects the fact that the higher stellar mass galaxies shown here also typically have higher SFRs, increasing $\sigma$.  

\begin{figure*}
                    \centering
                    \includegraphics[width=0.9\textwidth,trim={0cm 0.5cm 0cm 0.5cm},clip]{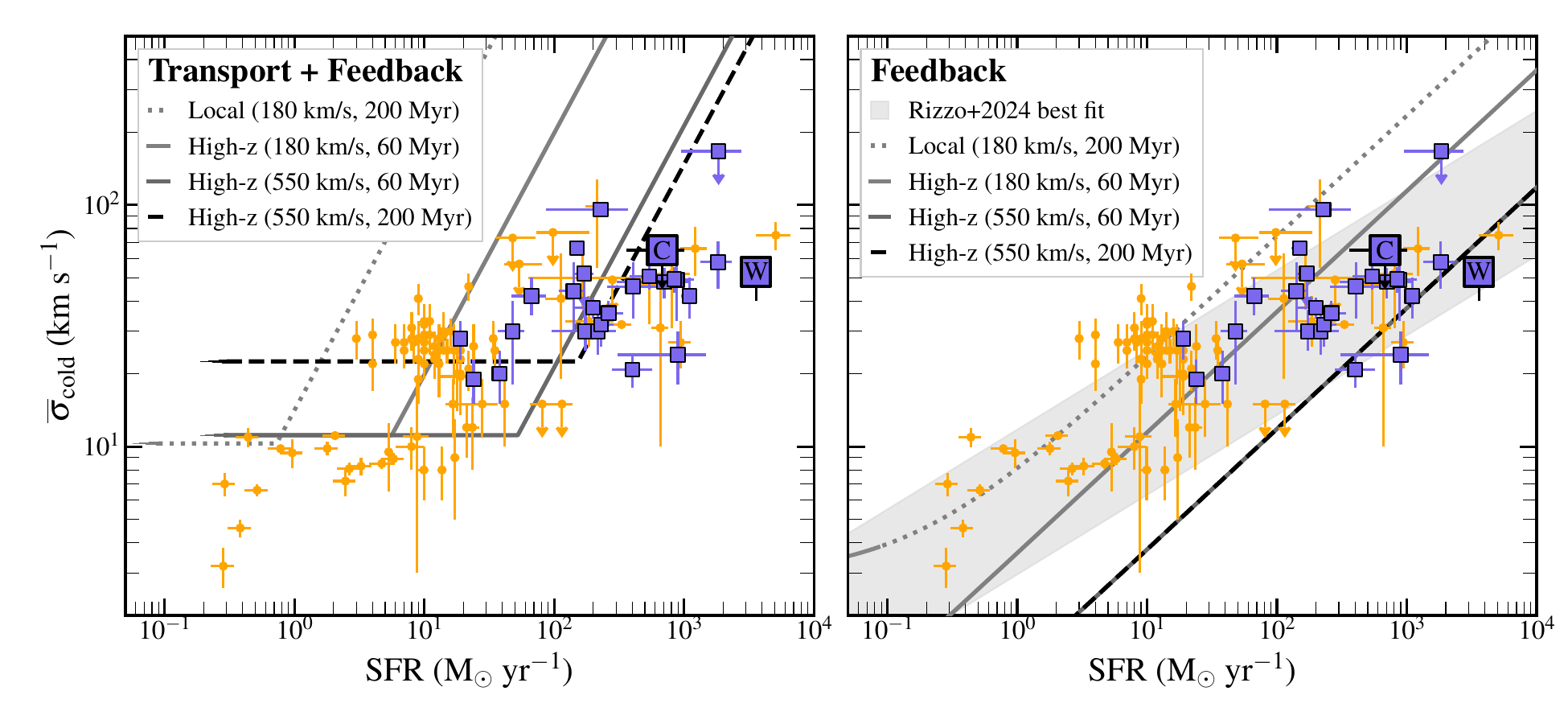}
                    \caption{Cold gas velocity dispersion vs global SFR as measured from observations (circles and squares) compared to predictions from star-formation models incorporating stellar feedback and radial gas transport (left) vs stellar feedback only (right). As in Fig.~\ref{fig:v_evolution}, we compare galaxies in overdensities (purple squares: including W and C, studied here) to galaxies in the field (orange circles). The dotted vs dashed/solid lines show the \protect\cite{krumholz_2018} prescriptions for low- and high-z galaxies (with different rotational support), whereas the grey shaded area in the right panel show the semi-empirical, SNe-regulated model from \protect\cite{rizzo_2024}.}
                    \label{fig:sigma_sfr}
                \end{figure*}

            Our results add to recent findings that rotation-dominated cold gas discs ($V/\sigma>5$) may be typical in overdense environments at $z\gtrsim1$ (left panel, Fig.~\ref{fig:sigma_sfr}).
            \begin{description}[leftmargin=0.6em,labelindent=0.6em,topsep=0.5pt, partopsep=0.5pt, parsep=0.5pt, itemsep=0.5pt]
                \item[J2215.9-1738:] Resolved CO(2--1) observations of six galaxies in the $z=1.5$ cluster XMM XCS J2215.9-1738 were first presented in \cite{ikeda_2022}. Based on these data, \cite{rizzo_2023} found that at least 4/6 are discs (the other were uncertain), with all discs being rotation-dominated. Unlike the starbursts in the following overdensities though, these galaxies appear to be on the main sequence.
                \item[ADF22:] Based on NIRCam and ALMA continuum imaging, \cite{umehata_2025_morph} found evidence that 6/9 of the dusty star-forming galaxies in the core of the $z=3.1$ protocluster SSA2, ADF22, were discs. \cite{rizzo_2023} previously analysed the resolved CO(3--2) emission of three of these galaxies, finding all three to be rotation-dominated discs, with \cite{umehata_2025_kin} also revealing one of these to be a giant spiral. 
                \item[SGP3826:] \cite{roman-oliveira_2023} analysed the \cii\ emission of two starbursts in a group at $z=4.2$, finding both to be rotation-dominated discs. 
                \item[SPT2349-56:] Using ALMA observations of \cii\ emission in seven galaxies within the $z=4.3$ protocluster SPT2349-56, \cite{venkateshwaran_2024} found evidence for six rotating discs, two of which are rotation-dominated. 
                \item[MQN01:] \cite{pensabene_2025} recently analysed a galaxy pair in a $z=3.3$ cosmic web node. The quasar companion is clearly a rotation-dominated disc.
            \end{description}
            \cite{rizzo_2023} find two more $z>1$ galaxies in overdensities with sufficient data quality to perform the kinematic classification and modelling -- COSMOS 3182 at $z=2.1$ and $CLJ1001-131077$ at z=2.5. Both were found to be rotation-dominated discs. Because no other galaxies in these overdensities were resolved, we omit them from the following statistical estimate.

            Combining these studies with our analysis of the five HATLAS J0849 member galaxies we conclude that when isolating gas-rich, star-forming galaxies in $z>1$ overdensities, $\gtrsim67\%$ are rotating discs and $\gtrsim42\%$ are rotation-dominated ($V/\sigma>5$) discs. The latter fraction is likely a lower limit as only three of the six discs in ADF22 had resolved cold gas observations and ULIRG-C may yet prove to be rotation-dominated. If we only consider the $z>2$ systems, these galaxies all have gas masses of $M_\mathrm{H_2} \gtrsim 2\times10^{10}$\,\msun, gas fractions of $M_\mathrm{H_2}/(M_\mathrm{H_2}+M_\mathrm{\star})=0.2-0.7$, and SFRs of $\gtrsim170$\,\myr. We acknowledge that these numbers do not fully reflect the sample bias involved in only including galaxies with resolved observations of cold gas. However, this additional ``data-quality'' or ``feasibility'' criteria should bias the sample to the most dust- and gas-rich, most highly star-forming, and highest surface brightness systems. It is unclear why this selection should systematically favour rotating discs; if anything, we would expect that the opposite would be true as high gas fractions drive discs to be less stable while high SFRs drive strong feedback \citep[e.g.][]{genzel_2011,krumholz_2018}. 
            
            The combination of high SFR, high gas fractions, and high occurrence of rotation-dominated disc galaxies in $z>1$ overdensities has yet to be systematically explored and reproduced by models. However, several zoom-in simulations imply that these galaxies should be rare, because starbursts are triggered through strong angular momentum loss (wet mergers, counter-rotating inflows, and/or torques driven by violent disc instabilities), which naturally drive perturbed kinematics and elevated $\sigma$ and make it hard to sustain high $V/\sigma$ \citep{danovich_2012,danovich_2015,zolotov_2015,dekel_2020a,dekel_2020b}. In this picture, high $V/\sigma$ discs are most likely to occur during unusually coherent coplanar accretion episodes. For example, for the single-galaxy zoom-in simulation presented by \cite{kretschmer_2022}, the rotation-dominated, gas-rich disc only persists during the 410 Myr period of coplanar accretion (around five orbital periods). Complementary to this, \cite{hafen_2022} (FIRE-2) argue that forming and maintaining a thin, rotation-supported disc requires angular-momentum alignment of accreting gas prior to joining the galaxy. However, for the most massive halos in these models, extended gas discs can survive many orbital times and remain rotation-supported because typical disc-disrupting processes (spin flips from gas-rich mergers and strong SNe-driven turbulence) are less effective \citep{dekel_2020a,dekel_2020b}. It is currently unclear to what extent the high masses (e.g. of HyLIRG-W) would serve to balance out the turbulent, compaction-starburst episodes.  
            
           The zoom-in simulations described above imply that high $V/\sigma$ phases should have a limited duty cycle unless the starbursts are driven by strong gas compression without angular momentum cancellation or the bursts coincide with periods of aligned coherent accretion. It is hard to imagine a scenario where this kind of transient, coplanar accretion would be sufficiently common to explain the observed high fraction of rotation-dominated starbursts in overdensities, unless inflows somehow become aligned once a large enough discs forms, as in the simulations of \cite{semenov_2024}. In this scenario, strong alignment is not a pre-requisite but instead a consequence of disc formation. But it is difficult to motivate a mechanism by which filamentary inflows -- set on virial scales -- become systematically aligned with a galaxy. Instead, our observations indicate another plausible route. The lopsidedness and putative bars suggests that starbursts like HyLIRG-W could be partially sustained by torque-driven, non-axisymmetric inflows, which redistribute angular momentum internally (instead of cancelling it through counter-rotation). We return to this argument in the next section.

            Several studies, based on simulations and semi-analytic models, argue that extreme starbursts\footnote{characterised by short depletion timescales} are at least half-triggered by mergers \citep{hopkins_2006,lagos_2020,lower_2023,araya_2025}. Although we find evidence of at least 2/5 ongoing mergers in HATLAS J0849, the collection of resolved observations of starbursts in overdensities do not support $\gtrsim50\%$ ongoing mergers. For $\gtrsim50\%$ of starbursts to be explained under this scenario, a significant fraction of rotation-dominated starbursts would need to have been triggered by mergers. These rotation-dominated cold gas discs would need to (re-)form while the merger-driven excess in the SFR is still present. Several zoom-in and cosmological simulations show that mergers (even major ones) can produce disc-dominated remnants, especially when these have strong stellar feedback and a high gas content \citep{springel_2005,robertson_2006,hopkins_2009,governato_2009,hopkins_2013,sotillo-ramos_2022}, as the feedback serves to make the gas more stable to fragmentation. But it is unclear what fraction of these become rotation-dominated, and whether this phase coincides with the period of excess SFR. 

            Several simulation-based studies find that merger-induced SFR enhancements are typically short-lived, lasting $10-100$ Myr \citep{jog_1992,di_matteo_2008,renaud_2014,hani_2020}. However, discs may also reform on the order of a dynamical time \citep{hafen_2022,semenov_2024}, which is $\sim65$ Myr for HyLIRG-W, but typically a few 100 Myr for less massive discs. This leaves enough time for discs and merger-triggered starbursts to coincide. Indeed, \cite{lotz_2008} find that the high SFR typically peaks at or just after the final merger of gas-rich discs and continues 0.5--1 Gyr, by which time many remnants appear disc-like. Complementary to this, \cite{renaud_2022} find that merger-induced compression only triggers starbursts after the formation of the disc, meaning that these phases must overlap. However, several studies indicate that the new stable disc state should lead to morphological quenching on timescales of 100s of Myr \citep{martig_2009,zolotov_2015,petersson_2023}, implying that we are capturing merger-induced rotation-dominated discs during a brief phase. In combination, these studies indicate that a non‑negligible subset of the rotation‑dominated starbursts in $z>1$ overdensities may be merger‑induced systems caught during the short overlap between rapid disc (re‑)settling and the fading merger‑driven SFR excess. Indeed, the challenge with reforming rotation-dominated starburst discs post-merger seems much the same as without the merger, in terms of whether or not coplanar accretion is necessary.

     \subsection{The cold gas turbulence of starbursts in overdensities}
        \label{sub:turbulence_driving}

            We test how the global SFR governs turbulence in Fig.~\ref{fig:sigma_sfr}, by showing HyLIRG-W and ULIRG-C in the context of literature measurements, again separating galaxies depending on whether they reside in the field or in overdensities. On average, we find that $z>1$ galaxies in overdensities have higher SFRs, and hence also higher $\sigma$ than most field galaxies. However, they occupy the same locus, indicating that the turbulence of cold gas is regulated in a similar manner within both populations; the increased accretion and/or merger rate expected for starbursts in overdensities \citep[e.g.][]{dave_2010,narayanan_2015,sparre_2016,lower_2023} do not serve to systematically increase the cold-gas turbulence relative to galaxies in the field.

            In Fig.~\ref{fig:sigma_sfr}, we compare our measurements, and those in the literature, to two different analytic frameworks describing how turbulence is regulated -- ``feedback-only'' and ``feedback plus transport'' models. We compare to the analytic framework of \cite{krumholz_2018} as it has become one of the standard reference models for turbulence-regulated star formation in gas-rich discs. Their theory provides a simple, closed set of predictions for the velocity dispersion, and SFR in terms of directly observable global quantities and explicitly separates the ``feedback-only'' and ``feedback plus transport'' branches. In the former, the gas turbulence is driven by the energy from stellar feedback (SNe, winds) only; in the latter, gravitational instabilities trigger mass transport (i.e. radial flows of gas), which convert gravitational into kinetic energy, thereby driving turbulence. The latter model was specifically developed to describe the high values of $\sigma$ measured from ionised gas studies at $z>1$ \citep[e.g.][]{wisnioski_2015,johnson_2018}. 
            
            We also compare to the feedback-regulated model calibrated by \cite{rizzo_2024}, in which the cold-gas velocity dispersion is set by an energy-balance condition: supernovae inject a small fraction ($\lesssim 10\%$) turbulent kinetic energy at the same rate it is dissipated, yielding a characteristic scaling between turbulence and global SFR (with a fixed log-slope of 1/3). Using the extended sample of 57 discs over $z=0-5$ (which makes up most of the literature we compare to here), \cite{rizzo_2024} calibrate the normalization of this supernova-driven relation and embed it in a semi-empirical framework that predicts the redshift evolution of the dispersion by combining it with the evolving main-sequence SFR. We compare to this model in particular because it is empirically tuned to, and shown to reproduce, the locus and redshift evolution of existing cold-gas dispersion measurements, making it a more direct baseline for interpreting our observations.

            \begin{figure*}
                    \centering
                    \includegraphics[width=0.95\textwidth,trim={0cm 0.5cm 0cm 0.2cm},clip]{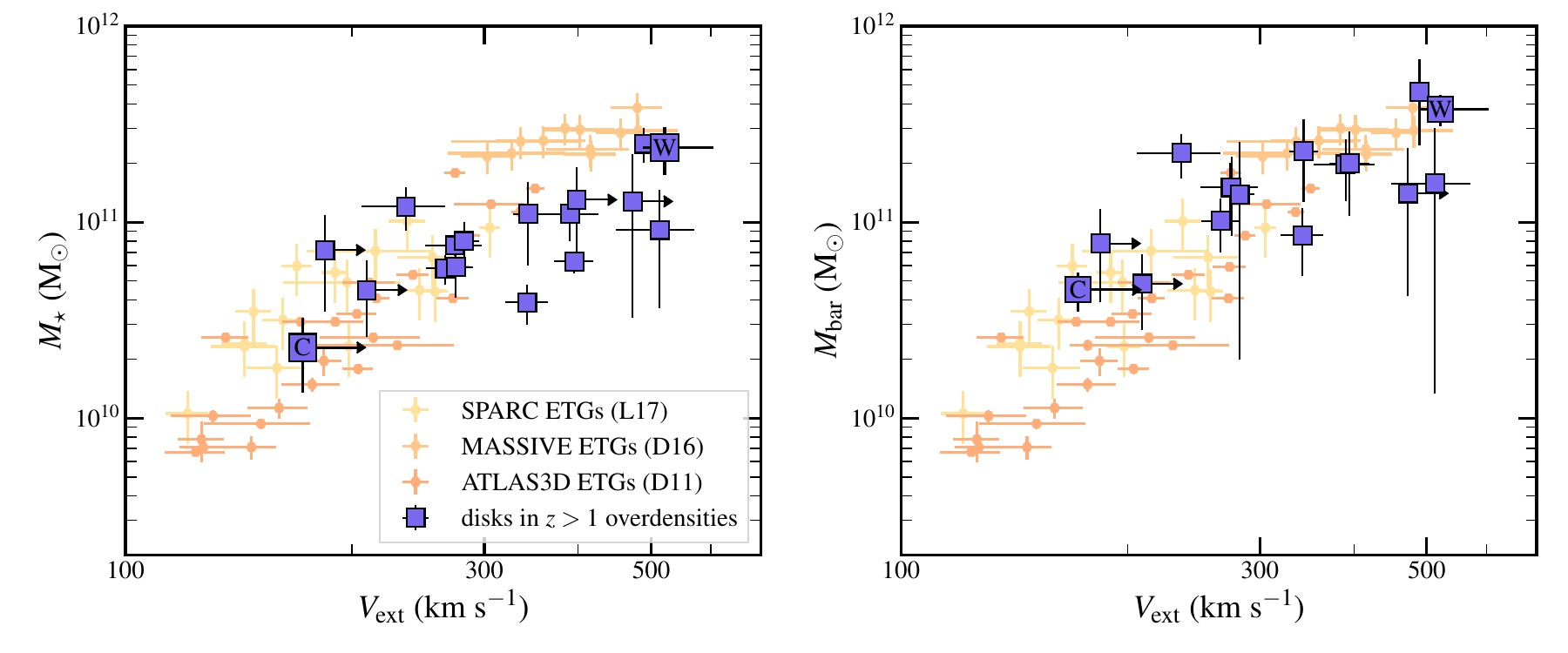}
                    \caption{Comparison of the circular velocities of the cold gas discs in local ETGs (orange circles) to $z>1$ disc galaxies in overdensities (purple squares). Left panel: the stellar masses vs rotation velocity. Right panel: same as the left panel but showing the total baryonic mass ($M_\mathrm{H_2}+M_\mathrm{\star}$) for the $z>1$ galaxies. For the $z>1$ literature sample, we assume  $V_\mathrm{ext}\sim V_\mathrm{circ}$ if $V/\sigma>5$. For $z>1$ discs with $V/\sigma < 5$ we show $V_\mathrm{ext}<V_\mathrm{circ}$ using lower limits.}
                    \label{fig:ETG}
                \end{figure*}

            The results for HyLIRG-W and ULIRG-C confirm the findings from several cold-gas studies over the last few years \citep[e.g.][]{girard_2021,roman-oliveira_2023,rizzo_2024}, namely that models incorporating both radial transport and feedback overpredict the cold-gas turbulence, whereas feedback-only models agree. In order for the feedback plus transport models to hold, we would need to systematically reduce the effective coupling between stellar feedback and the cold ISM (by at least $20\times$) and/or substantially suppress the amount of transport-driven stirring above the feedback floor (shifting it towards the feedback-only locus). Overall, the cold gas kinematics of these high SFR galaxies -- including HyLIRG-W and ULIRG-C -- imply that the effects of accretion (and additional tidal perturbations) primarily act to build up gas-rich, high–surface-density discs rather than sustaining strong, disc-wide random motions. This is consistent with a scenario in which most of the inflowing gas is transported inward through ordered streaming along the putative bars and/or other coherent structures, with only a small fraction of the associated kinetic energy converted into truly turbulent motions within the cold gas. 
            
            We acknowledge that there is a missing piece of the puzzle in our analysis -- direct tests of which regions may be undergoing the violent disc instabilities proposed to trigger extreme starbursts. Testing for these instabilities requires us to take into account that these galaxies are not razor-thin, nor gas-dominated, but are instead structures of gas embedded in a more massive stellar plus dark matter discs \citep{romeo_2011,elmegreen_2011,meidt_2022}. Recent work by \cite{bacchini_2024} has addressed this by modeling a vertically stratified gas disc in the combined potential of dark matter, stars, and gas, with the stability parameterised via a 3D Toomre-like criterion. When accounting for these effects, they find that the truly unstable regions decrease significantly (and are relegated to the outskirts of gas-rich discs at $z>1$). Similarly, \cite{meidt_2022} used a 3D dispersion relation including vertical structure and the stellar contribution to demonstrate ``partial 3D'' instabilities that can fragment gas and form molecular clouds even where the usual 2D Toomre analysis deems the total disc to be stable. Both these studies highlight how including the finite thickness and stellar potential is essential for correctly locating instabilities \citep[see also][]{nipoti_2023,nipoti_2024}. Because we have no constraints on the resolved stellar emission -- and hence densities -- we defer this analysis to upcoming work using the guaranteed JWST/NIRCam+MIRI observations.

        \subsection{Starbursts in $z\gtrsim 2$ overdensities as progenitors of today's most massive ellipticals}
            \label{sub:progenitors}

                Dusty star-forming galaxies in $z\gtrsim2$ overdensities are thought to evolve into today's most massive ellipticals based on three main arguments. (1) They are strongly clustered. Observations have revealed collections of submillimetre-bright galaxies that trace overdensities \citep{ivison_2000,zeballos_2018,calvi_2023}, which simulations predict to evolve into massive ellipticals in the centre of clusters and groups \citep{chiang_2013,muldrew_2018}. (2) These systems have sufficiently high gas masses and SFRs to form the large stellar mass of these ellipticals \citep{miller_2018,rotermund_2021}. (3) Some of the brightest systems already reside in massive halos. Galaxy evolution models indicate that galaxies in dense environments assemble mass earlier and more rapidly than field galaxies \citep{contini_2018} and have higher dynamical masses compared to the field \citep{delucia_2006}. Complementary to this, observations of several dust-rich, star-forming galaxies at $z\gtrsim4$ -- of which only a few are know to reside in overdensities -- have demonstrated that they follow scaling relations of local ETGs \citep{fraternali_2021,roman-oliveira_2024,amvrosiadis_2025}, indicating that they already reside in halos that are (or would become) as massive as those of $z\sim0$ ellipticals. Building on argument (3), we combine the literature sample from Sec.~\ref{sub:v_evolution} with galaxies W and C to test whether the gas- and dust-rich star-forming galaxies known to reside in $z\gtrsim2$ overdensities follow the same dynamical scaling relations as local ETGs, and hence whether they occupy similarly massive halos. 
                
                As discussed in Sec.~\ref{sub:v_evolution}, at $z\gtrsim2$ the disc galaxies with the highest rotation velocities reside in overdensities, supporting the argument that these galaxies assemble their mass earlier and more rapidly than field galaxies. In Fig.~\ref{fig:ETG}, we place the sample of rotating discs in $z>1.5$ overdensities in context with three sets of local ETGs, from \cite{davis_2011}, \cite{davis_2016}, and \cite{lelli_2017} (with the tables available in the supplementary materials). To ensure a physically consistent comparison across heterogeneous tracers, we adopt the least biased measurement available for the underlying gravitational potential in each galaxy, namely the outermost measured rotation velocities, $V_\mathrm{ext}$. We show $V_\mathrm{ext}$ versus stellar mass for all samples in the left panel and $V_\mathrm{ext}$ against $M_\mathrm{bar}$ for the $z>1$ galaxies only in the right panel (showing the stellar masses for the local ETGs).  Although these velocities arise at different absolute radii, using the largest accessible radius in each case minimises biases from central mass concentrations and non‑circular motions, and yields the closest approach to the asymptotic velocity each tracer can provide. 
                
                The external radii from which $V_\mathrm{ext}$ are derived differ across the ETG samples. For the ATLAS$^\mathrm{3D}$ sample of \cite{davis_2011,davis_2013} and the MASSIVE sample of \cite{davis_2016}, only global CO(2--1) line profiles are available. We therefore use the published values of the CO line width at 20\% of the peak flux, $W_\mathrm{20}$, and convert these to deprojected rotation velocities via $ W_\mathrm{20}/2\sin(i)$. Given the maximum CO-emitting radii measured for such ETGs, $r_\mathrm{max}=1-6$\,kpc \citep{young_2002,davis_2013,davis_2016,crocker_2011}, these measurements predominantly trace baryon-dominated regions and may be biased high relative to the outer disc due to the presence of central mass concentrations (e.g. bulges). The same is largely true for the $z>1$ discs, for which the measurements come from resolved CO, \ci, and \cii\ observations. Their maximum radii are comparable to the molecular gas discs in ETGs (HyLIRG-W being the largest $r_\mathrm{max}\sim6$\,kpc) and are also predominantly baryon-dominated. For rotation-dominated galaxies, these values will tend to $V_\mathrm{circ}$, whereas for dispersion-dominated discs, $V_\mathrm{ext} < V_\mathrm{circ}$. These are indicated by the lower limits in Fig.~\ref{fig:ETG}. To extend the dynamic range of the ETG sample, we also compare to the outer \hi\ rotation velocities for ETGs in the SPARC sample \cite{lelli_2017}, which are taken at 7--29\,kpc. These better probe the halo‑dominated regime, anchoring the high‑radius end of the relation.

                We find clear evidence for $z\gtrsim2$ disc galaxies in overdensities being suitable progenitors of todays' most massive ellipticals in overdensities. Fig.~\ref{fig:ETG} shows that these disc galaxies are mostly offset from ETGs when only considering their stellar mass. However, when adding the existing gas mass, they are consistent with these ETGs; that is, the external rotation velocities and total baryonic masses already match those of local ETGs. This implies that the most massive $z\gtrsim2$ discs in overdensities already reside in halos comparable to local ETGs. Indeed, many of these are comparable to the MASSIVE galaxy sample, of which most are Brightest Group Galaxies \citep{veale_2018}. HyLIRG-W and ADF22.1, in particular, are already consistent with the most massive of these BGGs. Given the short depletion and orbital times of these systems (a few 10s to a few 100s of Myr) there is ample time for these galaxies to quench and transform morphologically into the massive ellipticals in today's overdensities.

    \section{Conclusions}
        \label{sec:summary}

        We have presented new, $0\farcs15$ resolution ALMA observations of the CO(4--3), \ci\,1--0, and underlying dust-continuum emission in the five dusty starbursts (W, T, C, M, and E) known to reside in the $z\sim2.41$ protocluster HATLAS J0849. In this study we focus mostly on the kinematic classification and modelling using the high sensitivity CO(4--3) observations, leaving a detailed analysis of the morphologies and line excitation to a companion study. Our main results are as follows. 
        
        \begin{enumerate}[label=(\roman*),align=left, leftmargin=*, labelsep=.1em,topsep=0pt] 
            \item We find evidence of two rotating discs (W and C), two mergers (T and E), and one galaxy for which the classification is less certain (M). HyLIRG-W is a rotation-dominated, lopsided gas spiral with an external rotation velocity of $V_\mathrm{ext}\sim 520$\,\kms\ and a rotation-to-dispersion ratio of $V_\mathrm{ext}/\overline{\sigma}\sim10$. Contrary to previous studies, we find HyLIRG-T to be most consistent with a late-stage merger as it exhibits a kinematic major axis almost perpendicular to the morphological major axis, and significant asymmetries (which cannot be explained through lensing effects).  
            
            \item We find no clear evidence of molecular gas outflows, yet both rotating discs show deviations from a simple axisymmetric rotating disc. 
            Although some of these features could be explained by shocks, they are all consistent with the presence of $m=2$ modes, such as those induced by bars. 

            \item Combining the two rotating discs -- W and C -- with a growing sample of rotating cold gas discs in $z>1$ overdensities, we find that there is no systematic offset in the level of rotational support ($V/\sigma$) of galaxies in the field vs overdensities. However, the highest rotation velocities are exclusively occupied by massive starbursts in overdensities, supporting a picture in which galaxies in dense environments assemble mass more rapidly than field galaxies. 
            
            \item What is harder to explain with existing models is the high fraction ($\gtrsim42\%$) of rotation-dominated ($V/\sigma>5$) discs out of the population of massive, gas-rich, and highly star-forming galaxies in overdensities. These have yet to be systematically reproduced in simulations. A significant body of simulation-based results imply that starbursts triggered by vigorous accretion should only be rotation-dominated for short periods, while the accretion is coplanar. Moreover, although several simulations show that discs (re-)form on short enough timescales post-merger for the disc and starburst to coincide, there are no published simulations that reproduce highly rotation-dominated systems under this scenario. Thus, it is hard to gauge what fraction of such high $V/\sigma$ systems are accretion- vs merger-induced.

            \item The dominant drivers of turbulence do not appear to differ between $z>1$ galaxies in the field vs overdensities, with both occupying the same locus in the $\sigma-\mathrm{SFR}$ plane. As for other $z>1$ rotating discs with resolved cold gas observations, the turbulence within the cold gas of HyLIRG-W and ULIRG-C can be explained through feedback-driven models, without the need for a major contribution from the radial transport driven by violent disc instabilities -- which is surprising given theories of how such starbursts are triggered. We postulate that inflowing gas is effectively transported through ordered streaming (e.g. along the putative bars), such that only a small fraction of kinetic energy is converted into turbulent motions within the cold gas.  

            \item By comparing the rotation velocities and baryonic masses of $z\gtrsim2$ disc galaxies in ovderdensities with local early-type galaxies hosting cold gas discs, we show that these systems must already reside in halos of similar masses. Given the rapid timescales for quenching and morphological transformation, this provides strong support that the starbursts in $z\gtrsim2$ overdensities are the direct progenitors of today's most massive ellipticals (residing at the heart of local groups and clusters).
     
        \end{enumerate}

    \section*{Acknowledgements}

        We thank John McKean, Edward Taylor, and Lydia Haacke for helpful explanations on the impacts of strong and weak lensing. We also thank Filippo Fraternali, Toshiki Saito, and Tanio D\'{i}az Santos for useful discussions on local galaxies that may be analogous to those studied here and thank Eric Emsellem for his insights on bar formation. Special thanks to Enrico di Teodoro and Timothy Davis for creating well-documented public kinematic modelling tools and for providing advice on their use. MK acknowledges support from the Australian Research Council via the Discovery Early Career Researcher Award DE250100709. FR acknowledges support from the Dutch Research Council (NWO) through the Veni grant VI.Veni.222.146. CB acknowledges support from the Carlsberg Foundation Fellowship Programme by Carlsbergfondet. TT is supported by the JSPS Grant-in-Aid for Research Activity Start-up (25K23392) and the JSPS Core-to-Core Program (JPJSCCA20210003). FV acknowledges support from the Independent Research Fund Denmark (DFF) under grant 3120-00043B and the Danish National Research Foundation (DNRF) under grant 140.

    \section*{Data Availability}

        The ALMA data used in this work can be retrieved from the ALMA data archive\footnote{\url{https://almascience.nrao.edu/aq/}}\footnote{\url{https://almascience.eso.org/aq/}} using the program IDs 2023.1.00714.S and 2018.1.01146.S. The data products generated in this work will be made available upon request with the companion paper. The table of literature measurements used in this work are available in the online supplementary material. Both kinematic modelling tools used in this work are publicly available.


\bibliographystyle{mnras} 
\bibliography{bib_HyLIRG} 

@MISC{alma_handbook,
       author = {{Remijan}, A. and {Biggs}, A. and {Cortes}, P.~A. and {Dent}, B. and {Di Franceso}, J. and {Fomalont}, E. and {Hales}, A. and {Kameno}, S. and {Mason}, B. and {Philips}, N. and {Saini}, K. and {Vila Vilaro}, B. and {Villard}, E.},
        title = "{ALMA Technical Handbook,ALMA Doc. 7.3, ver. 1.1}",
 howpublished = {2019, ALMA Technical Handbook,ALMA Doc. 7.3, ver. 1.1ISBN 978-3-923524-66-2},
         year = 2019,
        month = jun,
          doi = {10.5281/zenodo.4511522},
       adsurl = {https://ui.adsabs.harvard.edu/abs/2019athb.rept.....R}
}

@ARTICLE{planck_2020,
       author = {{Planck Collaboration} and {Aghanim}, N. and {Akrami}, Y. and {Ashdown}, M. and {Aumont}, J. and {Baccigalupi}, C. and {Ballardini}, M. and {Banday}, A.~J. and {Barreiro}, R.~B. and {Bartolo}, N. and {Basak}, S. and {Battye}, R. and {Benabed}, K. and {Bernard}, J.-P. and {Bersanelli}, M. and {Bielewicz}, P. and {Bock}, J.~J. and {Bond}, J.~R. and {Borrill}, J. and {Bouchet}, F.~R. and {Boulanger}, F. and {Bucher}, M. and {Burigana}, C. and {Butler}, R.~C. and {Calabrese}, E. and {Cardoso}, J.-F. and {Carron}, J. and {Challinor}, A. and {Chiang}, H.~C. and {Chluba}, J. and {Colombo}, L.~P.~L. and {Combet}, C. and {Contreras}, D. and {Crill}, B.~P. and {Cuttaia}, F. and {de Bernardis}, P. and {de Zotti}, G. and {Delabrouille}, J. and {Delouis}, J.-M. and {Di Valentino}, E. and {Diego}, J.~M. and {Dor{\'e}}, O. and {Douspis}, M. and {Ducout}, A. and {Dupac}, X. and {Dusini}, S. and {Efstathiou}, G. and {Elsner}, F. and {En{\ss}lin}, T.~A. and {Eriksen}, H.~K. and {Fantaye}, Y. and {Farhang}, M. and {Fergusson}, J. and {Fernandez-Cobos}, R. and {Finelli}, F. and {Forastieri}, F. and {Frailis}, M. and {Fraisse}, A.~A. and {Franceschi}, E. and {Frolov}, A. and {Galeotta}, S. and {Galli}, S. and {Ganga}, K. and {G{\'e}nova-Santos}, R.~T. and {Gerbino}, M. and {Ghosh}, T. and {Gonz{\'a}lez-Nuevo}, J. and {G{\'o}rski}, K.~M. and {Gratton}, S. and {Gruppuso}, A. and {Gudmundsson}, J.~E. and {Hamann}, J. and {Handley}, W. and {Hansen}, F.~K. and {Herranz}, D. and {Hildebrandt}, S.~R. and {Hivon}, E. and {Huang}, Z. and {Jaffe}, A.~H. and {Jones}, W.~C. and {Karakci}, A. and {Keih{\"a}nen}, E. and {Keskitalo}, R. and {Kiiveri}, K. and {Kim}, J. and {Kisner}, T.~S. and {Knox}, L. and {Krachmalnicoff}, N. and {Kunz}, M. and {Kurki-Suonio}, H. and {Lagache}, G. and {Lamarre}, J.-M. and {Lasenby}, A. and {Lattanzi}, M. and {Lawrence}, C.~R. and {Le Jeune}, M. and {Lemos}, P. and {Lesgourgues}, J. and {Levrier}, F. and {Lewis}, A. and {Liguori}, M. and {Lilje}, P.~B. and {Lilley}, M. and {Lindholm}, V. and {L{\'o}pez-Caniego}, M. and {Lubin}, P.~M. and {Ma}, Y.-Z. and {Mac{\'\i}as-P{\'e}rez}, J.~F. and {Maggio}, G. and {Maino}, D. and {Mandolesi}, N. and {Mangilli}, A. and {Marcos-Caballero}, A. and {Maris}, M. and {Martin}, P.~G. and {Martinelli}, M. and {Mart{\'\i}nez-Gonz{\'a}lez}, E. and {Matarrese}, S. and {Mauri}, N. and {McEwen}, J.~D. and {Meinhold}, P.~R. and {Melchiorri}, A. and {Mennella}, A. and {Migliaccio}, M. and {Millea}, M. and {Mitra}, S. and {Miville-Desch{\^e}nes}, M.-A. and {Molinari}, D. and {Montier}, L. and {Morgante}, G. and {Moss}, A. and {Natoli}, P. and {N{\o}rgaard-Nielsen}, H.~U. and {Pagano}, L. and {Paoletti}, D. and {Partridge}, B. and {Patanchon}, G. and {Peiris}, H.~V. and {Perrotta}, F. and {Pettorino}, V. and {Piacentini}, F. and {Polastri}, L. and {Polenta}, G. and {Puget}, J.-L. and {Rachen}, J.~P. and {Reinecke}, M. and {Remazeilles}, M. and {Renzi}, A. and {Rocha}, G. and {Rosset}, C. and {Roudier}, G. and {Rubi{\~n}o-Mart{\'\i}n}, J.~A. and {Ruiz-Granados}, B. and {Salvati}, L. and {Sandri}, M. and {Savelainen}, M. and {Scott}, D. and {Shellard}, E.~P.~S. and {Sirignano}, C. and {Sirri}, G. and {Spencer}, L.~D. and {Sunyaev}, R. and {Suur-Uski}, A.-S. and {Tauber}, J.~A. and {Tavagnacco}, D. and {Tenti}, M. and {Toffolatti}, L. and {Tomasi}, M. and {Trombetti}, T. and {Valenziano}, L. and {Valiviita}, J. and {Van Tent}, B. and {Vibert}, L. and {Vielva}, P. and {Villa}, F. and {Vittorio}, N. and {Wandelt}, B.~D. and {Wehus}, I.~K. and {White}, M. and {White}, S.~D.~M. and {Zacchei}, A. and {Zonca}, A.},
        title = "{Planck 2018 results. VI. Cosmological parameters}",
      journal = {\aap},
         year = 2020,
        month = sep,
       volume = {641},
          eid = {A6},
        pages = {A6},
          doi = {10.1051/0004-6361/201833910},
archivePrefix = {arXiv},
       eprint = {1807.06209},
 primaryClass = {astro-ph.CO},
       adsurl = {https://ui.adsabs.harvard.edu/abs/2020A&A...641A...6P}
}

@book{BertsekasTsitsiklis2008,
  author    = {Dimitri P. Bertsekas and John N. Tsitsiklis},
  title     = {Introduction to Probability},
  edition   = {2nd},
  publisher = {Athena Scientific},
  address   = {Belmont, MA},
  year      = {2008},
  note      = {Reprinted 2019},
  isbn      = {978-1-886529-23-6}
}

@book{ThompsonMoranSwenson2017,
  author    = {Thompson, A. Richard and Moran, James M. and Swenson, George W.},
  title     = {Interferometry and Synthesis in Radio Astronomy},
  edition   = {3rd},
  publisher = {Springer},
  address   = {Cham},
  year      = {2017},
  isbn      = {978-3-319-44431-4},
  doi       = {10.1007/978-3-319-44431-4}
}

@ARTICLE{genzel_2011,
       author = {{Genzel}, R. and {Newman}, S. and {Jones}, T. and {F{\"o}rster Schreiber}, N.~M. and {Shapiro}, K. and {Genel}, S. and {Lilly}, S.~J. and {Renzini}, A. and {Tacconi}, L.~J. and {Bouch{\'e}}, N. and {Burkert}, A. and {Cresci}, G. and {Buschkamp}, P. and {Carollo}, C.~M. and {Ceverino}, D. and {Davies}, R. and {Dekel}, A. and {Eisenhauer}, F. and {Hicks}, E. and {Kurk}, J. and {Lutz}, D. and {Mancini}, C. and {Naab}, T. and {Peng}, Y. and {Sternberg}, A. and {Vergani}, D. and {Zamorani}, G.},
        title = "{The Sins Survey of z \raisebox{-0.5ex}\textasciitilde 2 Galaxy Kinematics: Properties of the Giant Star-forming Clumps}",
      journal = {\apj},
         year = 2011,
        month = jun,
       volume = {733},
       number = {2},
          eid = {101},
        pages = {101},
          doi = {10.1088/0004-637X/733/2/101},
archivePrefix = {arXiv},
       eprint = {1011.5360},
 primaryClass = {astro-ph.CO},
       adsurl = {https://ui.adsabs.harvard.edu/abs/2011ApJ...733..101G}
}

@ARTICLE{krumholz_2018,
       author = {{Krumholz}, Mark R. and {Burkhart}, Blakesley and {Forbes}, John C. and {Crocker}, Roland M.},
        title = "{A unified model for galactic discs: star formation, turbulence driving, and mass transport}",
      journal = {\mnras},
         year = 2018,
        month = jun,
       volume = {477},
       number = {2},
        pages = {2716-2740},
          doi = {10.1093/mnras/sty852},
archivePrefix = {arXiv},
       eprint = {1706.00106},
 primaryClass = {astro-ph.GA},
       adsurl = {https://ui.adsabs.harvard.edu/abs/2018MNRAS.477.2716K}
}

@ARTICLE{ee_1985,
       author = {{Elmegreen}, B.~G. and {Elmegreen}, D.~M.},
        title = "{Properties of barred spiral galaxies.}",
      journal = {\apj},
         year = 1985,
        month = jan,
       volume = {288},
        pages = {438-455},
          doi = {10.1086/162810},
       adsurl = {https://ui.adsabs.harvard.edu/abs/1985ApJ...288..438E}
}

@ARTICLE{ohta_1990,
       author = {{Ohta}, Kouji and {Hamabe}, Masaru and {Wakamatsu}, Ken-Ichi},
        title = "{Surface Photometry of Barred Spiral Galaxies}",
      journal = {\apj},
         year = 1990,
        month = jul,
       volume = {357},
        pages = {71},
          doi = {10.1086/168892},
       adsurl = {https://ui.adsabs.harvard.edu/abs/1990ApJ...357...71O}
}

@ARTICLE{laurikainen_2002,
       author = {{Laurikainen}, Eija and {Salo}, Heikki and {Rautiainen}, Pertti},
        title = "{Comparison of bar strengths in active and non-active galaxies}",
      journal = {\mnras},
         year = 2002,
        month = apr,
       volume = {331},
       number = {4},
        pages = {880-892},
          doi = {10.1046/j.1365-8711.2002.05243.x},
       adsurl = {https://ui.adsabs.harvard.edu/abs/2002MNRAS.331..880L}
}

@ARTICLE{elmegreen_1996,
       author = {{Elmegreen}, Debra Meloy and {Elmegreen}, Bruce G. and {Chromey}, Frederick R. and {Hasselbacher}, David Alan and {Bissell}, Bradley A.},
        title = "{Near-Infrared Observations of Isophotal Twists in Barred Spiral Galaxies}",
      journal = {\aj},
         year = 1996,
        month = may,
       volume = {111},
        pages = {1880},
          doi = {10.1086/117926},
       adsurl = {https://ui.adsabs.harvard.edu/abs/1996AJ....111.1880E}
}

@ARTICLE{laurikainen_2005,
       author = {{Laurikainen}, Eija and {Salo}, Heikki and {Buta}, Ronald},
        title = "{Multicomponent decompositions for a sample of S0 galaxies}",
      journal = {\mnras},
         year = 2005,
        month = oct,
       volume = {362},
       number = {4},
        pages = {1319-1347},
          doi = {10.1111/j.1365-2966.2005.09404.x},
archivePrefix = {arXiv},
       eprint = {astro-ph/0508097},
 primaryClass = {astro-ph},
       adsurl = {https://ui.adsabs.harvard.edu/abs/2005MNRAS.362.1319L}
}

@ARTICLE{gadotti_2007,
       author = {{Gadotti}, D.~A. and {Athanassoula}, E. and {Carrasco}, L. and {Bosma}, A. and {de Souza}, R.~E. and {Recillas}, E.},
        title = "{Near-infrared surface photometry of a sample of barred galaxies}",
      journal = {\mnras},
         year = 2007,
        month = nov,
       volume = {381},
       number = {3},
        pages = {943-961},
          doi = {10.1111/j.1365-2966.2007.12295.x},
archivePrefix = {arXiv},
       eprint = {0707.4599},
 primaryClass = {astro-ph},
       adsurl = {https://ui.adsabs.harvard.edu/abs/2007MNRAS.381..943G}
}

@ARTICLE{leconte_2024,
       author = {{Le Conte}, Zoe A. and {Gadotti}, Dimitri A. and {Ferreira}, Leonardo and {Conselice}, Christopher J. and {de S{\'a}-Freitas}, Camila and {Kim}, Taehyun and {Neumann}, Justus and {Fragkoudi}, Francesca and {Athanassoula}, E. and {Adams}, Nathan J.},
        title = "{A JWST investigation into the bar fraction at redshifts 1 {\ensuremath{\leq}} z {\ensuremath{\leq}} 3}",
      journal = {\mnras},
         year = 2024,
        month = may,
       volume = {530},
       number = {2},
        pages = {1984-2000},
          doi = {10.1093/mnras/stae921},
archivePrefix = {arXiv},
       eprint = {2309.10038},
 primaryClass = {astro-ph.GA},
       adsurl = {https://ui.adsabs.harvard.edu/abs/2024MNRAS.530.1984L}
}

@ARTICLE{guo_2025,
       author = {{Guo}, Yuchen and {Jogee}, Shardha and {Wise}, Eden and {Pritchett}, Keith and {McGrath}, Elizabeth J. and {Finkelstein}, Steven L. and {Iyer}, Kartheik G. and {Arrabal Haro}, Pablo and {Bagley}, Micaela B. and {Dickinson}, Mark and {Kartaltepe}, Jeyhan S. and {Koekemoer}, Anton M. and {Papovich}, Casey and {Pirzkal}, Nor and {Yung}, L.~Y. Aaron and {Backhaus}, Bren E. and {Bell}, Eric F. and {Bhatawdekar}, Rachana and {Cheng}, Yingjie and {Costantin}, Luca and {de la Vega}, Alexander and {Giavalisco}, Mauro and {Hathi}, Nimish P. and {Holwerda}, Benne W. and {Kurczynski}, Peter and {Lucas}, Ray A. and {Mobasher}, Bahram and {P{\'e}rez-Gonz{\'a}lez}, Pablo G. and {Pacucci}, Fabio},
        title = "{The Abundance and Properties of Barred Galaxies out to z {\ensuremath{\sim}} 4 Using JWST CEERS Data}",
      journal = {\apj},
         year = 2025,
        month = jun,
       volume = {985},
       number = {2},
          eid = {181},
        pages = {181},
          doi = {10.3847/1538-4357/adc8a7},
archivePrefix = {arXiv},
       eprint = {2409.06100},
 primaryClass = {astro-ph.GA},
       adsurl = {https://ui.adsabs.harvard.edu/abs/2025ApJ...985..181G}
}

@ARTICLE{gadotti_2011,
       author = {{Gadotti}, Dimitri A.},
        title = "{Secular evolution and structural properties of stellar bars in galaxies}",
      journal = {\mnras},
         year = 2011,
        month = aug,
       volume = {415},
       number = {4},
        pages = {3308-3318},
          doi = {10.1111/j.1365-2966.2011.18945.x},
archivePrefix = {arXiv},
       eprint = {1003.1719},
 primaryClass = {astro-ph.CO},
       adsurl = {https://ui.adsabs.harvard.edu/abs/2011MNRAS.415.3308G}
}

@ARTICLE{sheth_2012,
       author = {{Sheth}, Kartik and {Melbourne}, Jason and {Elmegreen}, Debra Meloy and {Elmegreen}, Bruce G. and {Athanassoula}, E. and {Abraham}, Roberto G. and {Weiner}, Benjamin J.},
        title = "{Hot Disks and Delayed Bar Formation}",
      journal = {\apj},
         year = 2012,
        month = oct,
       volume = {758},
       number = {2},
          eid = {136},
        pages = {136},
          doi = {10.1088/0004-637X/758/2/136},
archivePrefix = {arXiv},
       eprint = {1208.6304},
 primaryClass = {astro-ph.CO},
       adsurl = {https://ui.adsabs.harvard.edu/abs/2012ApJ...758..136S}
}

@ARTICLE{lokas_2016,
       author = {{{\L}okas}, Ewa L. and {Ebrov{\'a}}, Ivana and {del Pino}, Andr{\'e}s and {Sybilska}, Agnieszka and {Athanassoula}, E. and {Semczuk}, Marcin and {Gajda}, Grzegorz and {Fouquet}, Sylvain},
        title = "{Tidally Induced Bars of Galaxies in Clusters}",
      journal = {\apj},
         year = 2016,
        month = aug,
       volume = {826},
       number = {2},
          eid = {227},
        pages = {227},
          doi = {10.3847/0004-637X/826/2/227},
archivePrefix = {arXiv},
       eprint = {1601.07433},
 primaryClass = {astro-ph.GA},
       adsurl = {https://ui.adsabs.harvard.edu/abs/2016ApJ...826..227L}
}

@ARTICLE{lokas_2025,
       author = {{{\L}okas}, Ewa L.},
        title = "{High-redshift Merger-induced Bar-like Galaxies in IllustrisTNG}",
      journal = {\apjl},
         year = 2025,
        month = oct,
       volume = {991},
       number = {2},
          eid = {L52},
        pages = {L52},
          doi = {10.3847/2041-8213/ae07c2},
archivePrefix = {arXiv},
       eprint = {2509.22083},
 primaryClass = {astro-ph.GA},
       adsurl = {https://ui.adsabs.harvard.edu/abs/2025ApJ...991L..52L}
}

@ARTICLE{rosas_guevara_2022,
       author = {{Rosas-Guevara}, Yetli and {Bonoli}, Silvia and {Dotti}, Massimo and {Izquierdo-Villalba}, David and {Lupi}, Alessandro and {Zana}, Tommaso and {Bonetti}, Matteo and {Nelson}, Dylan and {Springel}, Volker and {Hernquist}, Lars and {Vogelsberger}, Mark},
        title = "{The evolution of the barred galaxy population in the TNG50 simulation}",
      journal = {\mnras},
         year = 2022,
        month = jun,
       volume = {512},
       number = {4},
        pages = {5339-5357},
          doi = {10.1093/mnras/stac816},
archivePrefix = {arXiv},
       eprint = {2110.04537},
 primaryClass = {astro-ph.GA},
       adsurl = {https://ui.adsabs.harvard.edu/abs/2022MNRAS.512.5339R}
}

@ARTICLE{rosas_guevara_2024,
       author = {{Rosas-Guevara}, Yetli and {Bonoli}, Silvia and {Misa Moreira}, Carmen and {Izquierdo-Villalba}, David},
        title = "{The rise and fall of bars in disc galaxies from z = 1 to z = 0. The role of environment}",
      journal = {\aap},
         year = 2024,
        month = apr,
       volume = {684},
          eid = {A179},
        pages = {A179},
          doi = {10.1051/0004-6361/202349003},
archivePrefix = {arXiv},
       eprint = {2401.15215},
 primaryClass = {astro-ph.GA},
       adsurl = {https://ui.adsabs.harvard.edu/abs/2024A&A...684A.179R}
}

@ARTICLE{melvin_2014,
       author = {{Melvin}, Thomas and {Masters}, Karen and {Lintott}, Chris and {Nichol}, Robert C. and {Simmons}, Brooke and {Bamford}, Steven P. and {Casteels}, Kevin R.~V. and {Cheung}, Edmond and {Edmondson}, Edward M. and {Fortson}, Lucy and {Schawinski}, Kevin and {Skibba}, Ramin A. and {Smith}, Arfon M. and {Willett}, Kyle W.},
        title = "{Galaxy Zoo: an independent look at the evolution of the bar fraction over the last eight billion years from HST-COSMOS}",
      journal = {\mnras},
         year = 2014,
        month = mar,
       volume = {438},
       number = {4},
        pages = {2882-2897},
          doi = {10.1093/mnras/stt2397},
archivePrefix = {arXiv},
       eprint = {1401.3334},
 primaryClass = {astro-ph.GA},
       adsurl = {https://ui.adsabs.harvard.edu/abs/2014MNRAS.438.2882M}
}

@ARTICLE{mendez-abreu_2012,
       author = {{M{\'e}ndez-Abreu}, J. and {S{\'a}nchez-Janssen}, R. and {Aguerri}, J.~A.~L. and {Corsini}, E.~M. and {Zarattini}, S.},
        title = "{The Nature and Nurture of Bars and Disks}",
      journal = {\apjl},
         year = 2012,
        month = dec,
       volume = {761},
       number = {1},
          eid = {L6},
        pages = {L6},
          doi = {10.1088/2041-8205/761/1/L6},
archivePrefix = {arXiv},
       eprint = {1211.1007},
 primaryClass = {astro-ph.CO},
       adsurl = {https://ui.adsabs.harvard.edu/abs/2012ApJ...761L...6M}
}

@ARTICLE{thompson_1981,
       author = {{Thompson}, L.~A.},
        title = "{Bar instabilities in Coma cluster galaxies}",
      journal = {\apjl},
         year = 1981,
        month = mar,
       volume = {244},
        pages = {L43-L45},
          doi = {10.1086/183476},
       adsurl = {https://ui.adsabs.harvard.edu/abs/1981ApJ...244L..43T}
}

@ARTICLE{fragkoudi_2025,
       author = {{Fragkoudi}, Francesca and {Grand}, Robert J.~J. and {Pakmor}, R{\"u}diger and {G{\'o}mez}, Facundo and {Marinacci}, Federico and {Springel}, Volker},
        title = "{Bar formation and evolution in the cosmological context: inputs from the Auriga simulations}",
      journal = {\mnras},
         year = 2025,
        month = apr,
       volume = {538},
       number = {3},
        pages = {1587-1608},
          doi = {10.1093/mnras/staf389},
archivePrefix = {arXiv},
       eprint = {2406.09453},
 primaryClass = {astro-ph.GA},
       adsurl = {https://ui.adsabs.harvard.edu/abs/2025MNRAS.538.1587F}
}

@ARTICLE{bournaud_2002,
       author = {{Bournaud}, F. and {Combes}, F.},
        title = "{Gas accretion on spiral galaxies: Bar formation and renewal}",
      journal = {\aap},
         year = 2002,
        month = sep,
       volume = {392},
        pages = {83-102},
          doi = {10.1051/0004-6361:20020920},
archivePrefix = {arXiv},
       eprint = {astro-ph/0206273},
 primaryClass = {astro-ph},
       adsurl = {https://ui.adsabs.harvard.edu/abs/2002A&A...392...83B}
}

@ARTICLE{bland-hawthorn_2024,
       author = {{Bland-Hawthorn}, Joss and {Tepper-Garcia}, Thor and {Agertz}, Oscar and {Federrath}, Christoph},
        title = "{Turbulent Gas-rich Disks at High Redshift: Bars and Bulges in a Radial Shear Flow}",
      journal = {\apj},
         year = 2024,
        month = jun,
       volume = {968},
       number = {2},
          eid = {86},
        pages = {86},
          doi = {10.3847/1538-4357/ad4118},
archivePrefix = {arXiv},
       eprint = {2402.06060},
 primaryClass = {astro-ph.GA},
       adsurl = {https://ui.adsabs.harvard.edu/abs/2024ApJ...968...86B}
}

@ARTICLE{tsukui_2024,
       author = {{Tsukui}, Takafumi and {Wisnioski}, Emily and {Bland-Hawthorn}, Joss and {Mai}, Yifan and {Iguchi}, Satoru and {Baba}, Junichi and {Freeman}, Ken},
        title = "{Detecting a disc bending wave in a barred-spiral galaxy at redshift 4.4}",
      journal = {\mnras},
         year = 2024,
        month = jan,
       volume = {527},
       number = {3},
        pages = {8941-8949},
          doi = {10.1093/mnras/stad3588},
archivePrefix = {arXiv},
       eprint = {2308.14798},
 primaryClass = {astro-ph.GA},
       adsurl = {https://ui.adsabs.harvard.edu/abs/2024MNRAS.527.8941T}
}

@ARTICLE{amvrosiadis_2025,
       author = {{Amvrosiadis}, A. and {Lange}, S. and {Nightingale}, J.~W. and {He}, Q. and {Frenk}, C.~S. and {Oman}, K.~A. and {Smail}, I. and {Swinbank}, A.~M. and {Fragkoudi}, F. and {Gadotti}, D.~A. and {Cole}, S. and {Borsato}, E. and {Robertson}, A. and {Massey}, R. and {Cao}, X. and {Li}, R.},
        title = "{The onset of bar formation in a massive galaxy at z \raisebox{-0.5ex}\textasciitilde 3.8}",
      journal = {\mnras},
         year = 2025,
        month = feb,
       volume = {537},
       number = {2},
        pages = {1163-1181},
          doi = {10.1093/mnras/staf048},
archivePrefix = {arXiv},
       eprint = {2404.01918},
 primaryClass = {astro-ph.GA},
       adsurl = {https://ui.adsabs.harvard.edu/abs/2025MNRAS.537.1163A}
}

@ARTICLE{huang_2025,
       author = {{Huang}, Shuo and {Kawabe}, Ryohei and {Umehata}, Hideki and {Kohno}, Kotaro and {Tamura}, Yoichi and {Saito}, Toshiki},
        title = "{Large gas inflow driven by a matured galactic bar in the early Universe}",
      journal = {\nat},
         year = 2025,
        month = may,
       volume = {641},
       number = {8064},
        pages = {861-865},
          doi = {10.1038/s41586-025-08914-2},
archivePrefix = {arXiv},
       eprint = {2511.16078},
 primaryClass = {astro-ph.GA},
       adsurl = {https://ui.adsabs.harvard.edu/abs/2025Natur.641..861H}
}

@ARTICLE{pastras_2025,
       author = {{Pastras}, Stavros and {Genzel}, Reinhard and {Tacconi}, Linda J. and {Schuster}, Karl and {Neri}, Roberto and {F{\"o}rster Schreiber}, Natascha M. and {Naab}, Thorsten and {Barfety}, Capucine and {Burkert}, Andreas and {Cao}, Yixian and {Chen}, Jianhang and {Combes}, Fran{\c{c}}oise and {Davies}, Ric and {Eisenhauer}, Frank and {Espejo Salcedo}, Juan M. and {Garc{\'\i}a-Burillo}, Santiago and {Herrera-Camus}, Rodrigo and {Jolly}, Jean-Baptiste and {Lee}, Lilian L. and {Lee}, Minju M. and {Liu}, Daizhong and {Lutz}, Dieter and {Nestor Shachar}, Amit and {Parlanti}, Eleonora and {Price}, Sedona H. and {Pulsoni}, Claudia and {Renzini}, Alvio and {Scaloni}, Letizia and {Shimizu}, Taro T. and {Springel}, Volker and {Sternberg}, Amiel and {Sturm}, Eckhard and {Tozzi}, Giulia and {Wuyts}, Stijn and {{\"U}bler}, Hannah},
        title = "{NOEMA$^{\rm 3D}$: A first kpc resolution study of a $z\sim1.5$ main sequence barred galaxy channeling gas into a growing bulge}",
      journal = {arXiv e-prints},
         year = 2025,
        month = may,
          eid = {arXiv:2505.07925},
        pages = {arXiv:2505.07925},
          doi = {10.48550/arXiv.2505.07925},
archivePrefix = {arXiv},
       eprint = {2505.07925},
 primaryClass = {astro-ph.GA},
       adsurl = {https://ui.adsabs.harvard.edu/abs/2025arXiv250507925P}
}

@ARTICLE{emsellem_2003,
       author = {{Emsellem}, Eric and {Goudfrooij}, Paul and {Ferruit}, Pierre},
        title = "{A two-arm gaseous spiral in the inner 200 pc of the early-type galaxy NGC 2974: signature of an inner bar}",
      journal = {\mnras},
         year = 2003,
        month = nov,
       volume = {345},
       number = {4},
        pages = {1297-1312},
          doi = {10.1046/j.1365-2966.2003.07050.x},
archivePrefix = {arXiv},
       eprint = {astro-ph/0308146},
 primaryClass = {astro-ph},
       adsurl = {https://ui.adsabs.harvard.edu/abs/2003MNRAS.345.1297E}
}

@ARTICLE{reynaud_1998,
       author = {{Reynaud}, D. and {Downes}, D.},
        title = "{Kinematics of the gas in a barred galaxy: do strong shocks inhibit star formation?}",
      journal = {\aap},
         year = 1998,
        month = sep,
       volume = {337},
        pages = {671-680},
       adsurl = {https://ui.adsabs.harvard.edu/abs/1998A&A...337..671R}
}

@ARTICLE{liang_2025,
       author = {{Laing}, Jennifer M. and {Wilson}, Christine D.},
        title = "{Increased Molecular Gas Velocity Dispersion and Star Formation Efficiency in Barred Galaxy Centres}",
      journal = {\aj},
         year = 2025,
        month = dec,
       volume = {170},
       number = {6},
          eid = {314},
        pages = {314},
          doi = {10.3847/1538-3881/ae0f8f},
archivePrefix = {arXiv},
       eprint = {2510.17768},
 primaryClass = {astro-ph.GA},
       adsurl = {https://ui.adsabs.harvard.edu/abs/2025AJ....170..314L}
}

@ARTICLE{chemin_2003,
       author = {{Chemin}, L. and {Cayatte}, V. and {Balkowski}, C. and {Marcelin}, M. and {Amram}, P. and {van Driel}, W. and {Flores}, H.},
        title = "{An H{\ensuremath{\alpha}}  study of the kinematics of NGC 3627}",
      journal = {\aap},
         year = 2003,
        month = jul,
       volume = {405},
        pages = {89-97},
          doi = {10.1051/0004-6361:20030590},
       adsurl = {https://ui.adsabs.harvard.edu/abs/2003A&A...405...89C}
}

@ARTICLE{pence_1984,
       author = {{Pence}, W.~D. and {Blackman}, C.~P.},
        title = "{Gas dynamics in barred spiral galaxies - II. NGC 7496 and 289.}",
      journal = {\mnras},
         year = 1984,
        month = oct,
       volume = {210},
        pages = {547-563},
          doi = {10.1093/mnras/210.3.547},
       adsurl = {https://ui.adsabs.harvard.edu/abs/1984MNRAS.210..547P}
}

@ARTICLE{lelli_2023,
       author = {{Lelli}, Federico and {Zhang}, Zhi-Yu and {Bisbas}, Thomas G. and {Lin}, Lingrui and {Papadopoulos}, Padelis and {Schombert}, James M. and {Di Teodoro}, Enrico and {Marasco}, Antonino and {McGaugh}, Stacy S.},
        title = "{Cold gas disks in main-sequence galaxies at cosmic noon: Low turbulence, flat rotation curves, and disk-halo degeneracy}",
      journal = {\aap},
         year = 2023,
        month = apr,
       volume = {672},
          eid = {A106},
        pages = {A106},
          doi = {10.1051/0004-6361/202245105},
archivePrefix = {arXiv},
       eprint = {2302.00030},
 primaryClass = {astro-ph.GA},
       adsurl = {https://ui.adsabs.harvard.edu/abs/2023A&A...672A.106L}
}

@ARTICLE{lelli_2016,
       author = {{Lelli}, Federico and {McGaugh}, Stacy S. and {Schombert}, James M.},
        title = "{SPARC: Mass Models for 175 Disk Galaxies with Spitzer Photometry and Accurate Rotation Curves}",
      journal = {\aj},
         year = 2016,
        month = dec,
       volume = {152},
       number = {6},
          eid = {157},
        pages = {157},
          doi = {10.3847/0004-6256/152/6/157},
archivePrefix = {arXiv},
       eprint = {1606.09251},
 primaryClass = {astro-ph.GA},
       adsurl = {https://ui.adsabs.harvard.edu/abs/2016AJ....152..157L}
}

@ARTICLE{wisnioski_2019,
       author = {{Wisnioski}, E. and {F{\"o}rster Schreiber}, N.~M. and {Fossati}, M. and {Mendel}, J.~T. and {Wilman}, D. and {Genzel}, R. and {Bender}, R. and {Wuyts}, S. and {Davies}, R.~L. and {{\"U}bler}, H. and {Bandara}, K. and {Beifiori}, A. and {Belli}, S. and {Brammer}, G. and {Chan}, J. and {Davies}, R.~I. and {Fabricius}, M. and {Galametz}, A. and {Lang}, P. and {Lutz}, D. and {Nelson}, E.~J. and {Momcheva}, I. and {Price}, S. and {Rosario}, D. and {Saglia}, R. and {Seitz}, S. and {Shimizu}, T. and {Tacconi}, L.~J. and {Tadaki}, K. and {van Dokkum}, P.~G. and {Wuyts}, E.},
        title = "{The KMOS$^{3D}$ Survey: Data Release and Final Survey Paper}",
      journal = {\apj},
         year = 2019,
        month = dec,
       volume = {886},
       number = {2},
          eid = {124},
        pages = {124},
          doi = {10.3847/1538-4357/ab4db8},
archivePrefix = {arXiv},
       eprint = {1909.11096},
 primaryClass = {astro-ph.GA},
       adsurl = {https://ui.adsabs.harvard.edu/abs/2019ApJ...886..124W}
}

@ARTICLE{kaasinen_2020,
       author = {{Kaasinen}, Melanie and {Walter}, Fabian and {Novak}, Mladen and {Neeleman}, Marcel and {Smail}, Ian and {Boogaard}, Leindert and {Cunha}, Elisabete da and {Weiss}, Axel and {Liu}, Daizhong and {Decarli}, Roberto and {Popping}, Gerg{\"o} and {Diaz-Santos}, Tanio and {Cort{\'e}s}, Paulo and {Aravena}, Manuel and {Werf}, Paul van der and {Riechers}, Dominik and {Inami}, Hanae and {Hodge}, Jacqueline A. and {Rix}, Hans-Walter and {Cox}, Pierre},
        title = "{A Comparison of the Stellar, CO, and Dust-continuum Emission from Three Star-forming HUDF Galaxies at z {\ensuremath{\sim}} 2}",
      journal = {\apj},
         year = 2020,
        month = aug,
       volume = {899},
       number = {1},
          eid = {37},
        pages = {37},
          doi = {10.3847/1538-4357/aba438},
archivePrefix = {arXiv},
       eprint = {2007.03697},
 primaryClass = {astro-ph.GA},
       adsurl = {https://ui.adsabs.harvard.edu/abs/2020ApJ...899...37K}
}

@ARTICLE{tadaki_2014,
       author = {{Tadaki}, Ken-ichi and {Kodama}, Tadayuki and {Tamura}, Yoichi and {Hayashi}, Masao and {Koyama}, Yusei and {Shimakawa}, Rhythm and {Tanaka}, Ichi and {Kohno}, Kotaro and {Hatsukade}, Bunyo and {Suzuki}, Kenta},
        title = "{Evidence for a Gas-rich Major Merger in a Proto-cluster at z = 2.5}",
      journal = {\apjl},
         year = 2014,
        month = jun,
       volume = {788},
       number = {2},
          eid = {L23},
        pages = {L23},
          doi = {10.1088/2041-8205/788/2/L23},
archivePrefix = {arXiv},
       eprint = {1403.0040},
 primaryClass = {astro-ph.GA},
       adsurl = {https://ui.adsabs.harvard.edu/abs/2014ApJ...788L..23T}
}

@ARTICLE{hine_2016,
       author = {{Hine}, N.~K. and {Geach}, J.~E. and {Alexander}, D.~M. and {Lehmer}, B.~D. and {Chapman}, S.~C. and {Matsuda}, Y.},
        title = "{An enhanced merger fraction within the galaxy population of the SSA22 protocluster at z = 3.1}",
      journal = {\mnras},
         year = 2016,
        month = jan,
       volume = {455},
       number = {3},
        pages = {2363-2370},
          doi = {10.1093/mnras/stv2448},
archivePrefix = {arXiv},
       eprint = {1506.05115},
 primaryClass = {astro-ph.GA},
       adsurl = {https://ui.adsabs.harvard.edu/abs/2016MNRAS.455.2363H}
}

@ARTICLE{coogan_2018,
       author = {{Coogan}, R.~T. and {Daddi}, E. and {Sargent}, M.~T. and {Strazzullo}, V. and {Valentino}, F. and {Gobat}, R. and {Magdis}, G. and {Bethermin}, M. and {Pannella}, M. and {Onodera}, M. and {Liu}, D. and {Cimatti}, A. and {Dannerbauer}, H. and {Carollo}, M. and {Renzini}, A. and {Tremou}, E.},
        title = "{Merger-driven star formation activity in Cl J1449+0856 at z = 1.99 as seen by ALMA and JVLA}",
      journal = {\mnras},
         year = 2018,
        month = sep,
       volume = {479},
       number = {1},
        pages = {703-729},
          doi = {10.1093/mnras/sty1446},
archivePrefix = {arXiv},
       eprint = {1805.09789},
 primaryClass = {astro-ph.GA},
       adsurl = {https://ui.adsabs.harvard.edu/abs/2018MNRAS.479..703C}
}

@ARTICLE{umehata_2021,
       author = {{Umehata}, Hideki and {Smail}, Ian and {Steidel}, Charles C. and {Hayes}, Matthew and {Scott}, Douglas and {Swinbank}, A.~M. and {Ivison}, R.~J. and {Nagao}, Toru and {Kubo}, Mariko and {Nakanishi}, Kouichiro and {Matsuda}, Yuichi and {Ikarashi}, Soh and {Tamura}, Yoichi and {Geach}, J.~E.},
        title = "{ALMA Observations of Ly{\ensuremath{\alpha}} Blob 1: Multiple Major Mergers and Widely Distributed Interstellar Media}",
      journal = {\apj},
         year = 2021,
        month = sep,
       volume = {918},
       number = {2},
          eid = {69},
        pages = {69},
          doi = {10.3847/1538-4357/ac1106},
archivePrefix = {arXiv},
       eprint = {2107.01162},
 primaryClass = {astro-ph.GA},
       adsurl = {https://ui.adsabs.harvard.edu/abs/2021ApJ...918...69U}
}

@ARTICLE{venkateshwaran_2024,
       author = {{Venkateshwaran}, Aparna and {Weiss}, Axel and {Sulzenauer}, Nikolaus and {Menten}, Karl and {Aravena}, Manuel and {Chapman}, Scott C. and {Gonzalez}, Anthony and {Gururajan}, Gayathri and {Hayward}, Christopher C. and {Hill}, Ryley and {Reuter}, Cassie and {Spilker}, Justin S. and {Vieira}, Joaquin D.},
        title = "{Kinematic Analysis of z = 4.3 Galaxies in the SPT2349{\textendash}56 Protocluster Core}",
      journal = {\apj},
         year = 2024,
        month = dec,
       volume = {977},
       number = {2},
          eid = {161},
        pages = {161},
          doi = {10.3847/1538-4357/ad7bb4},
archivePrefix = {arXiv},
       eprint = {2409.13823},
 primaryClass = {astro-ph.GA},
       adsurl = {https://ui.adsabs.harvard.edu/abs/2024ApJ...977..161V}
}

@ARTICLE{umehata_2025_morph,
       author = {{Umehata}, Hideki and {Kubo}, Mariko and {Smail}, Ian and {Lehmer}, Bret D. and {Monson}, Erik B. and {Nakanishi}, Kouichiro and {Matsuda}, Yuichi},
        title = "{ADF22-WEB: ALMA and JWST (sub)kpc-scale views of dusty star-forming galaxies in a $z\approx$3 proto-cluster}",
      journal = {arXiv e-prints},
         year = 2025,
        month = feb,
          eid = {arXiv:2502.01868},
        pages = {arXiv:2502.01868},
          doi = {10.48550/arXiv.2502.01868},
archivePrefix = {arXiv},
       eprint = {2502.01868},
 primaryClass = {astro-ph.GA},
       adsurl = {https://ui.adsabs.harvard.edu/abs/2025arXiv250201868U}
}

@ARTICLE{umehata_2025_kin,
       author = {{Umehata}, Hideki and {Steidel}, Charles C. and {Smail}, Ian and {Swinbank}, Mark and {Monson}, Erik B. and {Rosario}, David and {Lehmer}, Bret D. and {Nakanishi}, Kouichiro and {Kubo}, Mariko and {Iono}, Daisuke and {Alexander}, David M. and {Kohno}, Kotaro and {Tamura}, Yoichi and {Ivison}, Rob J. and {Saito}, Toshiki and {Mitsuhashi}, Ikki and {Huang}, Shuo and {Matsuda}, Yuichi},
        title = "{ADF22-WEB: A giant barred spiral starburst galaxy in the z = 3.1 SSA22 protocluster core}",
      journal = {\pasj},
         year = 2025,
        month = apr,
       volume = {77},
       number = {2},
        pages = {432-445},
          doi = {10.1093/pasj/psaf010},
       adsurl = {https://ui.adsabs.harvard.edu/abs/2025PASJ...77..432U}
}

@ARTICLE{wang_2025,
       author = {{Wang}, Weichen and {Cantalupo}, Sebastiano and {Pensabene}, Antonio and {Galbiati}, Marta and {Travascio}, Andrea and {Steidel}, Charles C. and {Maseda}, Michael V. and {Pezzulli}, Gabriele and {de Beer}, Stephanie and {Fossati}, Matteo and {Fumagalli}, Michele and {Gallego}, Sofia G. and {Lazeyras}, Titouan and {Mackenzie}, Ruari and {Matthee}, Jorryt and {Nanayakkara}, Themiya and {Quadri}, Giada},
        title = "{A giant disk galaxy two billion years after the Big Bang}",
      journal = {Nature Astronomy},
         year = 2025,
        month = may,
       volume = {9},
        pages = {710-719},
          doi = {10.1038/s41550-025-02500-2},
archivePrefix = {arXiv},
       eprint = {2409.17956},
 primaryClass = {astro-ph.GA},
       adsurl = {https://ui.adsabs.harvard.edu/abs/2025NatAs...9..710W}
}

@ARTICLE{lee_2019,
       author = {{Lee}, Minju M. and {Tanaka}, Ichi and {Kawabe}, Ryohei and {Aretxaga}, Itziar and {Hatsukade}, Bunyo and {Izumi}, Takuma and {Kajisawa}, Masaru and {Kodama}, Tadayuki and {Kohno}, Kotaro and {Nakanishi}, Kouichiro and {Saito}, Toshiki and {Tadaki}, Ken-ichi and {Tamura}, Yoichi and {Umehata}, Hideki and {Zeballos}, Milagros},
        title = "{A Radio-to-millimeter Census of Star-forming Galaxies in Protocluster 4C 23.56 at z = 2.5: Global and Local Gas Kinematics}",
      journal = {\apj},
         year = 2019,
        month = sep,
       volume = {883},
       number = {1},
          eid = {92},
        pages = {92},
          doi = {10.3847/1538-4357/ab3b5b},
archivePrefix = {arXiv},
       eprint = {1909.02028},
 primaryClass = {astro-ph.GA},
       adsurl = {https://ui.adsabs.harvard.edu/abs/2019ApJ...883...92L}
}

@ARTICLE{hernandez_2005,
       author = {{Hernandez}, O. and {Carignan}, C. and {Amram}, P. and {Chemin}, L. and {Daigle}, O.},
        title = "{BH{\ensuremath{\alpha}}BAR: big H{\ensuremath{\alpha}} kinematical sample of barred spiral galaxies - I. Fabry-Perot observations of 21 galaxies}",
      journal = {\mnras},
         year = 2005,
        month = jul,
       volume = {360},
       number = {4},
        pages = {1201-1230},
          doi = {10.1111/j.1365-2966.2005.09125.x},
archivePrefix = {arXiv},
       eprint = {astro-ph/0504393},
 primaryClass = {astro-ph},
       adsurl = {https://ui.adsabs.harvard.edu/abs/2005MNRAS.360.1201H}
}

@ARTICLE{bureau_1999_diagnostics,
       author = {{Bureau}, M. and {Athanassoula}, E.},
        title = "{Bar Diagnostics in Edge-on Spiral Galaxies. I. The Periodic Orbits Approach}",
      journal = {\apj},
         year = 1999,
        month = sep,
       volume = {522},
       number = {2},
        pages = {686-698},
          doi = {10.1086/307675},
archivePrefix = {arXiv},
       eprint = {astro-ph/9903061},
 primaryClass = {astro-ph},
       adsurl = {https://ui.adsabs.harvard.edu/abs/1999ApJ...522..686B}
}

@ARTICLE{merrifield_1999,
       author = {{Merrifield}, Michael R. and {Kuijken}, Konrad},
        title = "{Hidden bars and boxy bulges}",
      journal = {\aap},
         year = 1999,
        month = may,
       volume = {345},
        pages = {L47-L50},
          doi = {10.48550/arXiv.astro-ph/9904158},
archivePrefix = {arXiv},
       eprint = {astro-ph/9904158},
 primaryClass = {astro-ph},
       adsurl = {https://ui.adsabs.harvard.edu/abs/1999A&A...345L..47M}
}

@ARTICLE{randriamampandry_2015,
       author = {{Randriamampandry}, T.~H. and {Combes}, F. and {Carignan}, C. and {Deg}, N.},
        title = "{Estimating non-circular motions in barred galaxies using numerical N-body simulations}",
      journal = {\mnras},
         year = 2015,
        month = dec,
       volume = {454},
       number = {4},
        pages = {3743-3759},
          doi = {10.1093/mnras/stv2147},
archivePrefix = {arXiv},
       eprint = {1509.04881},
 primaryClass = {astro-ph.GA},
       adsurl = {https://ui.adsabs.harvard.edu/abs/2015MNRAS.454.3743R}
}

@ARTICLE{huettemeister_2000,
       author = {{H{\"u}ttemeister}, S. and {Aalto}, S. and {Das}, M. and {Wall}, W.~F.},
        title = "{Changing molecular gas properties in the bar and center of NGC 7479}",
      journal = {\aap},
         year = 2000,
        month = nov,
       volume = {363},
        pages = {93-107},
       adsurl = {https://ui.adsabs.harvard.edu/abs/2000A&A...363...93H}
}

@ARTICLE{lundgren_2004,
       author = {{Lundgren}, A.~A. and {Olofsson}, H. and {Wiklind}, T. and {Rydbeck}, G.},
        title = "{Molecular gas in the galaxy M 83.  II. Kinematics of the molecular gas}",
      journal = {\aap},
         year = 2004,
        month = aug,
       volume = {422},
        pages = {865-881},
          doi = {10.1051/0004-6361:20040230},
archivePrefix = {arXiv},
       eprint = {astro-ph/0404026},
 primaryClass = {astro-ph},
       adsurl = {https://ui.adsabs.harvard.edu/abs/2004A&A...422..865L}
}

@ARTICLE{rizzo_2022,
       author = {{Rizzo}, F. and {Kohandel}, M. and {Pallottini}, A. and {Zanella}, A. and {Ferrara}, A. and {Vallini}, L. and {Toft}, S.},
        title = "{Dynamical characterization of galaxies up to z {\ensuremath{\sim}} 7}",
      journal = {\aap},
         year = 2022,
        month = nov,
       volume = {667},
          eid = {A5},
        pages = {A5},
          doi = {10.1051/0004-6361/202243582},
archivePrefix = {arXiv},
       eprint = {2204.05325},
 primaryClass = {astro-ph.GA},
       adsurl = {https://ui.adsabs.harvard.edu/abs/2022A&A...667A...5R}
}

@ARTICLE{davis_2017,
       author = {{Davis}, Timothy A. and {Bureau}, Martin and {Onishi}, Kyoko and {Cappellari}, Michele and {Iguchi}, Satoru and {Sarzi}, Marc},
        title = "{WISDOM Project - II. Molecular gas measurement of the supermassive black hole mass in NGC 4697}",
      journal = {\mnras},
         year = 2017,
        month = jul,
       volume = {468},
       number = {4},
        pages = {4675-4690},
          doi = {10.1093/mnras/stw3217},
archivePrefix = {arXiv},
       eprint = {1703.05248},
 primaryClass = {astro-ph.GA},
       adsurl = {https://ui.adsabs.harvard.edu/abs/2017MNRAS.468.4675D}
}

@software{gastimator,
       author = {{Davis}, Timothy A.},
        title = "{GAStimator: Python MCMC gibbs-sampler with adaptive stepping}",
 howpublished = {Astrophysics Source Code Library, record ascl:2406.001},
         year = 2024,
        month = jun,
          eid = {ascl:2406.001},
archivePrefix = {ascl},
       eprint = {2406.001},
       adsurl = {https://ui.adsabs.harvard.edu/abs/2024ascl.soft06001D}
}

@software{kinms,
       author = {{Davis}, Timothy A. and {Zabel}, Nikki and {Dawson}, James M.},
        title = "{KinMS: Three-dimensional kinematic modeling of arbitrary gas distributions}",
 howpublished = {Astrophysics Source Code Library, record ascl:2006.003},
         year = 2020,
        month = jun,
          eid = {ascl:2006.003},
       adsurl = {https://ui.adsabs.harvard.edu/abs/2020ascl.soft06003D}
}

@ARTICLE{3dbarolo,
       author = {{Di Teodoro}, E.~M. and {Fraternali}, F.},
        title = "{$^{3D}$ BAROLO: a new 3D algorithm to derive rotation curves of galaxies}",
      journal = {\mnras},
         year = 2015,
        month = aug,
       volume = {451},
       number = {3},
        pages = {3021-3033},
          doi = {10.1093/mnras/stv1213},
archivePrefix = {arXiv},
       eprint = {1505.07834},
 primaryClass = {astro-ph.GA},
       adsurl = {https://ui.adsabs.harvard.edu/abs/2015MNRAS.451.3021D}
}

@ARTICLE{pensabene_2025,
       author = {{Pensabene}, A. and {Cantalupo}, S. and {Wang}, W. and {Bacchini}, C. and {Fraternali}, F. and {Bischetti}, M. and {Cicone}, C. and {Decarli}, R. and {Pezzulli}, G. and {Galbiati}, M. and {Lazeyras}, T. and {Ledos}, N. and {Quadri}, G. and {Travascio}, A.},
        title = "{ALMA survey of a massive node of the Cosmic Web at $z\sim 3$. II. A dynamically cold and massive disk galaxy in the proximity of a hyperluminous quasar}",
      journal = {arXiv e-prints},
         year = 2025,
        month = jul,
          eid = {arXiv:2507.16921},
        pages = {arXiv:2507.16921},
          doi = {10.48550/arXiv.2507.16921},
archivePrefix = {arXiv},
       eprint = {2507.16921},
 primaryClass = {astro-ph.GA},
       adsurl = {https://ui.adsabs.harvard.edu/abs/2025arXiv250716921P}
}

@ARTICLE{leroy_2015,
       author = {{Leroy}, Adam K. and {Walter}, Fabian and {Martini}, Paul and {Roussel}, H{\'e}l{\`e}ne and {Sandstrom}, Karin and {Ott}, J{\"u}rgen and {Weiss}, Axel and {Bolatto}, Alberto D. and {Schuster}, Karl and {Dessauges-Zavadsky}, Miroslava},
        title = "{The Multi-phase Cold Fountain in M82 Revealed by a Wide, Sensitive Map of the Molecular Interstellar Medium}",
      journal = {\apj},
         year = 2015,
        month = dec,
       volume = {814},
       number = {2},
          eid = {83},
        pages = {83},
          doi = {10.1088/0004-637X/814/2/83},
archivePrefix = {arXiv},
       eprint = {1509.02932},
 primaryClass = {astro-ph.GA},
       adsurl = {https://ui.adsabs.harvard.edu/abs/2015ApJ...814...83L}
}

@ARTICLE{leroy_2021,
       author = {{Leroy}, Adam K. and {Hughes}, Annie and {Liu}, Daizhong and {Pety}, J{\'e}r{\^o}me and {Rosolowsky}, Erik and {Saito}, Toshiki and {Schinnerer}, Eva and {Schruba}, Andreas and {Usero}, Antonio and {Faesi}, Christopher M. and {Herrera}, Cinthya N. and {Chevance}, M{\'e}lanie and {Hygate}, Alexander P.~S. and {Kepley}, Amanda A. and {Koch}, Eric W. and {Querejeta}, Miguel and {Sliwa}, Kazimierz and {Will}, David and {Wilson}, Christine D. and {Anand}, Gagandeep S. and {Barnes}, Ashley and {Belfiore}, Francesco and {Be{\v{s}}li{\'c}}, Ivana and {Bigiel}, Frank and {Blanc}, Guillermo A. and {Bolatto}, Alberto D. and {Boquien}, M{\`e}d{\`e}ric and {Cao}, Yixian and {Chandar}, Rupali and {Chastenet}, J{\'e}r{\'e}my and {Chiang}, I-Da and {Congiu}, Enrico and {Dale}, Daniel A. and {Deger}, Sinan and {den Brok}, Jakob S. and {Eibensteiner}, Cosima and {Emsellem}, Eric and {Garc{\'\i}a-Rodr{\'\i}guez}, Axel and {Glover}, Simon C.~O. and {Grasha}, Kathryn and {Groves}, Brent and {Henshaw}, Jonathan D. and {Jim{\'e}nez Donaire}, Mar{\'\i}a J. and {Kim}, Jaeyeon and {Klessen}, Ralf S. and {Kreckel}, Kathryn and {Kruijssen}, J.~M. Diederik and {Larson}, Kirsten L. and {Lee}, Janice C. and {Mayker}, Ness and {McElroy}, Rebecca and {Meidt}, Sharon E. and {Mok}, Angus and {Pan}, Hsi-An and {Puschnig}, Johannes and {Razza}, Alessandro and {S{\'a}nchez-Bl'azquez}, Patricia and {Sandstrom}, Karin M. and {Santoro}, Francesco and {Sardone}, Amy and {Scheuermann}, Fabian and {Sun}, Jiayi and {Thilker}, David A. and {Turner}, Jordan A. and {Ubeda}, Leonardo and {Utomo}, Dyas and {Watkins}, Elizabeth J. and {Williams}, Thomas G.},
        title = "{PHANGS-ALMA Data Processing and Pipeline}",
      journal = {\apjs},
         year = 2021,
        month = jul,
       volume = {255},
       number = {1},
          eid = {19},
        pages = {19},
          doi = {10.3847/1538-4365/abec80},
archivePrefix = {arXiv},
       eprint = {2104.07665},
 primaryClass = {astro-ph.IM},
       adsurl = {https://ui.adsabs.harvard.edu/abs/2021ApJS..255...19L}
}

@ARTICLE{roman-oliveira_2023,
       author = {{Roman-Oliveira}, Fernanda and {Fraternali}, Filippo and {Rizzo}, Francesca},
        title = "{Regular rotation and low turbulence in a diverse sample of z {\ensuremath{\sim}} 4.5 galaxies observed with ALMA}",
      journal = {\mnras},
         year = 2023,
        month = may,
       volume = {521},
       number = {1},
        pages = {1045-1065},
          doi = {10.1093/mnras/stad530},
archivePrefix = {arXiv},
       eprint = {2302.03049},
 primaryClass = {astro-ph.GA},
       adsurl = {https://ui.adsabs.harvard.edu/abs/2023MNRAS.521.1045R}
}

@ARTICLE{fraternali_2021,
       author = {{Fraternali}, F. and {Karim}, A. and {Magnelli}, B. and {G{\'o}mez-Guijarro}, C. and {Jim{\'e}nez-Andrade}, E.~F. and {Posses}, A.~C.},
        title = "{Fast rotating and low-turbulence discs at z ≃ 4.5: Dynamical evidence of their evolution into local early-type galaxies}",
      journal = {\aap},
         year = 2021,
        month = mar,
       volume = {647},
          eid = {A194},
        pages = {A194},
          doi = {10.1051/0004-6361/202039807},
archivePrefix = {arXiv},
       eprint = {2011.05340},
 primaryClass = {astro-ph.GA},
       adsurl = {https://ui.adsabs.harvard.edu/abs/2021A&A...647A.194F}
}

@ARTICLE{roman-oliveira_2024,
       author = {{Roman-Oliveira}, F. and {Rizzo}, F. and {Fraternali}, F.},
        title = "{Dynamical modelling and the origin of gas turbulence in z {\ensuremath{\sim}} 4.5 galaxies}",
      journal = {\aap},
         year = 2024,
        month = jul,
       volume = {687},
          eid = {A35},
        pages = {A35},
          doi = {10.1051/0004-6361/202348828},
archivePrefix = {arXiv},
       eprint = {2403.00904},
 primaryClass = {astro-ph.GA},
       adsurl = {https://ui.adsabs.harvard.edu/abs/2024A&A...687A..35R}
}

@ARTICLE{hopkins_2006,
       author = {{Hopkins}, Philip F. and {Hernquist}, Lars and {Cox}, Thomas J. and {Di Matteo}, Tiziana and {Robertson}, Brant and {Springel}, Volker},
        title = "{A Unified, Merger-driven Model of the Origin of Starbursts, Quasars, the Cosmic X-Ray Background, Supermassive Black Holes, and Galaxy Spheroids}",
      journal = {\apjs},
         year = 2006,
        month = mar,
       volume = {163},
       number = {1},
        pages = {1-49},
          doi = {10.1086/499298},
archivePrefix = {arXiv},
       eprint = {astro-ph/0506398},
 primaryClass = {astro-ph},
       adsurl = {https://ui.adsabs.harvard.edu/abs/2006ApJS..163....1H}
}

@ARTICLE{dave_2010,
       author = {{Dav{\'e}}, Romeel and {Finlator}, Kristian and {Oppenheimer}, Benjamin D. and {Fardal}, Mark and {Katz}, Neal and {Kere{\v{s}}}, Du{\v{s}}an and {Weinberg}, David H.},
        title = "{The nature of submillimetre galaxies in cosmological hydrodynamic simulations}",
      journal = {\mnras},
         year = 2010,
        month = may,
       volume = {404},
       number = {3},
        pages = {1355-1368},
          doi = {10.1111/j.1365-2966.2010.16395.x},
archivePrefix = {arXiv},
       eprint = {0909.4078},
 primaryClass = {astro-ph.CO},
       adsurl = {https://ui.adsabs.harvard.edu/abs/2010MNRAS.404.1355D}
}

@ARTICLE{narayanan_2015,
       author = {{Narayanan}, Desika and {Turk}, Matthew and {Feldmann}, Robert and {Robitaille}, Thomas and {Hopkins}, Philip and {Thompson}, Robert and {Hayward}, Christopher and {Ball}, David and {Faucher-Gigu{\`e}re}, Claude-Andr{\'e} and {Kere{\v{s}}}, Du{\v{s}}an},
        title = "{The formation of submillimetre-bright galaxies from gas infall over a billion years}",
      journal = {\nat},
         year = 2015,
        month = sep,
       volume = {525},
       number = {7570},
        pages = {496-499},
          doi = {10.1038/nature15383},
archivePrefix = {arXiv},
       eprint = {1509.06377},
 primaryClass = {astro-ph.GA},
       adsurl = {https://ui.adsabs.harvard.edu/abs/2015Natur.525..496N}
}

@ARTICLE{hopkins_2013,
       author = {{Hopkins}, Philip F. and {Cox}, Thomas J. and {Hernquist}, Lars and {Narayanan}, Desika and {Hayward}, Christopher C. and {Murray}, Norman},
        title = "{Star formation in galaxy mergers with realistic models of stellar feedback and the interstellar medium}",
      journal = {\mnras},
         year = 2013,
        month = apr,
       volume = {430},
       number = {3},
        pages = {1901-1927},
          doi = {10.1093/mnras/stt017},
archivePrefix = {arXiv},
       eprint = {1206.0011},
 primaryClass = {astro-ph.CO},
       adsurl = {https://ui.adsabs.harvard.edu/abs/2013MNRAS.430.1901H}
}

@ARTICLE{sparre_2016,
       author = {{Sparre}, Martin and {Springel}, Volker},
        title = "{Zooming in on major mergers: dense, starbursting gas in cosmological simulations}",
      journal = {\mnras},
         year = 2016,
        month = nov,
       volume = {462},
       number = {3},
        pages = {2418-2430},
          doi = {10.1093/mnras/stw1793},
archivePrefix = {arXiv},
       eprint = {1604.08205},
 primaryClass = {astro-ph.GA},
       adsurl = {https://ui.adsabs.harvard.edu/abs/2016MNRAS.462.2418S}
}

@ARTICLE{lower_2023,
       author = {{Lower}, Sidney and {Narayanan}, Desika and {Li}, Qi and {Dav{\'e}}, Romeel},
        title = "{Cosmic Sands: The Origin of Dusty, Star-forming Galaxies in the Epoch of Reionization}",
      journal = {\apj},
         year = 2023,
        month = jun,
       volume = {950},
       number = {2},
          eid = {94},
        pages = {94},
          doi = {10.3847/1538-4357/accf8c},
archivePrefix = {arXiv},
       eprint = {2212.02636},
 primaryClass = {astro-ph.GA},
       adsurl = {https://ui.adsabs.harvard.edu/abs/2023ApJ...950...94L}
}

@ARTICLE{robertson_2006,
       author = {{Robertson}, Brant and {Bullock}, James S. and {Cox}, Thomas J. and {Di Matteo}, Tiziana and {Hernquist}, Lars and {Springel}, Volker and {Yoshida}, Naoki},
        title = "{A Merger-driven Scenario for Cosmological Disk Galaxy Formation}",
      journal = {\apj},
         year = 2006,
        month = jul,
       volume = {645},
       number = {2},
        pages = {986-1000},
          doi = {10.1086/504412},
archivePrefix = {arXiv},
       eprint = {astro-ph/0503369},
 primaryClass = {astro-ph},
       adsurl = {https://ui.adsabs.harvard.edu/abs/2006ApJ...645..986R}
}

@ARTICLE{hopkins_2009,
       author = {{Hopkins}, Philip F. and {Cox}, Thomas J. and {Younger}, Joshua D. and {Hernquist}, Lars},
        title = "{How do Disks Survive Mergers?}",
      journal = {\apj},
         year = 2009,
        month = feb,
       volume = {691},
       number = {2},
        pages = {1168-1201},
          doi = {10.1088/0004-637X/691/2/1168},
archivePrefix = {arXiv},
       eprint = {0806.1739},
 primaryClass = {astro-ph},
       adsurl = {https://ui.adsabs.harvard.edu/abs/2009ApJ...691.1168H}
}

@ARTICLE{governato_2009,
       author = {{Governato}, F. and {Brook}, C.~B. and {Brooks}, A.~M. and {Mayer}, L. and {Willman}, B. and {Jonsson}, P. and {Stilp}, A.~M. and {Pope}, L. and {Christensen}, C. and {Wadsley}, J. and {Quinn}, T.},
        title = "{Forming a large disc galaxy from a z < 1 major merger}",
      journal = {\mnras},
         year = 2009,
        month = sep,
       volume = {398},
       number = {1},
        pages = {312-320},
          doi = {10.1111/j.1365-2966.2009.15143.x},
archivePrefix = {arXiv},
       eprint = {0812.0379},
 primaryClass = {astro-ph},
       adsurl = {https://ui.adsabs.harvard.edu/abs/2009MNRAS.398..312G}
}

@ARTICLE{sotillo-ramos_2022,
       author = {{Sotillo-Ramos}, Diego and {Pillepich}, Annalisa and {Donnari}, Martina and {Nelson}, Dylan and {Eisert}, Lukas and {Rodriguez-Gomez}, Vicente and {Joshi}, Gandhali and {Vogelsberger}, Mark and {Hernquist}, Lars},
        title = "{The merger and assembly histories of Milky Way- and M31-like galaxies with TNG50: disc survival through mergers}",
      journal = {\mnras},
         year = 2022,
        month = nov,
       volume = {516},
       number = {4},
        pages = {5404-5427},
          doi = {10.1093/mnras/stac2586},
archivePrefix = {arXiv},
       eprint = {2211.00036},
 primaryClass = {astro-ph.GA},
       adsurl = {https://ui.adsabs.harvard.edu/abs/2022MNRAS.516.5404S}
}

@ARTICLE{lotz_2008,
       author = {{Lotz}, Jennifer M. and {Jonsson}, Patrik and {Cox}, T.~J. and {Primack}, Joel R.},
        title = "{Galaxy merger morphologies and time-scales from simulations of equal-mass gas-rich disc mergers}",
      journal = {\mnras},
         year = 2008,
        month = dec,
       volume = {391},
       number = {3},
        pages = {1137-1162},
          doi = {10.1111/j.1365-2966.2008.14004.x},
archivePrefix = {arXiv},
       eprint = {0805.1246},
 primaryClass = {astro-ph},
       adsurl = {https://ui.adsabs.harvard.edu/abs/2008MNRAS.391.1137L}
}

@ARTICLE{renaud_2022,
       author = {{Renaud}, Florent and {Segovia Otero}, {\'A}lvaro and {Agertz}, Oscar},
        title = "{The merger-starburst connection across cosmic times}",
      journal = {\mnras},
         year = 2022,
        month = nov,
       volume = {516},
       number = {4},
        pages = {4922-4931},
          doi = {10.1093/mnras/stac2557},
archivePrefix = {arXiv},
       eprint = {2209.03983},
 primaryClass = {astro-ph.GA},
       adsurl = {https://ui.adsabs.harvard.edu/abs/2022MNRAS.516.4922R}
}

@ARTICLE{martig_2009,
       author = {{Martig}, Marie and {Bournaud}, Fr{\'e}d{\'e}ric and {Teyssier}, Romain and {Dekel}, Avishai},
        title = "{Morphological Quenching of Star Formation: Making Early-Type Galaxies Red}",
      journal = {\apj},
         year = 2009,
        month = dec,
       volume = {707},
       number = {1},
        pages = {250-267},
          doi = {10.1088/0004-637X/707/1/250},
archivePrefix = {arXiv},
       eprint = {0905.4669},
 primaryClass = {astro-ph.CO},
       adsurl = {https://ui.adsabs.harvard.edu/abs/2009ApJ...707..250M}
}

@ARTICLE{petersson_2023,
       author = {{Petersson}, Jonathan and {Renaud}, Florent and {Agertz}, Oscar and {Dekel}, Avishai and {Duc}, Pierre-Alain},
        title = "{From starburst to quenching: merger-driven evolution of the star formation regimes in a shell galaxy}",
      journal = {\mnras},
         year = 2023,
        month = jan,
       volume = {518},
       number = {3},
        pages = {3261-3273},
          doi = {10.1093/mnras/stac3136},
archivePrefix = {arXiv},
       eprint = {2210.16333},
 primaryClass = {astro-ph.GA},
       adsurl = {https://ui.adsabs.harvard.edu/abs/2023MNRAS.518.3261P}
}

@ARTICLE{jog_1992,
       author = {{Jog}, Chanda J. and {Solomon}, P.~M.},
        title = "{A Triggering Mechanism for Enhanced Star Formation in Colliding Galaxies}",
      journal = {\apj},
         year = 1992,
        month = mar,
       volume = {387},
        pages = {152},
          doi = {10.1086/171067},
       adsurl = {https://ui.adsabs.harvard.edu/abs/1992ApJ...387..152J}
}

@ARTICLE{di_matteo_2008,
       author = {{Di Matteo}, P. and {Bournaud}, F. and {Martig}, M. and {Combes}, F. and {Melchior}, A.-L. and {Semelin}, B.},
        title = "{On the frequency, intensity, and duration of starburst episodes triggered by galaxy interactions and mergers}",
      journal = {\aap},
         year = 2008,
        month = dec,
       volume = {492},
       number = {1},
        pages = {31-49},
          doi = {10.1051/0004-6361:200809480},
archivePrefix = {arXiv},
       eprint = {0809.2592},
 primaryClass = {astro-ph},
       adsurl = {https://ui.adsabs.harvard.edu/abs/2008A&A...492...31D}
}

@ARTICLE{renaud_2014,
       author = {{Renaud}, F. and {Bournaud}, F. and {Kraljic}, K. and {Duc}, P.-A.},
        title = "{Starbursts triggered by intergalactic tides andinterstellar compressive turbulence.}",
      journal = {\mnras},
         year = 2014,
        month = jul,
       volume = {442},
        pages = {L33-L37},
          doi = {10.1093/mnrasl/slu050},
archivePrefix = {arXiv},
       eprint = {1403.7316},
 primaryClass = {astro-ph.GA},
       adsurl = {https://ui.adsabs.harvard.edu/abs/2014MNRAS.442L..33R}
}

@ARTICLE{hani_2020,
       author = {{Hani}, Maan H. and {Gosain}, Hayman and {Ellison}, Sara L. and {Patton}, David R. and {Torrey}, Paul},
        title = "{Interacting galaxies in the IllustrisTNG simulations - II: star formation in the post-merger stage}",
      journal = {\mnras},
         year = 2020,
        month = apr,
       volume = {493},
       number = {3},
        pages = {3716-3731},
          doi = {10.1093/mnras/staa459},
archivePrefix = {arXiv},
       eprint = {2001.04472},
 primaryClass = {astro-ph.GA},
       adsurl = {https://ui.adsabs.harvard.edu/abs/2020MNRAS.493.3716H}
}

@ARTICLE{lagos_2020,
       author = {{Lagos}, Claudia del P. and {da Cunha}, Elisabete and {Robotham}, Aaron S.~G. and {Obreschkow}, Danail and {Valentino}, Francesco and {Fujimoto}, Seiji and {Magdis}, Georgios E. and {Tobar}, Rodrigo},
        title = "{Physical properties and evolution of (sub-)millimetre-selected galaxies in the galaxy formation simulation SHARK}",
      journal = {\mnras},
         year = 2020,
        month = dec,
       volume = {499},
       number = {2},
        pages = {1948-1971},
          doi = {10.1093/mnras/staa2861},
archivePrefix = {arXiv},
       eprint = {2007.09853},
 primaryClass = {astro-ph.GA},
       adsurl = {https://ui.adsabs.harvard.edu/abs/2020MNRAS.499.1948L}
}

@ARTICLE{araya_2025,
       author = {{Araya-Araya}, Pablo and {Cochrane}, Rachel K. and {Hayward}, Christopher C. and {Sodr{\'e}}, Jr., Laerte and {Yates}, Robert M. and {van Daalen}, Marcel P. and {Vicentin}, Marcelo C.},
        title = "{Simultaneously modelling dusty star-forming galaxies and massive quiescents: a calibration framework for galaxy formation models}",
      journal = {\mnras},
         year = 2025,
        month = oct,
       volume = {542},
       number = {4},
        pages = {2808-2829},
          doi = {10.1093/mnras/staf1389},
archivePrefix = {arXiv},
       eprint = {2504.15283},
 primaryClass = {astro-ph.GA},
       adsurl = {https://ui.adsabs.harvard.edu/abs/2025MNRAS.542.2808A}
}

@ARTICLE{alberts_noble_2022,
       author = {{Alberts}, Stacey and {Noble}, Allison},
        title = "{From Clusters to Proto-Clusters: The Infrared Perspective on Environmental Galaxy Evolution}",
      journal = {Universe},
         year = 2022,
        month = oct,
       volume = {8},
       number = {11},
          eid = {554},
        pages = {554},
          doi = {10.3390/universe8110554},
archivePrefix = {arXiv},
       eprint = {2209.02781},
 primaryClass = {astro-ph.GA},
       adsurl = {https://ui.adsabs.harvard.edu/abs/2022Univ....8..554A}
}

@ARTICLE{chabrier_2003,
   author = {{Chabrier}, G.},
    title = "{Galactic Stellar and Substellar Initial Mass Function}",
  journal = {\pasp},
   eprint = {astro-ph/0304382},
     year = 2003,
    month = jul,
   volume = 115,
    pages = {763-795},
      doi = {10.1086/376392},
   adsurl = {http://adsabs.harvard.edu/abs/2003PASP..115..763C}
}

@ARTICLE{dekel_2009l,
   author = {{Dekel}, A. and {Birnboim}, Y. and {Engel}, G. and {Freundlich}, J. and 
	{Goerdt}, T. and {Mumcuoglu}, M. and {Neistein}, E. and {Pichon}, C. and 
	{Teyssier}, R. and {Zinger}, E.},
    title = "{Cold streams in early massive hot haloes as the main mode of galaxy formation}",
  journal = {\nat},
archivePrefix = "arXiv",
   eprint = {0808.0553},
     year = 2009,
    month = jan,
   volume = 457,
    pages = {451-454},
      doi = {10.1038/nature07648},
   adsurl = {http://adsabs.harvard.edu/abs/2009Natur.457..451D}
}

@ARTICLE{danovich_2012,
       author = {{Danovich}, Mark and {Dekel}, Avishai and {Hahn}, Oliver and {Teyssier}, Romain},
        title = "{Coplanar streams, pancakes and angular-momentum exchange in high-z disc galaxies}",
      journal = {\mnras},
         year = 2012,
        month = may,
       volume = {422},
       number = {2},
        pages = {1732-1749},
          doi = {10.1111/j.1365-2966.2012.20751.x},
archivePrefix = {arXiv},
       eprint = {1110.6209},
 primaryClass = {astro-ph.CO},
       adsurl = {https://ui.adsabs.harvard.edu/abs/2012MNRAS.422.1732D}
}

@ARTICLE{danovich_2015,
       author = {{Danovich}, Mark and {Dekel}, Avishai and {Hahn}, Oliver and {Ceverino}, Daniel and {Primack}, Joel},
        title = "{Four phases of angular-momentum buildup in high-z galaxies: from cosmic-web streams through an extended ring to disc and bulge}",
      journal = {\mnras},
         year = 2015,
        month = may,
       volume = {449},
       number = {2},
        pages = {2087-2111},
          doi = {10.1093/mnras/stv270},
archivePrefix = {arXiv},
       eprint = {1407.7129},
 primaryClass = {astro-ph.GA},
       adsurl = {https://ui.adsabs.harvard.edu/abs/2015MNRAS.449.2087D}
}

@ARTICLE{zolotov_2015,
       author = {{Zolotov}, Adi and {Dekel}, Avishai and {Mandelker}, Nir and {Tweed}, Dylan and {Inoue}, Shigeki and {DeGraf}, Colin and {Ceverino}, Daniel and {Primack}, Joel R. and {Barro}, Guillermo and {Faber}, Sandra M.},
        title = "{Compaction and quenching of high-z galaxies in cosmological simulations: blue and red nuggets}",
      journal = {\mnras},
         year = 2015,
        month = jul,
       volume = {450},
       number = {3},
        pages = {2327-2353},
          doi = {10.1093/mnras/stv740},
archivePrefix = {arXiv},
       eprint = {1412.4783},
 primaryClass = {astro-ph.GA},
       adsurl = {https://ui.adsabs.harvard.edu/abs/2015MNRAS.450.2327Z}
}

@ARTICLE{dekel_2020a,
       author = {{Dekel}, Avishai and {Ginzburg}, Omri and {Jiang}, Fangzhou and {Freundlich}, Jonathan and {Lapiner}, Sharon and {Ceverino}, Daniel and {Primack}, Joel},
        title = "{A mass threshold for galactic gas discs by spin flips}",
      journal = {\mnras},
         year = 2020,
        month = apr,
       volume = {493},
       number = {3},
        pages = {4126-4142},
          doi = {10.1093/mnras/staa470},
archivePrefix = {arXiv},
       eprint = {1912.08213},
 primaryClass = {astro-ph.GA},
       adsurl = {https://ui.adsabs.harvard.edu/abs/2020MNRAS.493.4126D}
}

@ARTICLE{dekel_2020b,
       author = {{Dekel}, Avishai and {Lapiner}, Sharon and {Ginzburg}, Omri and {Freundlich}, Jonathan and {Jiang}, Fangzhou and {Finish}, Bar and {Kretschmer}, Michael and {Lin}, Doug and {Ceverino}, Daniel and {Primack}, Joel and {Giavalisco}, Mauro and {Ji}, Zhiyuan},
        title = "{Origin of star-forming rings around massive centres in massive galaxies at z < 4}",
      journal = {\mnras},
         year = 2020,
        month = aug,
       volume = {496},
       number = {4},
        pages = {5372-5398},
          doi = {10.1093/mnras/staa1713},
archivePrefix = {arXiv},
       eprint = {2003.08984},
 primaryClass = {astro-ph.GA},
       adsurl = {https://ui.adsabs.harvard.edu/abs/2020MNRAS.496.5372D}
}

@ARTICLE{kretschmer_2022,
       author = {{Kretschmer}, Michael and {Dekel}, Avishai and {Teyssier}, Romain},
        title = "{On the origin of surprisingly cold gas discs in galaxies at high redshift}",
      journal = {\mnras},
         year = 2022,
        month = mar,
       volume = {510},
       number = {3},
        pages = {3266-3275},
          doi = {10.1093/mnras/stab3648},
archivePrefix = {arXiv},
       eprint = {2103.06882},
 primaryClass = {astro-ph.GA},
       adsurl = {https://ui.adsabs.harvard.edu/abs/2022MNRAS.510.3266K}
}

@ARTICLE{hafen_2022,
       author = {{Hafen}, Zachary and {Stern}, Jonathan and {Bullock}, James and {Gurvich}, Alexander B. and {Yu}, Sijie and {Faucher-Gigu{\`e}re}, Claude-Andr{\'e} and {Fielding}, Drummond B. and {Angl{\'e}s-Alc{\'a}zar}, Daniel and {Quataert}, Eliot and {Wetzel}, Andrew and {Starkenburg}, Tjitske and {Boylan-Kolchin}, Michael and {Moreno}, Jorge and {Feldmann}, Robert and {El-Badry}, Kareem and {Chan}, T.~K. and {Trapp}, Cameron and {Kere{\v{s}}}, Du{\v{s}}an and {Hopkins}, Philip F.},
        title = "{Hot-mode accretion and the physics of thin-disc galaxy formation}",
      journal = {\mnras},
         year = 2022,
        month = aug,
       volume = {514},
       number = {4},
        pages = {5056-5073},
          doi = {10.1093/mnras/stac1603},
archivePrefix = {arXiv},
       eprint = {2201.07235},
 primaryClass = {astro-ph.GA},
       adsurl = {https://ui.adsabs.harvard.edu/abs/2022MNRAS.514.5056H}
}

@ARTICLE{semenov_2024,
       author = {{Semenov}, Vadim A. and {Conroy}, Charlie and {Chandra}, Vedant and {Hernquist}, Lars and {Nelson}, Dylan},
        title = "{Formation of Galactic Disks. II. The Physical Drivers of Disk Spin-up}",
      journal = {\apj},
         year = 2024,
        month = sep,
       volume = {972},
       number = {1},
          eid = {73},
        pages = {73},
          doi = {10.3847/1538-4357/ad57ba},
archivePrefix = {arXiv},
       eprint = {2306.13125},
 primaryClass = {astro-ph.GA},
       adsurl = {https://ui.adsabs.harvard.edu/abs/2024ApJ...972...73S}
}

@ARTICLE{bassini_2020,
       author = {{Bassini}, L. and {Rasia}, E. and {Borgani}, S. and {Granato}, G.~L. and {Ragone-Figueroa}, C. and {Biffi}, V. and {Ragagnin}, A. and {Dolag}, K. and {Lin}, W. and {Murante}, G. and {Napolitano}, N.~R. and {Taffoni}, G. and {Tornatore}, L. and {Wang}, Y.},
        title = "{The DIANOGA simulations of galaxy clusters: characterising star formation in protoclusters}",
      journal = {\aap},
         year = 2020,
        month = oct,
       volume = {642},
          eid = {A37},
        pages = {A37},
          doi = {10.1051/0004-6361/202038396},
archivePrefix = {arXiv},
       eprint = {2006.13951},
 primaryClass = {astro-ph.GA},
       adsurl = {https://ui.adsabs.harvard.edu/abs/2020A&A...642A..37B}
}

@ARTICLE{lim_2021,
       author = {{Lim}, Seunghwan and {Scott}, Douglas and {Babul}, Arif and {Barnes}, David J. and {Kay}, Scott T. and {McCarthy}, Ian G. and {Rennehan}, Douglas and {Vogelsberger}, Mark},
        title = "{Is there enough star formation in simulated protoclusters?}",
      journal = {\mnras},
         year = 2021,
        month = feb,
       volume = {501},
       number = {2},
        pages = {1803-1822},
          doi = {10.1093/mnras/staa3693},
archivePrefix = {arXiv},
       eprint = {2010.02259},
 primaryClass = {astro-ph.GA},
       adsurl = {https://ui.adsabs.harvard.edu/abs/2021MNRAS.501.1803L}
}

@ARTICLE{remus_2023,
       author = {{Remus}, Rhea-Silvia and {Dolag}, Klaus and {Dannerbauer}, Helmut},
        title = "{The Young and the Wild: What Happens to Protoclusters Forming at Redshift z {\ensuremath{\approx}} 4?}",
      journal = {\apj},
         year = 2023,
        month = jun,
       volume = {950},
       number = {2},
          eid = {191},
        pages = {191},
          doi = {10.3847/1538-4357/accb91},
archivePrefix = {arXiv},
       eprint = {2208.01053},
 primaryClass = {astro-ph.CO},
       adsurl = {https://ui.adsabs.harvard.edu/abs/2023ApJ...950..191R}
}

@ARTICLE{kapferer_2008,
       author = {{Kapferer}, W. and {Kronberger}, T. and {Ferrari}, C. and {Riser}, T. and {Schindler}, S.},
        title = "{On the influence of ram-pressure stripping on interacting galaxies in clusters}",
      journal = {\mnras},
         year = 2008,
        month = sep,
       volume = {389},
       number = {3},
        pages = {1405-1413},
          doi = {10.1111/j.1365-2966.2008.13665.x},
archivePrefix = {arXiv},
       eprint = {0807.0083},
 primaryClass = {astro-ph},
       adsurl = {https://ui.adsabs.harvard.edu/abs/2008MNRAS.389.1405K}
}

@ARTICLE{bekki_2014,
       author = {{Bekki}, Kenji},
        title = "{Galactic star formation enhanced and quenched by ram pressure in groups and clusters}",
      journal = {\mnras},
         year = 2014,
        month = feb,
       volume = {438},
       number = {1},
        pages = {444-462},
          doi = {10.1093/mnras/stt2216},
archivePrefix = {arXiv},
       eprint = {1311.3010},
 primaryClass = {astro-ph.CO},
       adsurl = {https://ui.adsabs.harvard.edu/abs/2014MNRAS.438..444B}
}

@ARTICLE{lee_2020,
       author = {{Lee}, Jaehyun and {Kimm}, Taysun and {Katz}, Harley and {Rosdahl}, Joakim and {Devriendt}, Julien and {Slyz}, Adrianne},
        title = "{Dual Effects of Ram Pressure on Star Formation in Multiphase Disk Galaxies with Strong Stellar Feedback}",
      journal = {\apj},
         year = 2020,
        month = dec,
       volume = {905},
       number = {1},
          eid = {31},
        pages = {31},
          doi = {10.3847/1538-4357/abc3b8},
archivePrefix = {arXiv},
       eprint = {2010.11028},
 primaryClass = {astro-ph.GA},
       adsurl = {https://ui.adsabs.harvard.edu/abs/2020ApJ...905...31L}
}

@ARTICLE{kennicutt_1998,
   author = {{Kennicutt}, Jr., R.~C.},
    title = "{Star Formation in Galaxies Along the Hubble Sequence}",
  journal = {\araa},
   eprint = {astro-ph/9807187},
     year = 1998,
   volume = 36,
    pages = {189-232},
      doi = {10.1146/annurev.astro.36.1.189},
   adsurl = {http://adsabs.harvard.edu/abs/1998ARA%26A..36..189K}
}

@ARTICLE{tacconi_2020,
       author = {{Tacconi}, Linda J. and {Genzel}, Reinhard and {Sternberg}, Amiel},
        title = "{The Evolution of the Star-Forming Interstellar Medium Across Cosmic Time}",
      journal = {\araa},
         year = 2020,
        month = aug,
       volume = {58},
        pages = {157-203},
          doi = {10.1146/annurev-astro-082812-141034},
archivePrefix = {arXiv},
       eprint = {2003.06245},
 primaryClass = {astro-ph.GA},
       adsurl = {https://ui.adsabs.harvard.edu/abs/2020ARA&A..58..157T}
}

@ARTICLE{springel_2005,
   author = {{Springel}, V. and {White}, S.~D.~M. and {Jenkins}, A. and {Frenk}, C.~S. and 
	{Yoshida}, N. and {Gao}, L. and {Navarro}, J. and {Thacker}, R. and 
	{Croton}, D. and {Helly}, J. and {Peacock}, J.~A. and {Cole}, S. and 
	{Thomas}, P. and {Couchman}, H. and {Evrard}, A. and {Colberg}, J. and 
	{Pearce}, F.},
    title = "{Simulations of the formation, evolution and clustering of galaxies and quasars}",
  journal = {\nat},
   eprint = {astro-ph/0504097},
     year = 2005,
    month = jun,
   volume = 435,
    pages = {629-636},
      doi = {10.1038/nature03597},
   adsurl = {http://adsabs.harvard.edu/abs/2005Natur.435..629S}
}

@ARTICLE{lovell_2021,
       author = {{Lovell}, Christopher C. and {Vijayan}, Aswin P. and {Thomas}, Peter A. and {Wilkins}, Stephen M. and {Barnes}, David J. and {Irodotou}, Dimitrios and {Roper}, Will},
        title = "{First Light And Reionization Epoch Simulations (FLARES) - I. Environmental dependence of high-redshift galaxy evolution}",
      journal = {\mnras},
         year = 2021,
        month = jan,
       volume = {500},
       number = {2},
        pages = {2127-2145},
          doi = {10.1093/mnras/staa3360},
archivePrefix = {arXiv},
       eprint = {2004.07283},
 primaryClass = {astro-ph.GA},
       adsurl = {https://ui.adsabs.harvard.edu/abs/2021MNRAS.500.2127L}
}

@ARTICLE{delucia_2006,
       author = {{De Lucia}, Gabriella and {Springel}, Volker and {White}, Simon D.~M. and {Croton}, Darren and {Kauffmann}, Guinevere},
        title = "{The formation history of elliptical galaxies}",
      journal = {\mnras},
         year = 2006,
        month = feb,
       volume = {366},
       number = {2},
        pages = {499-509},
          doi = {10.1111/j.1365-2966.2005.09879.x},
archivePrefix = {arXiv},
       eprint = {astro-ph/0509725},
 primaryClass = {astro-ph},
       adsurl = {https://ui.adsabs.harvard.edu/abs/2006MNRAS.366..499D}
}

@ARTICLE{smirnov_2024,
       author = {{Smirnov}, Daniil V. and {Mosenkov}, Aleksandr V. and {Reshetnikov}, Vladimir P.},
        title = "{Polar-ring galaxies in the Illustris TNG50 simulation}",
      journal = {\mnras},
         year = 2024,
        month = jan,
       volume = {527},
       number = {2},
        pages = {4112-4128},
          doi = {10.1093/mnras/stad3368},
archivePrefix = {arXiv},
       eprint = {2310.18597},
 primaryClass = {astro-ph.GA},
       adsurl = {https://ui.adsabs.harvard.edu/abs/2024MNRAS.527.4112S}
}

@ARTICLE{koenig_2014,
       author = {{K{\"o}nig}, S. and {Aalto}, S. and {Lindroos}, L. and {Muller}, S. and {Gallagher}, J.~S. and {Beswick}, R.~J. and {Petitpas}, G. and {J{\"u}tte}, E.},
        title = "{Molecular tendrils feeding star formation in the Eye of the Medusa. The Medusa merger in high resolution $^{12}$CO 2-1 maps}",
      journal = {\aap},
         year = 2014,
        month = sep,
       volume = {569},
          eid = {A6},
        pages = {A6},
          doi = {10.1051/0004-6361/201423548},
archivePrefix = {arXiv},
       eprint = {1407.8347},
 primaryClass = {astro-ph.GA},
       adsurl = {https://ui.adsabs.harvard.edu/abs/2014A&A...569A...6K}
}

@ARTICLE{walter_2002,
       author = {{Walter}, F. and {Weiss}, A. and {Scoville}, N.},
        title = "{Molecular Gas in M82: Resolving the Outflow and Streamers}",
      journal = {\apjl},
         year = 2002,
        month = nov,
       volume = {580},
       number = {1},
        pages = {L21-L25},
          doi = {10.1086/345287},
archivePrefix = {arXiv},
       eprint = {astro-ph/0210602},
 primaryClass = {astro-ph},
       adsurl = {https://ui.adsabs.harvard.edu/abs/2002ApJ...580L..21W}
}

@ARTICLE{stuber_2021,
       author = {{Stuber}, Sophia K. and {Saito}, Toshiki and {Schinnerer}, Eva and {Emsellem}, Eric and {Querejeta}, Miguel and {Williams}, Thomas G. and {Barnes}, Ashley T. and {Bigiel}, Frank and {Blanc}, Guillermo and {Dale}, Daniel A. and {Grasha}, Kathryn and {Klessen}, Ralf and {Kruijssen}, J.~M. Diederik and {Leroy}, Adam K. and {Meidt}, Sharon and {Pan}, Hsi-An and {Rosolowsky}, Erik and {Schruba}, Andreas and {Sun}, Jiayi and {Usero}, Antonio},
        title = "{Frequency and nature of central molecular outflows in nearby star-forming disk galaxies}",
      journal = {\aap},
         year = 2021,
        month = sep,
       volume = {653},
          eid = {A172},
        pages = {A172},
          doi = {10.1051/0004-6361/202141093},
archivePrefix = {arXiv},
       eprint = {2107.11227},
 primaryClass = {astro-ph.GA},
       adsurl = {https://ui.adsabs.harvard.edu/abs/2021A&A...653A.172S}
}

@ARTICLE{zanchettin_2023,
       author = {{Zanchettin}, M.~V. and {Feruglio}, C. and {Massardi}, M. and {Lapi}, A. and {Bischetti}, M. and {Cantalupo}, S. and {Fiore}, F. and {Bongiorno}, A. and {Malizia}, A. and {Marinucci}, A. and {Molina}, M. and {Piconcelli}, E. and {Tombesi}, F. and {Travascio}, A. and {Tozzi}, G. and {Tripodi}, R.},
        title = "{NGC 2992: Interplay between the multiphase disc, wind, and radio bubbles}",
      journal = {\aap},
         year = 2023,
        month = nov,
       volume = {679},
          eid = {A88},
        pages = {A88},
          doi = {10.1051/0004-6361/202245729},
archivePrefix = {arXiv},
       eprint = {2308.04108},
 primaryClass = {astro-ph.GA},
       adsurl = {https://ui.adsabs.harvard.edu/abs/2023A&A...679A..88Z}
}

@ARTICLE{esposito_2024,
       author = {{Esposito}, Federico and {Alonso-Herrero}, Almudena and {Garc{\'\i}a-Burillo}, Santiago and {Casasola}, Viviana and {Combes}, Fran{\c{c}}oise and {Dallacasa}, Daniele and {Davies}, Richard and {Garc{\'\i}a-Bernete}, Ismael and {Garc{\'\i}a-Lorenzo}, Bego{\~n}a and {Hermosa Mu{\~n}oz}, Laura and {de Arriba}, Luis Peralta and {Pereira-Santaella}, Miguel and {Pozzi}, Francesca and {Ramos Almeida}, Cristina and {Shimizu}, Thomas Taro and {Vallini}, Livia and {Bellocchi}, Enrica and {Gonz{\'a}lez-Mart{\'\i}n}, Omaira and {Hicks}, Erin K.~S. and {H{\"o}nig}, Sebastian and {Labiano}, Alvaro and {Levenson}, Nancy A. and {Ricci}, Claudio and {Rosario}, David J.},
        title = "{AGN feedback in the Local Universe: Multiphase outflow of the Seyfert galaxy NGC 5506}",
      journal = {\aap},
         year = 2024,
        month = jun,
       volume = {686},
          eid = {A46},
        pages = {A46},
          doi = {10.1051/0004-6361/202449245},
archivePrefix = {arXiv},
       eprint = {2403.03981},
 primaryClass = {astro-ph.GA},
       adsurl = {https://ui.adsabs.harvard.edu/abs/2024A&A...686A..46E}
}

@ARTICLE{shimizu_2019,
       author = {{Shimizu}, T. Taro and {Davies}, R.~I. and {Lutz}, D. and {Burtscher}, L. and {Lin}, M. and {Baron}, D. and {Davies}, R.~L. and {Genzel}, R. and {Hicks}, E.~K.~S. and {Koss}, M. and {Maciejewski}, W. and {M{\"u}ller-S{\'a}nchez}, F. and {Orban de Xivry}, G. and {Price}, S.~H. and {Ricci}, C. and {Riffel}, R. and {Riffel}, R.~A. and {Rosario}, D. and {Schartmann}, M. and {Schnorr-M{\"u}ller}, A. and {Sternberg}, A. and {Sturm}, E. and {Storchi-Bergmann}, T. and {Tacconi}, L. and {Veilleux}, S.},
        title = "{The multiphase gas structure and kinematics in the circumnuclear region of NGC 5728}",
      journal = {\mnras},
         year = 2019,
        month = dec,
       volume = {490},
       number = {4},
        pages = {5860-5887},
          doi = {10.1093/mnras/stz2802},
archivePrefix = {arXiv},
       eprint = {1907.03801},
 primaryClass = {astro-ph.GA},
       adsurl = {https://ui.adsabs.harvard.edu/abs/2019MNRAS.490.5860S}
}

@ARTICLE{oosterloo_2017,
       author = {{Oosterloo}, Tom and {Raymond Oonk}, J.~B. and {Morganti}, Raffaella and {Combes}, Fran{\c{c}}oise and {Dasyra}, Kalliopi and {Salom{\'e}}, Philippe and {Vlahakis}, Nektarios and {Tadhunter}, Clive},
        title = "{Properties of the molecular gas in the fast outflow in the Seyfert galaxy IC 5063}",
      journal = {\aap},
         year = 2017,
        month = dec,
       volume = {608},
          eid = {A38},
        pages = {A38},
          doi = {10.1051/0004-6361/201731781},
archivePrefix = {arXiv},
       eprint = {1710.01570},
 primaryClass = {astro-ph.GA},
       adsurl = {https://ui.adsabs.harvard.edu/abs/2017A&A...608A..38O}
}

@ARTICLE{aalto_2020,
       author = {{Aalto}, S. and {Falstad}, N. and {Muller}, S. and {Wada}, K. and {Gallagher}, J.~S. and {K{\"o}nig}, S. and {Sakamoto}, K. and {Vlemmings}, W. and {Ceccobello}, C. and {Dasyra}, K. and {Combes}, F. and {Garc{\'\i}a-Burillo}, S. and {Oya}, Y. and {Mart{\'\i}n}, S. and {van der Werf}, P. and {Evans}, A.~S. and {Kotilainen}, J.},
        title = "{ALMA resolves the remarkable molecular jet and rotating wind in the extremely radio-quiet galaxy NGC 1377}",
      journal = {\aap},
         year = 2020,
        month = aug,
       volume = {640},
          eid = {A104},
        pages = {A104},
          doi = {10.1051/0004-6361/202038282},
archivePrefix = {arXiv},
       eprint = {2007.07824},
 primaryClass = {astro-ph.GA},
       adsurl = {https://ui.adsabs.harvard.edu/abs/2020A&A...640A.104A}
}

@ARTICLE{fluetsch_2019,
       author = {{Fluetsch}, A. and {Maiolino}, R. and {Carniani}, S. and {Marconi}, A. and {Cicone}, C. and {Bourne}, M.~A. and {Costa}, T. and {Fabian}, A.~C. and {Ishibashi}, W. and {Venturi}, G.},
        title = "{Cold molecular outflows in the local Universe and their feedback effect on galaxies}",
      journal = {\mnras},
         year = 2019,
        month = mar,
       volume = {483},
       number = {4},
        pages = {4586-4614},
          doi = {10.1093/mnras/sty3449},
archivePrefix = {arXiv},
       eprint = {1805.05352},
 primaryClass = {astro-ph.GA},
       adsurl = {https://ui.adsabs.harvard.edu/abs/2019MNRAS.483.4586F}
}

@ARTICLE{cicone_2014,
       author = {{Cicone}, C. and {Maiolino}, R. and {Sturm}, E. and {Graci{\'a}-Carpio}, J. and {Feruglio}, C. and {Neri}, R. and {Aalto}, S. and {Davies}, R. and {Fiore}, F. and {Fischer}, J. and {Garc{\'\i}a-Burillo}, S. and {Gonz{\'a}lez-Alfonso}, E. and {Hailey-Dunsheath}, S. and {Piconcelli}, E. and {Veilleux}, S.},
        title = "{Massive molecular outflows and evidence for AGN feedback from CO observations}",
      journal = {\aap},
         year = 2014,
        month = feb,
       volume = {562},
          eid = {A21},
        pages = {A21},
          doi = {10.1051/0004-6361/201322464},
archivePrefix = {arXiv},
       eprint = {1311.2595},
 primaryClass = {astro-ph.CO},
       adsurl = {https://ui.adsabs.harvard.edu/abs/2014A&A...562A..21C}
}

@ARTICLE{watson_1994,
       author = {{Watson}, Dan M. and {Guptill}, Matthew T. and {Buchholz}, Leah M.},
        title = "{Detection of CO J = 2 1 Emission from the Polar Rings of NGC 2685 and NGC 4650A}",
      journal = {\apjl},
         year = 1994,
        month = jan,
       volume = {420},
        pages = {L21},
          doi = {10.1086/187153},
       adsurl = {https://ui.adsabs.harvard.edu/abs/1994ApJ...420L..21W}
}

@ARTICLE{schinnerer_2002,
       author = {{Schinnerer}, Eva and {Scoville}, Nick},
        title = "{First Interferometric Observations of Molecular Gas in a Polar Ring: The Helix Galaxy NGC 2685}",
      journal = {\apjl},
         year = 2002,
        month = oct,
       volume = {577},
       number = {2},
        pages = {L103-L106},
          doi = {10.1086/344242},
archivePrefix = {arXiv},
       eprint = {astro-ph/0209004},
 primaryClass = {astro-ph},
       adsurl = {https://ui.adsabs.harvard.edu/abs/2002ApJ...577L.103S}
}

@ARTICLE{vandriel_1995,
       author = {{van Driel}, W. and {Combes}, F. and {Casoli}, F. and {Gerin}, M. and {Nakai}, N. and {Miyaji}, T. and {Hamabe}, M. and {Sofue}, Y. and {Ichikawa}, T. and {Yoshida}, S. and {Kobayashi}, Y. and {Geng}, F. and {Minezaki}, T. and {Arimoto}, N. and {Kodama}, T. and {Goudfrooij}, P. and {Mulder}, P.~S. and {Wakamatsu}, K. and {Yanagisawa}, K.},
        title = "{Polar Ring Spiral Galaxy NGC 660}",
      journal = {\aj},
         year = 1995,
        month = mar,
       volume = {109},
        pages = {942},
          doi = {10.1086/117333},
       adsurl = {https://ui.adsabs.harvard.edu/abs/1995AJ....109..942V}
}

@ARTICLE{combes_1992,
       author = {{Combes}, F. and {Braine}, J. and {Casoli}, F. and {Gerin}, M. and {van Driel}, W.},
        title = "{Molecular clouds in a polar ring.}",
      journal = {\aap},
         year = 1992,
        month = jun,
       volume = {259},
        pages = {L65-L68},
       adsurl = {https://ui.adsabs.harvard.edu/abs/1992A&A...259L..65C}
}

@ARTICLE{iodice_2015,
       author = {{Iodice}, E. and {Coccato}, L. and {Combes}, F. and {de Zeeuw}, T. and {Arnaboldi}, M. and {Weilbacher}, P.~M. and {Bacon}, R. and {Kuntschner}, H. and {Spavone}, M.},
        title = "{Mapping the inner regions of the polar disk galaxy NGC 4650A with MUSE}",
      journal = {\aap},
         year = 2015,
        month = nov,
       volume = {583},
          eid = {A48},
        pages = {A48},
          doi = {10.1051/0004-6361/201526446},
archivePrefix = {arXiv},
       eprint = {1509.01112},
 primaryClass = {astro-ph.GA},
       adsurl = {https://ui.adsabs.harvard.edu/abs/2015A&A...583A..48I}
}

@ARTICLE{simons_2019,
       author = {{Simons}, Raymond C. and {Kassin}, Susan A. and {Snyder}, Gregory F. and {Primack}, Joel R. and {Ceverino}, Daniel and {Dekel}, Avishai and {Hayward}, Christopher C. and {Mandelker}, Nir and {Mantha}, Kameswara Bharadwaj and {Pacifici}, Camilla and {de la Vega}, Alexander and {Wang}, Weichen},
        title = "{Distinguishing Mergers and Disks in High-redshift Observations of Galaxy Kinematics}",
      journal = {\apj},
         year = 2019,
        month = mar,
       volume = {874},
       number = {1},
          eid = {59},
        pages = {59},
          doi = {10.3847/1538-4357/ab07c9},
archivePrefix = {arXiv},
       eprint = {1902.06762},
 primaryClass = {astro-ph.GA},
       adsurl = {https://ui.adsabs.harvard.edu/abs/2019ApJ...874...59S}
}

@ARTICLE{loiacono_2019,
       author = {{Loiacono}, Federica and {Talia}, Margherita and {Fraternali}, Filippo and {Cimatti}, Andrea and {Di Teodoro}, Enrico M. and {Caminha}, Gabriel B.},
        title = "{A multiwavelength study of a massive, active galaxy at z {\ensuremath{\sim}} 2: coupling the kinematics of the ionized and molecular gas}",
      journal = {\mnras},
         year = 2019,
        month = oct,
       volume = {489},
       number = {1},
        pages = {681-698},
          doi = {10.1093/mnras/stz2170},
archivePrefix = {arXiv},
       eprint = {1908.01358},
 primaryClass = {astro-ph.GA},
       adsurl = {https://ui.adsabs.harvard.edu/abs/2019MNRAS.489..681L}
}

@ARTICLE{rizzo_2020,
       author = {{Rizzo}, F. and {Vegetti}, S. and {Powell}, D. and {Fraternali}, F. and {McKean}, J.~P. and {Stacey}, H.~R. and {White}, S.~D.~M.},
        title = "{A dynamically cold disk galaxy in the early Universe}",
      journal = {\nat},
         year = 2020,
        month = aug,
       volume = {584},
       number = {7820},
        pages = {201-204},
          doi = {10.1038/s41586-020-2572-6},
archivePrefix = {arXiv},
       eprint = {2009.01251},
 primaryClass = {astro-ph.GA},
       adsurl = {https://ui.adsabs.harvard.edu/abs/2020Natur.584..201R}
}

@ARTICLE{neeleman_2020,
       author = {{Neeleman}, Marcel and {Prochaska}, J. Xavier and {Kanekar}, Nissim and {Rafelski}, Marc},
        title = "{A cold, massive, rotating disk galaxy 1.5 billion years after the Big Bang}",
      journal = {\nat},
         year = 2020,
        month = may,
       volume = {581},
       number = {7808},
        pages = {269-272},
          doi = {10.1038/s41586-020-2276-y},
archivePrefix = {arXiv},
       eprint = {2005.09661},
 primaryClass = {astro-ph.GA},
       adsurl = {https://ui.adsabs.harvard.edu/abs/2020Natur.581..269N}
}

@ARTICLE{tsukui_2021,
       author = {{Tsukui}, Takafumi and {Iguchi}, Satoru},
        title = "{Spiral morphology in an intensely star-forming disk galaxy more than 12 billion years ago}",
      journal = {Science},
         year = 2021,
        month = jun,
       volume = {372},
       number = {6547},
        pages = {1201-1205},
          doi = {10.1126/science.abe9680},
archivePrefix = {arXiv},
       eprint = {2108.02206},
 primaryClass = {astro-ph.GA},
       adsurl = {https://ui.adsabs.harvard.edu/abs/2021Sci...372.1201T}
}

@ARTICLE{lelli_2021,
       author = {{Lelli}, Federico and {Di Teodoro}, Enrico M. and {Fraternali}, Filippo and {Man}, Allison W.~S. and {Zhang}, Zhi-Yu and {De Breuck}, Carlos and {Davis}, Timothy A. and {Maiolino}, Roberto},
        title = "{A massive stellar bulge in a regularly rotating galaxy 1.2 billion years after the Big Bang}",
      journal = {Science},
         year = 2021,
        month = feb,
       volume = {371},
       number = {6530},
        pages = {713-716},
          doi = {10.1126/science.abc1893},
archivePrefix = {arXiv},
       eprint = {2102.05957},
 primaryClass = {astro-ph.GA},
       adsurl = {https://ui.adsabs.harvard.edu/abs/2021Sci...371..713L}
}

@ARTICLE{rizzo_2021,
       author = {{Rizzo}, Francesca and {Vegetti}, Simona and {Fraternali}, Filippo and {Stacey}, Hannah R. and {Powell}, Devon},
        title = "{Dynamical properties of z  4.5 dusty star-forming galaxies and their connection with local early-type galaxies}",
      journal = {\mnras},
         year = 2021,
        month = nov,
       volume = {507},
       number = {3},
        pages = {3952-3984},
          doi = {10.1093/mnras/stab2295},
archivePrefix = {arXiv},
       eprint = {2102.05671},
 primaryClass = {astro-ph.GA},
       adsurl = {https://ui.adsabs.harvard.edu/abs/2021MNRAS.507.3952R}
}

@ARTICLE{gomez_2025,
       author = {{G{\'o}mez}, Jonathan S. and {Messias}, Hugo and {Nagar}, Neil M. and {Orellana-Gonz{\'a}lez}, Gustavo and {Ivison}, R.~J. and {van der Werf}, Paul},
        title = "{Resolved Schmidt-Kennicutt relation in a binary hyperluminous infrared galaxy at $z=2.41$}",
      journal = {arXiv e-prints},
         year = 2025,
        month = nov,
          eid = {arXiv:2511.06537},
        pages = {arXiv:2511.06537},
          doi = {10.48550/arXiv.2511.06537},
archivePrefix = {arXiv},
       eprint = {2511.06537},
 primaryClass = {astro-ph.GA},
       adsurl = {https://ui.adsabs.harvard.edu/abs/2025arXiv251106537G}
}

@ARTICLE{ivison_2019,
       author = {{Ivison}, R.~J. and {Page}, M.~J. and {Cirasuolo}, M. and {Harrison}, C.~M. and {Mainieri}, V. and {Arumugam}, V. and {Dudzevi{\v{c}}i{\={u}}t{\.{e}}}, U.},
        title = "{Hyperluminous starburst gives up its secrets}",
      journal = {\mnras},
         year = 2019,
        month = oct,
       volume = {489},
       number = {1},
        pages = {427-436},
          doi = {10.1093/mnras/stz2180},
archivePrefix = {arXiv},
       eprint = {1908.03199},
 primaryClass = {astro-ph.GA},
       adsurl = {https://ui.adsabs.harvard.edu/abs/2019MNRAS.489..427I}
}

@ARTICLE{ivison_2013,
       author = {{Ivison}, R.~J. and {Swinbank}, A.~M. and {Smail}, Ian and {Harris}, A.~I. and {Bussmann}, R.~S. and {Cooray}, A. and {Cox}, P. and {Fu}, H. and {Kov{\'a}cs}, A. and {Krips}, M. and {Narayanan}, D. and {Negrello}, M. and {Neri}, R. and {Pe{\~n}arrubia}, J. and {Richard}, J. and {Riechers}, D.~A. and {Rowlands}, K. and {Staguhn}, J.~G. and {Targett}, T.~A. and {Amber}, S. and {Baker}, A.~J. and {Bourne}, N. and {Bertoldi}, F. and {Bremer}, M. and {Calanog}, J.~A. and {Clements}, D.~L. and {Dannerbauer}, H. and {Dariush}, A. and {De Zotti}, G. and {Dunne}, L. and {Eales}, S.~A. and {Farrah}, D. and {Fleuren}, S. and {Franceschini}, A. and {Geach}, J.~E. and {George}, R.~D. and {Helly}, J.~C. and {Hopwood}, R. and {Ibar}, E. and {Jarvis}, M.~J. and {Kneib}, J. -P. and {Maddox}, S. and {Omont}, A. and {Scott}, D. and {Serjeant}, S. and {Smith}, M.~W.~L. and {Thompson}, M.~A. and {Valiante}, E. and {Valtchanov}, I. and {Vieira}, J. and {van der Werf}, P.},
        title = "{Herschel-ATLAS: A Binary HyLIRG Pinpointing a Cluster of Starbursting Protoellipticals}",
      journal = {\apj},
         year = 2013,
        month = aug,
       volume = {772},
       number = {2},
          eid = {137},
        pages = {137},
          doi = {10.1088/0004-637X/772/2/137},
archivePrefix = {arXiv},
       eprint = {1302.4436},
 primaryClass = {astro-ph.CO},
       adsurl = {https://ui.adsabs.harvard.edu/abs/2013ApJ...772..137I}
}

@ARTICLE{bussmann_2013,
       author = {{Bussmann}, R.~S. and {P{\'e}rez-Fournon}, I. and {Amber}, S. and {Calanog}, J. and {Gurwell}, M.~A. and {Dannerbauer}, H. and {De Bernardis}, F. and {Fu}, Hai and {Harris}, A.~I. and {Krips}, M. and {Lapi}, A. and {Maiolino}, R. and {Omont}, A. and {Riechers}, D. and {Wardlow}, J. and {Baker}, A.~J. and {Birkinshaw}, M. and {Bock}, J. and {Bourne}, N. and {Clements}, D.~L. and {Cooray}, A. and {De Zotti}, G. and {Dunne}, L. and {Dye}, S. and {Eales}, S. and {Farrah}, D. and {Gavazzi}, R. and {Gonz{\'a}lez Nuevo}, J. and {Hopwood}, R. and {Ibar}, E. and {Ivison}, R.~J. and {Laporte}, N. and {Maddox}, S. and {Mart{\'\i}nez-Navajas}, P. and {Michalowski}, M. and {Negrello}, M. and {Oliver}, S.~J. and {Roseboom}, I.~G. and {Scott}, Douglas and {Serjeant}, S. and {Smith}, A.~J. and {Smith}, Matthew and {Streblyanska}, A. and {Valiante}, E. and {van der Werf}, P. and {Verma}, A. and {Vieira}, J.~D. and {Wang}, L. and {Wilner}, D.},
        title = "{Gravitational Lens Models Based on Submillimeter Array Imaging of Herschel-selected Strongly Lensed Sub-millimeter Galaxies at z > 1.5}",
      journal = {\apj},
         year = 2013,
        month = dec,
       volume = {779},
       number = {1},
          eid = {25},
        pages = {25},
          doi = {10.1088/0004-637X/779/1/25},
archivePrefix = {arXiv},
       eprint = {1309.0836},
 primaryClass = {astro-ph.CO},
       adsurl = {https://ui.adsabs.harvard.edu/abs/2013ApJ...779...25B}
}

@ARTICLE{gomez_2018,
       author = {{G{\'o}mez}, Jonathan S. and {Messias}, Hugo and {Nagar}, Neil M. and {Orellana}, Gustavo and {Ivison}, Rob J. and {van der Werf}, Paul},
        title = "{A resolved warm/dense gas Schmidt-Kennicutt relationship in a binary HyLIRG at $z=2.41$}",
      journal = {arXiv e-prints},
         year = 2018,
        month = jun,
          eid = {arXiv:1806.01951},
        pages = {arXiv:1806.01951},
          doi = {10.48550/arXiv.1806.01951},
archivePrefix = {arXiv},
       eprint = {1806.01951},
 primaryClass = {astro-ph.GA},
       adsurl = {https://ui.adsabs.harvard.edu/abs/2018arXiv180601951G}
}

@ARTICLE{ivison_2000,
       author = {{Ivison}, R.~J. and {Dunlop}, J.~S. and {Smail}, Ian and {Dey}, Arjun and {Liu}, Michael C. and {Graham}, J.~R.},
        title = "{An Excess of Submillimeter Sources near 4C 41.17: A Candidate Protocluster at Z = 3.8?}",
      journal = {\apj},
         year = 2000,
        month = oct,
       volume = {542},
       number = {1},
        pages = {27-34},
          doi = {10.1086/309536},
archivePrefix = {arXiv},
       eprint = {astro-ph/0005234},
 primaryClass = {astro-ph},
       adsurl = {https://ui.adsabs.harvard.edu/abs/2000ApJ...542...27I}
}

@ARTICLE{zeballos_2018,
       author = {{Zeballos}, M. and {Aretxaga}, I. and {Hughes}, D.~H. and {Humphrey}, A. and {Wilson}, G.~W. and {Austermann}, J. and {Dunlop}, J.~S. and {Ezawa}, H. and {Ferrusca}, D. and {Hatsukade}, B. and {Ivison}, R.~J. and {Kawabe}, R. and {Kim}, S. and {Kodama}, T. and {Kohno}, K. and {Monta{\~n}a}, A. and {Nakanishi}, K. and {Plionis}, M. and {S{\'a}nchez-Arg{\"u}elles}, D. and {Stevens}, J.~A. and {Tamura}, Y. and {Velazquez}, M. and {Yun}, M.~S.},
        title = "{AzTEC 1.1 mm observations of high-z protocluster environments: SMG overdensities and misalignment between AGN jets and SMG distribution}",
      journal = {\mnras},
         year = 2018,
        month = oct,
       volume = {479},
       number = {4},
        pages = {4577-4632},
          doi = {10.1093/mnras/sty1714},
archivePrefix = {arXiv},
       eprint = {1806.10291},
 primaryClass = {astro-ph.GA},
       adsurl = {https://ui.adsabs.harvard.edu/abs/2018MNRAS.479.4577Z}
}

@ARTICLE{calvi_2023,
       author = {{Calvi}, Rosa and {Castignani}, Gianluca and {Dannerbauer}, Helmut},
        title = "{Bright submillimeter galaxies do trace galaxy protoclusters}",
      journal = {\aap},
         year = 2023,
        month = oct,
       volume = {678},
          eid = {A15},
        pages = {A15},
          doi = {10.1051/0004-6361/202346200},
archivePrefix = {arXiv},
       eprint = {2302.10323},
 primaryClass = {astro-ph.GA},
       adsurl = {https://ui.adsabs.harvard.edu/abs/2023A&A...678A..15C}
}

@ARTICLE{muldrew_2018,
       author = {{Muldrew}, Stuart I. and {Hatch}, Nina A. and {Cooke}, Elizabeth A.},
        title = "{Galaxy evolution in protoclusters}",
      journal = {\mnras},
         year = 2018,
        month = jan,
       volume = {473},
       number = {2},
        pages = {2335-2347},
          doi = {10.1093/mnras/stx2454},
archivePrefix = {arXiv},
       eprint = {1709.07009},
 primaryClass = {astro-ph.GA},
       adsurl = {https://ui.adsabs.harvard.edu/abs/2018MNRAS.473.2335M}
}

@ARTICLE{chiang_2013,
       author = {{Chiang}, Yi-Kuan and {Overzier}, Roderik and {Gebhardt}, Karl},
        title = "{Ancient Light from Young Cosmic Cities: Physical and Observational Signatures of Galaxy Proto-clusters}",
      journal = {\apj},
         year = 2013,
        month = dec,
       volume = {779},
       number = {2},
          eid = {127},
        pages = {127},
          doi = {10.1088/0004-637X/779/2/127},
archivePrefix = {arXiv},
       eprint = {1310.2938},
 primaryClass = {astro-ph.CO},
       adsurl = {https://ui.adsabs.harvard.edu/abs/2013ApJ...779..127C}
}

@ARTICLE{miller_2018,
       author = {{Miller}, T.~B. and {Chapman}, S.~C. and {Aravena}, M. and {Ashby}, M.~L.~N. and {Hayward}, C.~C. and {Vieira}, J.~D. and {Wei{\ss}}, A. and {Babul}, A. and {B{\'e}thermin}, M. and {Bradford}, C.~M. and {Brodwin}, M. and {Carlstrom}, J.~E. and {Chen}, Chian-Chou and {Cunningham}, D.~J.~M. and {De Breuck}, C. and {Gonzalez}, A.~H. and {Greve}, T.~R. and {Harnett}, J. and {Hezaveh}, Y. and {Lacaille}, K. and {Litke}, K.~C. and {Ma}, J. and {Malkan}, M. and {Marrone}, D.~P. and {Morningstar}, W. and {Murphy}, E.~J. and {Narayanan}, D. and {Pass}, E. and {Perry}, R. and {Phadke}, K.~A. and {Rennehan}, D. and {Rotermund}, K.~M. and {Simpson}, J. and {Spilker}, J.~S. and {Sreevani}, J. and {Stark}, A.~A. and {Strandet}, M.~L. and {Strom}, A.~L.},
        title = "{A massive core for a cluster of galaxies at a redshift of 4.3}",
      journal = {\nat},
         year = 2018,
        month = apr,
       volume = {556},
       number = {7702},
        pages = {469-472},
          doi = {10.1038/s41586-018-0025-2},
archivePrefix = {arXiv},
       eprint = {1804.09231},
 primaryClass = {astro-ph.GA},
       adsurl = {https://ui.adsabs.harvard.edu/abs/2018Natur.556..469M}
}

@ARTICLE{rotermund_2021,
       author = {{Rotermund}, K.~M. and {Chapman}, S.~C. and {Phadke}, K.~A. and {Hill}, R. and {Pass}, E. and {Aravena}, M. and {Ashby}, M.~L.~N. and {Babul}, A. and {B{\'e}thermin}, M. and {Canning}, R. and {de Breuck}, C. and {Dong}, C. and {Gonzalez}, A.~H. and {Hayward}, C.~C. and {Jarugula}, S. and {Marrone}, D.~P. and {Narayanan}, D. and {Reuter}, C. and {Scott}, D. and {Spilker}, J.~S. and {Vieira}, J.~D. and {Wang}, G. and {Weiss}, A.},
        title = "{Optical and near-infrared observations of the SPT2349-56 proto-cluster core at z = 4.3}",
      journal = {\mnras},
         year = 2021,
        month = apr,
       volume = {502},
       number = {2},
        pages = {1797-1815},
          doi = {10.1093/mnras/stab103},
archivePrefix = {arXiv},
       eprint = {2006.15345},
 primaryClass = {astro-ph.GA},
       adsurl = {https://ui.adsabs.harvard.edu/abs/2021MNRAS.502.1797R}
}

@ARTICLE{dannerbauer_2014,
       author = {{Dannerbauer}, H. and {Kurk}, J.~D. and {De Breuck}, C. and {Wylezalek}, D. and {Santos}, J.~S. and {Koyama}, Y. and {Seymour}, N. and {Tanaka}, M. and {Hatch}, N. and {Altieri}, B. and {Coia}, D. and {Galametz}, A. and {Kodama}, T. and {Miley}, G. and {R{\"o}ttgering}, H. and {Sanchez-Portal}, M. and {Valtchanov}, I. and {Venemans}, B. and {Ziegler}, B.},
        title = "{An excess of dusty starbursts related to the Spiderweb galaxy}",
      journal = {\aap},
         year = 2014,
        month = oct,
       volume = {570},
          eid = {A55},
        pages = {A55},
          doi = {10.1051/0004-6361/201423771},
archivePrefix = {arXiv},
       eprint = {1410.3730},
 primaryClass = {astro-ph.GA},
       adsurl = {https://ui.adsabs.harvard.edu/abs/2014A&A...570A..55D}
}

@ARTICLE{oteo_2018,
       author = {{Oteo}, I. and {Ivison}, R.~J. and {Dunne}, L. and {Manilla-Robles}, A. and {Maddox}, S. and {Lewis}, A.~J.~R. and {de Zotti}, G. and {Bremer}, M. and {Clements}, D.~L. and {Cooray}, A. and {Dannerbauer}, H. and {Eales}, S. and {Greenslade}, J. and {Omont}, A. and {Perez─Fourn{\'o}n}, I. and {Riechers}, D. and {Scott}, D. and {van der Werf}, P. and {Weiss}, A. and {Zhang}, Z.-Y.},
        title = "{An Extreme Protocluster of Luminous Dusty Starbursts in the Early Universe}",
      journal = {\apj},
         year = 2018,
        month = mar,
       volume = {856},
       number = {1},
          eid = {72},
        pages = {72},
          doi = {10.3847/1538-4357/aaa1f1},
archivePrefix = {arXiv},
       eprint = {1709.02809},
 primaryClass = {astro-ph.GA},
       adsurl = {https://ui.adsabs.harvard.edu/abs/2018ApJ...856...72O}
}

@ARTICLE{kamieneski_2024,
       author = {{Kamieneski}, Patrick S. and {Yun}, Min S. and {Harrington}, Kevin C. and {Lowenthal}, James D. and {Wang}, Q. Daniel and {Frye}, Brenda L. and {Jim{\'e}nez-Andrade}, Eric F. and {Vishwas}, Amit and {Cooper}, Olivia and {Pascale}, Massimo and {Foo}, Nicholas and {Berman}, Derek and {Englert}, Anthony and {Garcia Diaz}, Carlos},
        title = "{PASSAGES: The Wide-ranging, Extreme Intrinsic Properties of Planck-selected, Lensed Dusty Star-forming Galaxies}",
      journal = {\apj},
         year = 2024,
        month = jan,
       volume = {961},
       number = {1},
          eid = {2},
        pages = {2},
          doi = {10.3847/1538-4357/acf930},
archivePrefix = {arXiv},
       eprint = {2301.09746},
 primaryClass = {astro-ph.GA},
       adsurl = {https://ui.adsabs.harvard.edu/abs/2024ApJ...961....2K}
}

@ARTICLE{shapiro_2008,
       author = {{Shapiro}, Kristen L. and {Genzel}, Reinhard and {F{\"o}rster Schreiber}, Natascha M. and {Tacconi}, Linda J. and {Bouch{\'e}}, Nicolas and {Cresci}, Giovanni and {Davies}, Richard and {Eisenhauer}, Frank and {Johansson}, Peter H. and {Krajnovi{\'c}}, Davor and {Lutz}, Dieter and {Naab}, Thorsten and {Arimoto}, Nobuo and {Arribas}, Santiago and {Cimatti}, Andrea and {Colina}, Luis and {Daddi}, Emanuele and {Daigle}, Olivier and {Erb}, Dawn and {Hernandez}, Olivier and {Kong}, Xu and {Mignoli}, Marco and {Onodera}, Masato and {Renzini}, Alvio and {Shapley}, Alice and {Steidel}, Charles},
        title = "{Kinemetry of SINS High-Redshift Star-Forming Galaxies: Distinguishing Rotating Disks from Major Mergers}",
      journal = {\apj},
         year = 2008,
        month = jul,
       volume = {682},
       number = {1},
        pages = {231-251},
          doi = {10.1086/587133},
archivePrefix = {arXiv},
       eprint = {0802.0879},
 primaryClass = {astro-ph},
       adsurl = {https://ui.adsabs.harvard.edu/abs/2008ApJ...682..231S}
}

@ARTICLE{bellocchi_2012,
       author = {{Bellocchi}, E. and {Arribas}, S. and {Colina}, L.},
        title = "{Studying the kinematic asymmetries of disks and post-coalescence mergers using a new ``kinemetry'' criterion}",
      journal = {\aap},
         year = 2012,
        month = jun,
       volume = {542},
          eid = {A54},
        pages = {A54},
          doi = {10.1051/0004-6361/201117894},
archivePrefix = {arXiv},
       eprint = {1203.1930},
 primaryClass = {astro-ph.CO},
       adsurl = {https://ui.adsabs.harvard.edu/abs/2012A&A...542A..54B}
}

@ARTICLE{girard_2019,
       author = {{Girard}, M. and {Dessauges-Zavadsky}, M. and {Combes}, F. and {Chisholm}, J. and {Patr{\'\i}cio}, V. and {Richard}, J. and {Schaerer}, D.},
        title = "{Towards sub-kpc scale kinematics of molecular and ionized gas of star-forming galaxies at z {\ensuremath{\sim}} 1}",
      journal = {\aap},
         year = 2019,
        month = nov,
       volume = {631},
          eid = {A91},
        pages = {A91},
          doi = {10.1051/0004-6361/201935896},
archivePrefix = {arXiv},
       eprint = {1909.07400},
 primaryClass = {astro-ph.GA},
       adsurl = {https://ui.adsabs.harvard.edu/abs/2019A&A...631A..91G}
}

@ARTICLE{girard_2021,
       author = {{Girard}, M. and {Fisher}, D.~B. and {Bolatto}, A.~D. and {Abraham}, R. and {Bassett}, R. and {Glazebrook}, K. and {Herrera-Camus}, R. and {Jim{\'e}nez}, E. and {Lenki{\'c}}, L. and {Obreschkow}, D.},
        title = "{Systematic Difference between Ionized and Molecular Gas Velocity Dispersions in z {\ensuremath{\sim}} 1-2 Disks and Local Analogs}",
      journal = {\apj},
         year = 2021,
        month = mar,
       volume = {909},
       number = {1},
          eid = {12},
        pages = {12},
          doi = {10.3847/1538-4357/abd5b9},
archivePrefix = {arXiv},
       eprint = {2101.04122},
 primaryClass = {astro-ph.GA},
       adsurl = {https://ui.adsabs.harvard.edu/abs/2021ApJ...909...12G}
}

@ARTICLE{ikeda_2022,
       author = {{Ikeda}, Ryota and {Tadaki}, Ken-ichi and {Iono}, Daisuke and {Kodama}, Tadayuki and {Chan}, Jeffrey C.~C. and {Hatsukade}, Bunyo and {Hayashi}, Masao and {Izumi}, Takuma and {Kohno}, Kotaro and {Koyama}, Yusei and {Shimakawa}, Rhythm and {Suzuki}, Tomoko L. and {Tamura}, Yoichi and {Tanaka}, Ichi},
        title = "{High-resolution ALMA Study of CO J = 2-1 Line and Dust Continuum Emissions in Cluster Galaxies at z = 1.46}",
      journal = {\apj},
         year = 2022,
        month = jul,
       volume = {933},
       number = {1},
          eid = {11},
        pages = {11},
          doi = {10.3847/1538-4357/ac6cdc},
archivePrefix = {arXiv},
       eprint = {2205.05731},
 primaryClass = {astro-ph.GA},
       adsurl = {https://ui.adsabs.harvard.edu/abs/2022ApJ...933...11I}
}

@ARTICLE{wisnioski_2015,
       author = {{Wisnioski}, E. and {F{\"o}rster Schreiber}, N.~M. and {Wuyts}, S. and {Wuyts}, E. and {Bandara}, K. and {Wilman}, D. and {Genzel}, R. and {Bender}, R. and {Davies}, R. and {Fossati}, M. and {Lang}, P. and {Mendel}, J.~T. and {Beifiori}, A. and {Brammer}, G. and {Chan}, J. and {Fabricius}, M. and {Fudamoto}, Y. and {Kulkarni}, S. and {Kurk}, J. and {Lutz}, D. and {Nelson}, E.~J. and {Momcheva}, I. and {Rosario}, D. and {Saglia}, R. and {Seitz}, S. and {Tacconi}, L.~J. and {van Dokkum}, P.~G.},
        title = "{The KMOS$^{3D}$ Survey: Design, First Results, and the Evolution of Galaxy Kinematics from 0.7 <= z <= 2.7}",
      journal = {\apj},
         year = 2015,
        month = feb,
       volume = {799},
       number = {2},
          eid = {209},
        pages = {209},
          doi = {10.1088/0004-637X/799/2/209},
archivePrefix = {arXiv},
       eprint = {1409.6791},
 primaryClass = {astro-ph.GA},
       adsurl = {https://ui.adsabs.harvard.edu/abs/2015ApJ...799..209W}
}

@ARTICLE{johnson_2018,
       author = {{Johnson}, H.~L. and {Harrison}, C.~M. and {Swinbank}, A.~M. and {Tiley}, A.~L. and {Stott}, J.~P. and {Bower}, R.~G. and {Smail}, Ian and {Bunker}, A.~J. and {Sobral}, D. and {Turner}, O.~J. and {Best}, P. and {Bureau}, M. and {Cirasuolo}, M. and {Jarvis}, M.~J. and {Magdis}, G. and {Sharples}, R.~M. and {Bland-Hawthorn}, J. and {Catinella}, B. and {Cortese}, L. and {Croom}, S.~M. and {Federrath}, C. and {Glazebrook}, K. and {Sweet}, S.~M. and {Bryant}, J.~J. and {Goodwin}, M. and {Konstantopoulos}, I.~S. and {Lawrence}, J.~S. and {Medling}, A.~M. and {Owers}, M.~S. and {Richards}, S.},
        title = "{The KMOS Redshift One Spectroscopic Survey (KROSS): the origin of disc turbulence in z {\ensuremath{\approx}} 1 star-forming galaxies}",
      journal = {\mnras},
         year = 2018,
        month = mar,
       volume = {474},
       number = {4},
        pages = {5076-5104},
          doi = {10.1093/mnras/stx3016},
archivePrefix = {arXiv},
       eprint = {1707.02302},
 primaryClass = {astro-ph.GA},
       adsurl = {https://ui.adsabs.harvard.edu/abs/2018MNRAS.474.5076J}
}

@ARTICLE{rizzo_2023,
       author = {{Rizzo}, F. and {Roman-Oliveira}, F. and {Fraternali}, F. and {Frickmann}, D. and {Valentino}, F.~M. and {Brammer}, G. and {Zanella}, A. and {Kokorev}, V. and {Popping}, G. and {Whitaker}, K.~E. and {Kohandel}, M. and {Magdis}, G.~E. and {Di Mascolo}, L. and {Ikeda}, R. and {Jin}, S. and {Toft}, S.},
        title = "{The ALMA-ALPAKA survey. I. High-resolution CO and [CI] kinematics of star-forming galaxies at z = 0.5-3.5}",
      journal = {\aap},
         year = 2023,
        month = nov,
       volume = {679},
          eid = {A129},
        pages = {A129},
          doi = {10.1051/0004-6361/202346444},
archivePrefix = {arXiv},
       eprint = {2303.16227},
 primaryClass = {astro-ph.GA},
       adsurl = {https://ui.adsabs.harvard.edu/abs/2023A&A...679A.129R}
}

@ARTICLE{rizzo_2024,
       author = {{Rizzo}, F. and {Bacchini}, C. and {Kohandel}, M. and {Di Mascolo}, L. and {Fraternali}, F. and {Roman-Oliveira}, F. and {Zanella}, A. and {Popping}, G. and {Valentino}, F. and {Magdis}, G. and {Whitaker}, K.},
        title = "{The ALMA-ALPAKA survey: II. Evolution of turbulence in galaxy disks across cosmic time: Difference between cold and warm gas}",
      journal = {\aap},
         year = 2024,
        month = sep,
       volume = {689},
          eid = {A273},
        pages = {A273},
          doi = {10.1051/0004-6361/202450455},
archivePrefix = {arXiv},
       eprint = {2407.06261},
 primaryClass = {astro-ph.GA},
       adsurl = {https://ui.adsabs.harvard.edu/abs/2024A&A...689A.273R}
}

@ARTICLE{bacchini_2020a,
       author = {{Bacchini}, Cecilia and {Fraternali}, Filippo and {Iorio}, Giuliano and {Pezzulli}, Gabriele and {Marasco}, Antonino and {Nipoti}, Carlo},
        title = "{Evidence for supernova feedback sustaining gas turbulence in nearby star-forming galaxies}",
      journal = {\aap},
         year = 2020,
        month = sep,
       volume = {641},
          eid = {A70},
        pages = {A70},
          doi = {10.1051/0004-6361/202038223},
archivePrefix = {arXiv},
       eprint = {2006.10764},
 primaryClass = {astro-ph.GA},
       adsurl = {https://ui.adsabs.harvard.edu/abs/2020A&A...641A..70B}
}

@ARTICLE{romeo_2011,
       author = {{Romeo}, Alessandro B. and {Wiegert}, Joachim},
        title = "{The effective stability parameter for two-component galactic discs: is Q$^{-1}$ {\ensuremath{\approx}} Q$^{-1}$$_{stars}$ + Q$^{-1}$$_{gas}$?}",
      journal = {\mnras},
         year = 2011,
        month = sep,
       volume = {416},
       number = {2},
        pages = {1191-1196},
          doi = {10.1111/j.1365-2966.2011.19120.x},
archivePrefix = {arXiv},
       eprint = {1101.4519},
 primaryClass = {astro-ph.CO},
       adsurl = {https://ui.adsabs.harvard.edu/abs/2011MNRAS.416.1191R}
}

@ARTICLE{elmegreen_2011,
       author = {{Elmegreen}, Bruce G.},
        title = "{Gravitational Instabilities in Two-component Galaxy Disks with Gas Dissipation}",
      journal = {\apj},
         year = 2011,
        month = aug,
       volume = {737},
       number = {1},
          eid = {10},
        pages = {10},
          doi = {10.1088/0004-637X/737/1/10},
archivePrefix = {arXiv},
       eprint = {1106.1580},
 primaryClass = {astro-ph.GA},
       adsurl = {https://ui.adsabs.harvard.edu/abs/2011ApJ...737...10E}
}

@ARTICLE{meidt_2022,
       author = {{Meidt}, Sharon E.},
        title = "{Molecular Clouds as Gravitational Instabilities in Rotating Disks: A Modified Stability Criterion}",
      journal = {\apj},
         year = 2022,
        month = oct,
       volume = {937},
       number = {2},
          eid = {88},
        pages = {88},
          doi = {10.3847/1538-4357/ac86ce},
archivePrefix = {arXiv},
       eprint = {2208.01888},
 primaryClass = {astro-ph.GA},
       adsurl = {https://ui.adsabs.harvard.edu/abs/2022ApJ...937...88M}
}

@ARTICLE{nipoti_2023,
       author = {{Nipoti}, Carlo},
        title = "{Local gravitational instability of stratified rotating fluids: three-dimensional criteria for gaseous discs}",
      journal = {\mnras},
         year = 2023,
        month = feb,
       volume = {518},
       number = {4},
        pages = {5154-5162},
          doi = {10.1093/mnras/stac3403},
archivePrefix = {arXiv},
       eprint = {2211.09831},
 primaryClass = {astro-ph.GA},
       adsurl = {https://ui.adsabs.harvard.edu/abs/2023MNRAS.518.5154N}
}

@ARTICLE{nipoti_2024,
       author = {{Nipoti}, Carlo and {Caprioglio}, Cristina and {Bacchini}, Cecilia},
        title = "{Local gravitational instability of two-component thick discs in three dimensions}",
      journal = {\aap},
         year = 2024,
        month = sep,
       volume = {689},
          eid = {A61},
        pages = {A61},
          doi = {10.1051/0004-6361/202450462},
archivePrefix = {arXiv},
       eprint = {2405.13123},
 primaryClass = {astro-ph.GA},
       adsurl = {https://ui.adsabs.harvard.edu/abs/2024A&A...689A..61N}
}

@ARTICLE{bacchini_2024,
       author = {{Bacchini}, C. and {Nipoti}, C. and {Iorio}, G. and {Roman-Oliveira}, F. and {Rizzo}, F. and {Mancera Pi{\~n}a}, P.~E. and {Marasco}, A. and {Zanella}, A. and {Lelli}, F.},
        title = "{A 3D view on the local gravitational instability of cold gas discs in star-forming galaxies at 0 {\ensuremath{\lesssim}} z {\ensuremath{\lesssim}} 5}",
      journal = {\aap},
         year = 2024,
        month = jul,
       volume = {687},
          eid = {A115},
        pages = {A115},
          doi = {10.1051/0004-6361/202449925},
archivePrefix = {arXiv},
       eprint = {2405.00103},
 primaryClass = {astro-ph.GA},
       adsurl = {https://ui.adsabs.harvard.edu/abs/2024A&A...687A.115B}
}

@ARTICLE{kohandel_2020,
       author = {{Kohandel}, M. and {Pallottini}, A. and {Ferrara}, A. and {Carniani}, S. and {Gallerani}, S. and {Vallini}, L. and {Zanella}, A. and {Behrens}, C.},
        title = "{Velocity dispersion in the interstellar medium of early galaxies}",
      journal = {\mnras},
         year = 2020,
        month = nov,
       volume = {499},
       number = {1},
        pages = {1250-1265},
          doi = {10.1093/mnras/staa2792},
archivePrefix = {arXiv},
       eprint = {2009.05049},
 primaryClass = {astro-ph.GA},
       adsurl = {https://ui.adsabs.harvard.edu/abs/2020MNRAS.499.1250K}
}

@ARTICLE{Eales_2010,
       author = {{Eales}, S. and {Dunne}, L. and {Clements}, D. and {Cooray}, A. and {De Zotti}, G. and {Dye}, S. and {Ivison}, R. and {Jarvis}, M. and {Lagache}, G. and {Maddox}, S. and {Negrello}, M. and {Serjeant}, S. and {Thompson}, M.~A. and {Van Kampen}, E. and {Amblard}, A. and {Andreani}, P. and {Baes}, M. and {Beelen}, A. and {Bendo}, G.~J. and {Benford}, D. and {Bertoldi}, F. and {Bock}, J. and {Bonfield}, D. and {Boselli}, A. and {Bridge}, C. and {Buat}, V. and {Burgarella}, D. and {Carlberg}, R. and {Cava}, A. and {Chanial}, P. and {Charlot}, S. and {Christopher}, N. and {Coles}, P. and {Cortese}, L. and {Dariush}, A. and {da Cunha}, E. and {Dalton}, G. and {Danese}, L. and {Dannerbauer}, H. and {Driver}, S. and {Dunlop}, J. and {Fan}, L. and {Farrah}, D. and {Frayer}, D. and {Frenk}, C. and {Geach}, J. and {Gardner}, J. and {Gomez}, H. and {Gonz{\'a}lez-Nuevo}, J. and {Gonz{\'a}lez-Solares}, E. and {Griffin}, M. and {Hardcastle}, M. and {Hatziminaoglou}, E. and {Herranz}, D. and {Hughes}, D. and {Ibar}, E. and {Jeong}, Woong-Seob and {Lacey}, C. and {Lapi}, A. and {Lawrence}, A. and {Lee}, M. and {Leeuw}, L. and {Liske}, J. and {L{\'o}pez-Caniego}, M. and {M{\"u}ller}, T. and {Nandra}, K. and {Panuzzo}, P. and {Papageorgiou}, A. and {Patanchon}, G. and {Peacock}, J. and {Pearson}, C. and {Phillipps}, S. and {Pohlen}, M. and {Popescu}, C. and {Rawlings}, S. and {Rigby}, E. and {Rigopoulou}, M. and {Robotham}, A. and {Rodighiero}, G. and {Sansom}, A. and {Schulz}, B. and {Scott}, D. and {Smith}, D.~J.~B. and {Sibthorpe}, B. and {Smail}, I. and {Stevens}, J. and {Sutherland}, W. and {Takeuchi}, T. and {Tedds}, J. and {Temi}, P. and {Tuffs}, R. and {Trichas}, M. and {Vaccari}, M. and {Valtchanov}, I. and {van der Werf}, P. and {Verma}, A. and {Vieria}, J. and {Vlahakis}, C. and {White}, Glenn J.},
        title = "{The Herschel ATLAS}",
      journal = {\pasp},
         year = 2010,
        month = may,
       volume = {122},
       number = {891},
        pages = {499},
          doi = {10.1086/653086},
archivePrefix = {arXiv},
       eprint = {0910.4279},
 primaryClass = {astro-ph.CO},
       adsurl = {https://ui.adsabs.harvard.edu/abs/2010PASP..122..499E}
}

\appendix

\section{Data modelling}

    \subsection{First-pass emission line fits}

     To determine the best-fit redshifts and line widths, we fit the global CO(4--3) line profile extracted from the 2023 data, imaged with a robust weighting of 2 and cleaning down to $1\sigma$. We extract the spectra from within circular apertures, determining the most suitable aperture through a curve-of-growth analysis (i.e. taking the aperture at which the contained flux reaches a maximum). The CO(4--3) spectra, best-fit Gaussians, and region spanning 90\% of the line emission are shown in the left column of Fig.~\ref{fig:spectra}, with the best-fit redshift presented in Table~\ref{tab:source_prop}. The spectroscopic redshifts of W, T, C, and M match the spectroscopic redshifts derived from CO(3--2)in \cite{ivison_2013} within uncertainties. However, we find a slightly lower spectroscopic redshift for ULIRG-E than that measured from CO(1--0) in \cite{ivison_2019}, of $z=2.4121\pm0.0004$ vs $2.415\pm0.002$. This could be partly due to the faintness of this galaxy compared to the others. However, we also find significant offsets between the CO(4--3), \ci\ 1--0, and dust-continuum peaks, which could indicate a more complex multi-component system.

 \begin{figure*}
       \raggedright
      \includegraphics[width=0.49\textwidth,trim={1cm 0.5cm 1cm 0.5cm},clip]{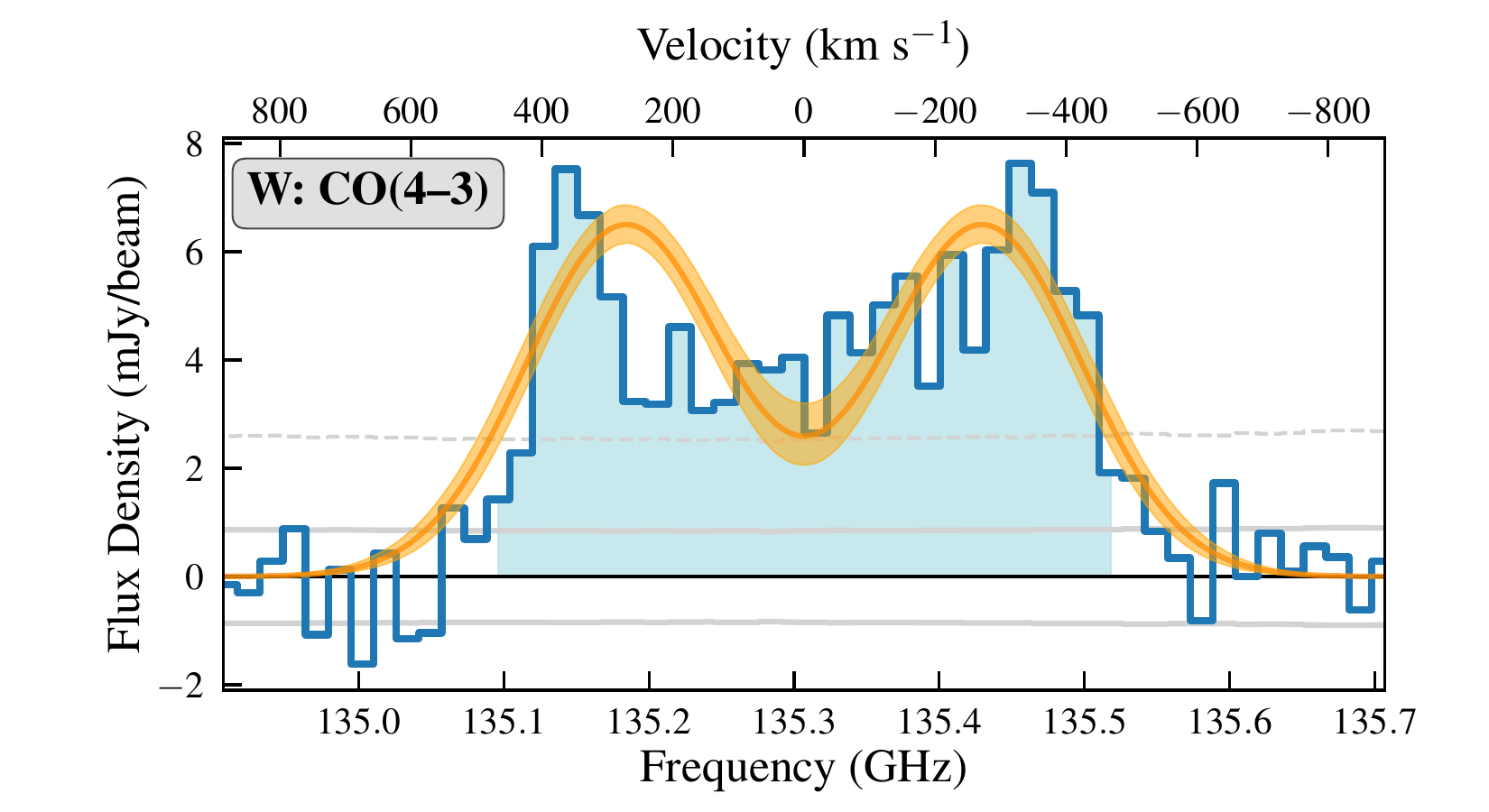}
      ~
      \includegraphics[width=0.49\textwidth,trim={1cm 0.5cm 1cm 0.5cm},clip]{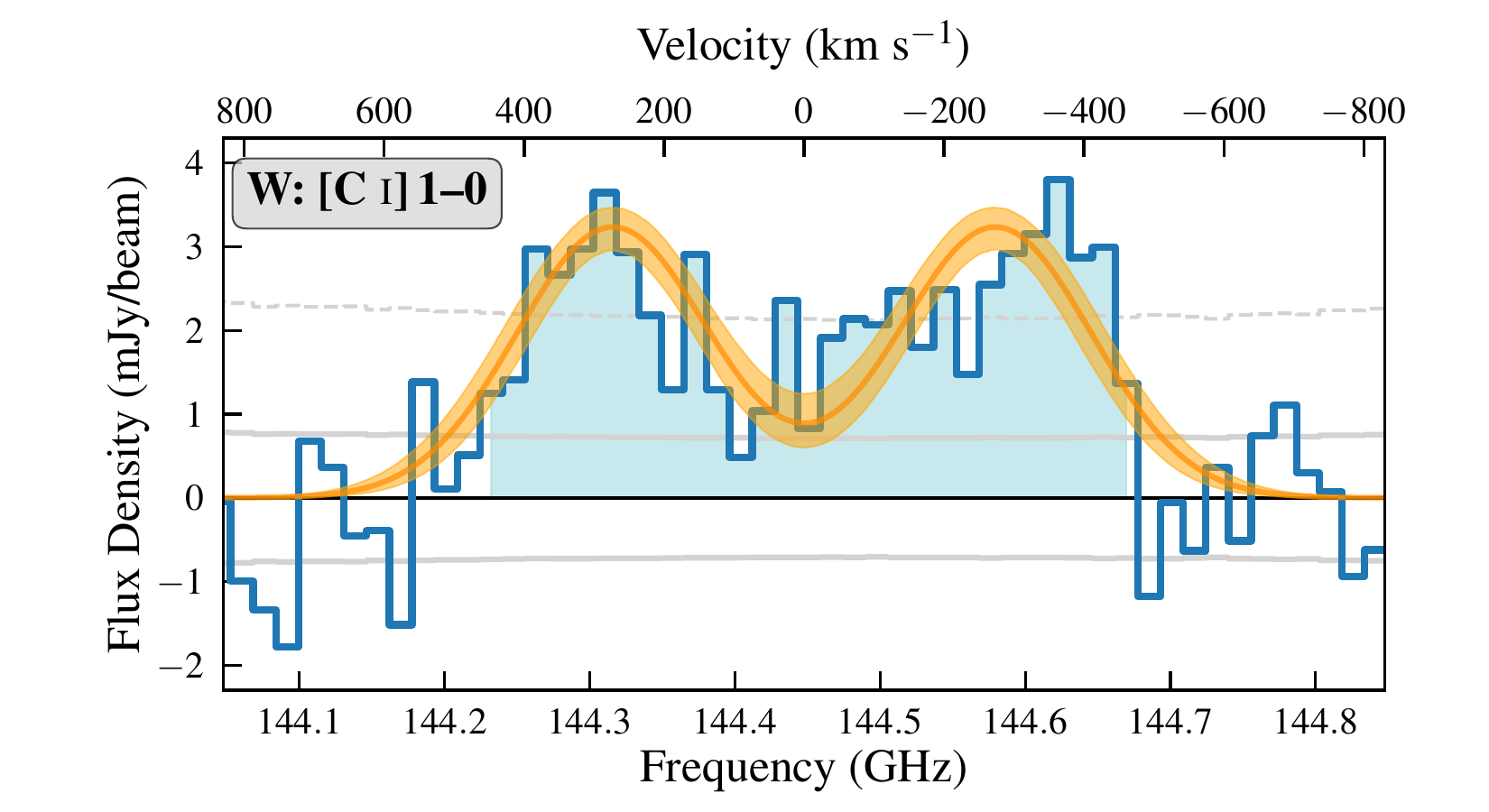}
      \\
      \includegraphics[width=0.49\textwidth,trim={1cm 0.1cm 1cm 0.1cm},clip]{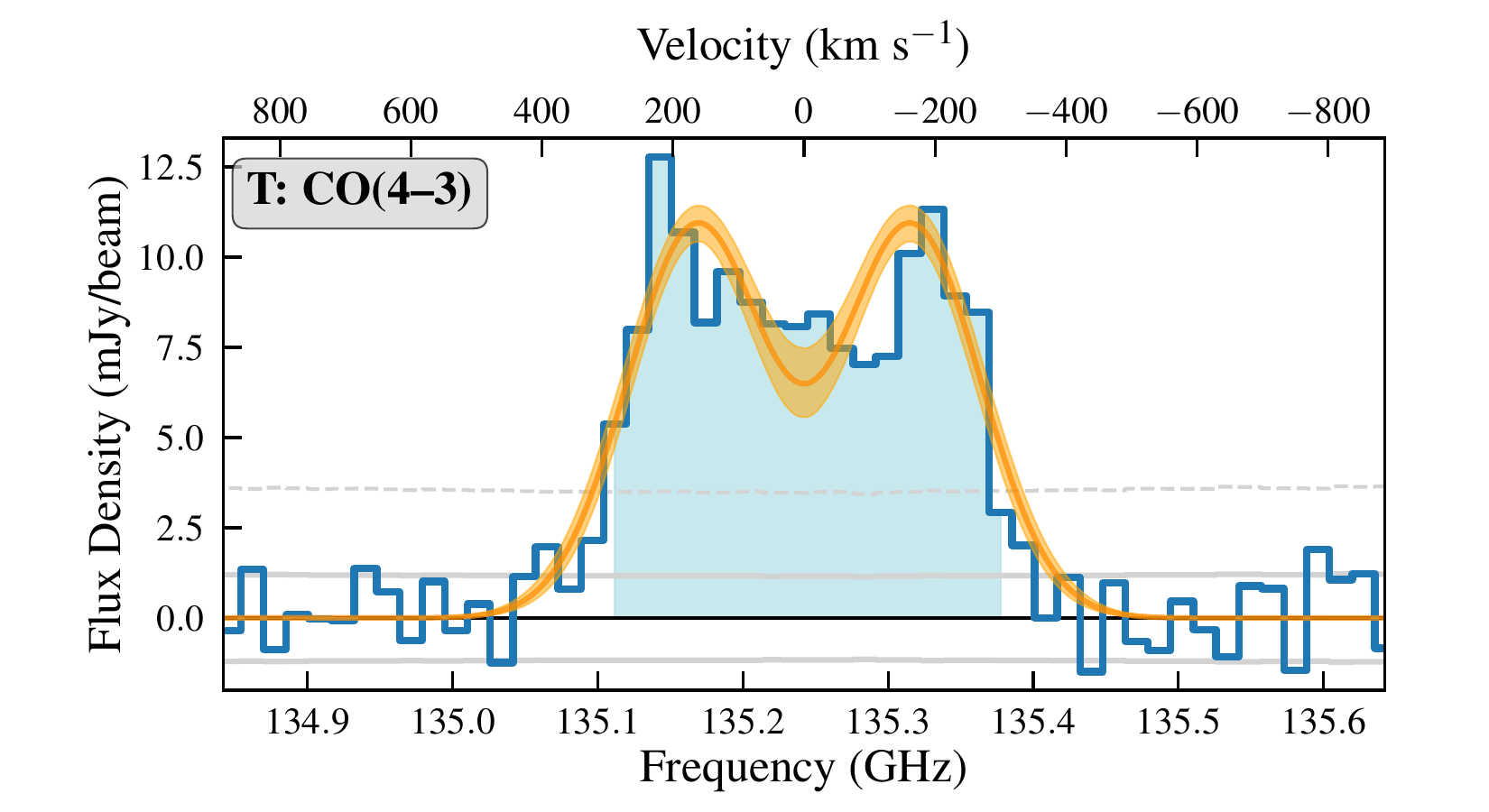}
      ~
      \includegraphics[width=0.49\textwidth,trim={1cm 0.1cm 1cm 0.1cm},clip]{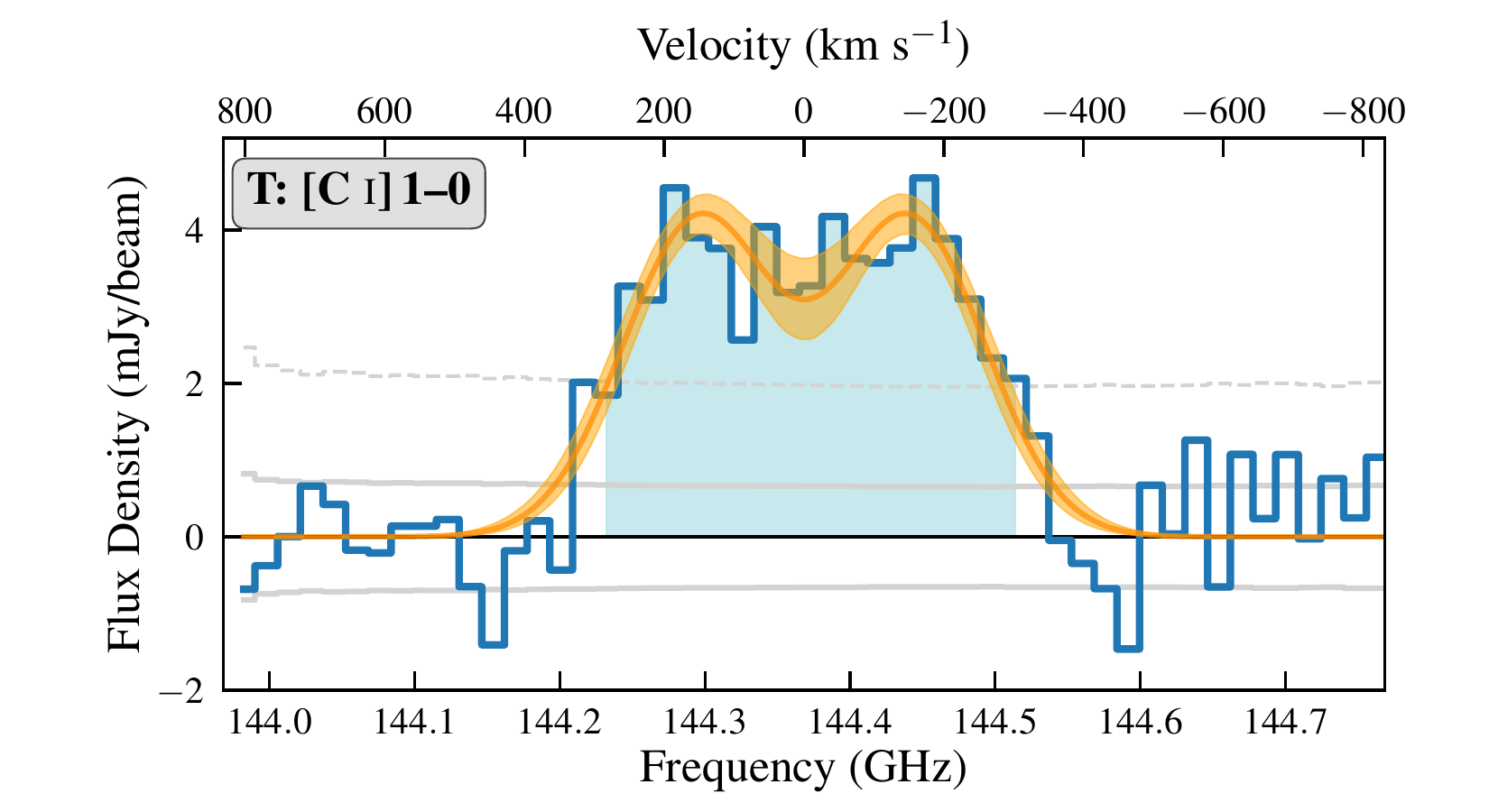}
      \\
       \includegraphics[width=0.49\textwidth,trim={1cm 0.1cm 1cm 0.1cm},clip]{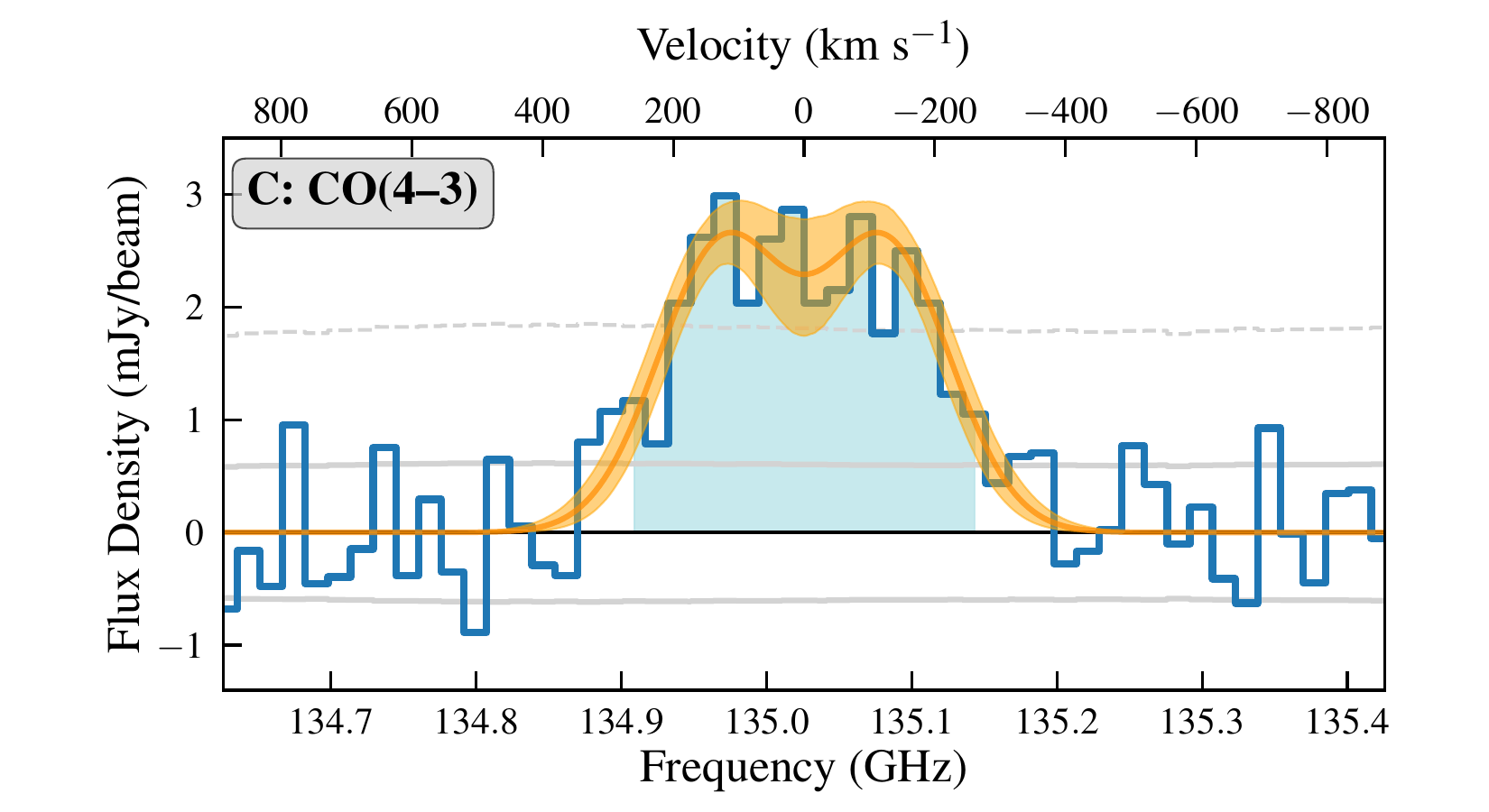}
       ~
       \includegraphics[width=0.49\textwidth,trim={1cm 0.1cm 1cm 0.1cm},clip]{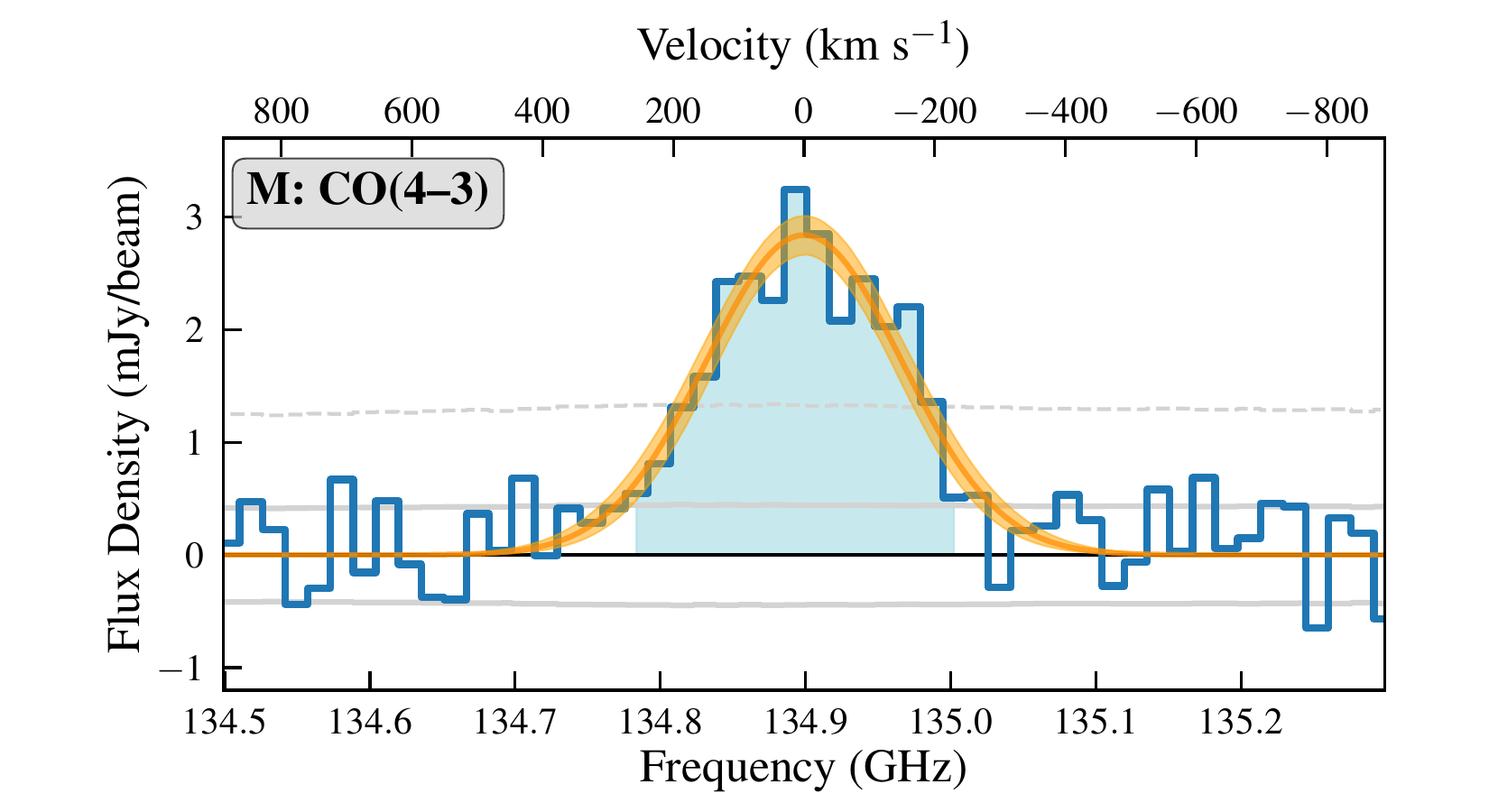}
       \\
       \includegraphics[width=0.49\textwidth,trim={0.1cm 0.1cm 0.2cm 0.1cm},clip]{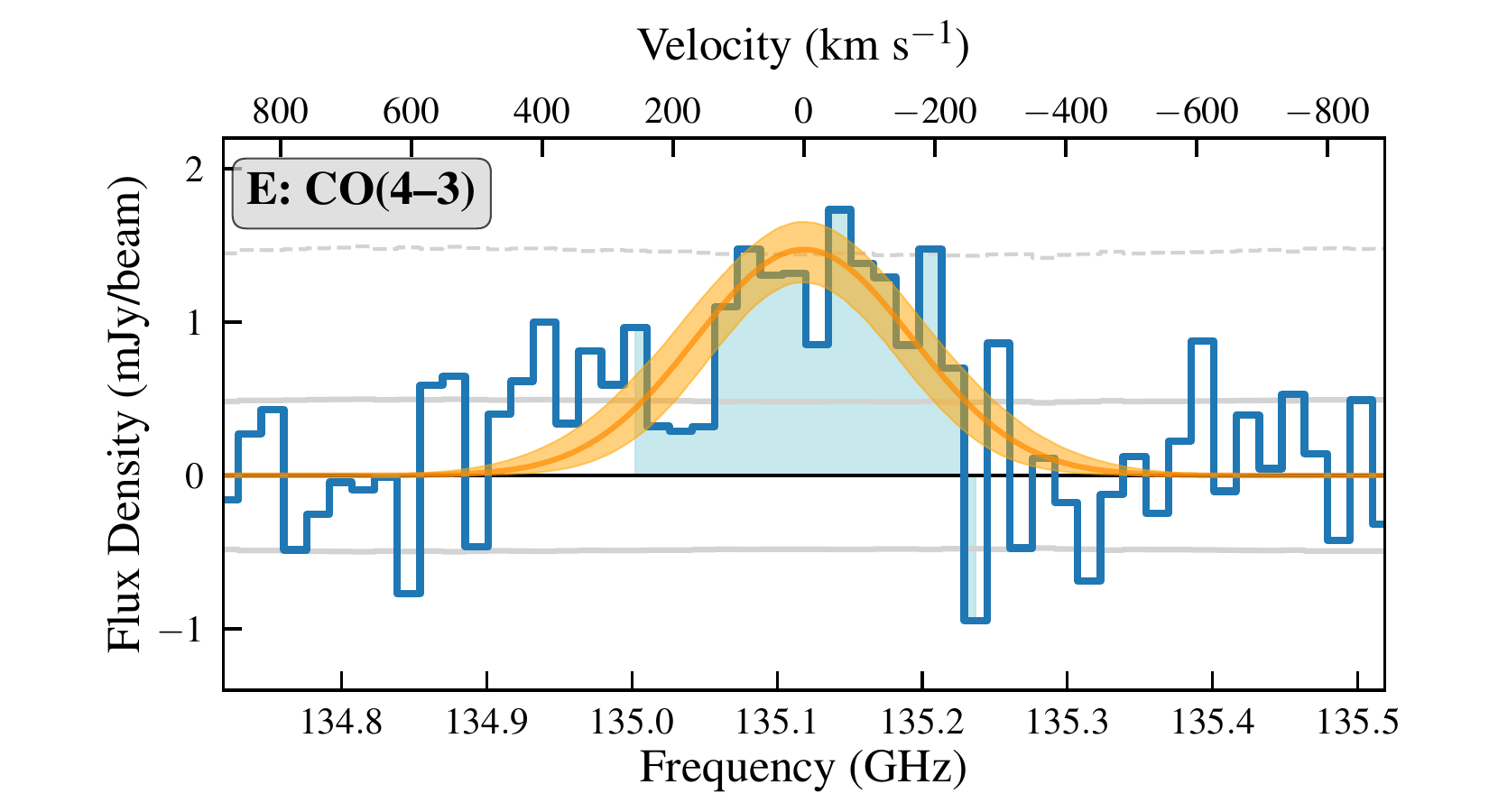}
       \caption{CO(4--3) vs \ci 1--0 global line profiles (blue lines, left vs right panels) and best-fit Gaussian profiles (orange lines and filled areas denoting 16th to 84th percentiles). All CO(4--3) spectra, as well as the \ci 1--0 spectra of HyLIRGs-W and T, are extracted from the $\sim35$ \kms cubes generated from program 2023.1.00714.S. To extract these spectra we used circular apertures defined by the maximum in a curve-of-growth analysis (between $0\farcs9$ and $1\farcs2$). The best-fit Gaussian profiles were used to determine the spectroscopic redshifts and uncertainties, which define the systemic velocities in the PV diagrams and kinematic modelling.}
       \label{fig:spectra}
   \end{figure*}

       

            \subsection{A note on the spectral response}
                \label{sub:spectra_response}

                Kinematic modelling tools need to account for the spectral response to model the intrinsic line broadening. To quantify the effective spectral response of the interferometric data (critical to recovering $\sigma_\mathrm{gas}$), we modelled the line spread function (LSF) resulting from the combination of Hanning smoothing and subsequent channel averaging. The original visibilities were Hanning smoothed at the correlator level, corresponding to convolution with a 3-point kernel of weights (0.25, 0.5, 0.25) \citep{alma_handbook}. By using \texttt{tclean} to image these data over multiple channels, we effectively applied a boxcar kernel of $\mathrm{width}=N_\mathrm{input~ channels}$, where $N_\mathrm{input~ channels}$ is the number of channels being combined. By taking the discrete convolution of the Hanning and boxcar kernels, we compute the spectral response function, normalising to the unit area and centering on the mean. We characterise both the standard deviation, $\sigma_\mathrm{inst}$ and the $FWHM_\mathrm{inst}$, expressing the result in output-channel units, $N_\mathrm{output~ channels}$. This procedure yields the exact discrete LSF and avoids assuming Gaussianity, as the convolution of a Hanning LSF (triangular kernel) and boxcar (tophat) results in a triangular-topped trapezoid. We compare the analytical results for individual kernels: the Hanning filter alone has $\sigma_\mathrm{Hanning}=\sqrt{0.5}$ \citep[derivation based on weights from e.g.][]{ThompsonMoranSwenson2017} and the boxcar of length $N$, has $\sigma_\mathrm{boxcar}=\sqrt{(N^2-1)/12}$ \citep[e.g.][for the latter derivation]{BertsekasTsitsiklis2008}. This yields $FWHM_\mathrm{inst}=1$ output channel for $N_\mathrm{input~ channels}\geq 3$ and $\sigma_{inst}=0.331 (0.300)$ output channels for $4\times$ (8$\times$) native channel averaging. Although the line spread function for our data are not Gaussian, existing kinematic modelling tools like \bbarolo\ assume a Gaussian profile. Given this in-built assumption, we convert $FWHM_\mathrm{inst}$ to $\sigma_\mathrm{inst}$ by taking $1/(2\sqrt{2 \ln(2)})=0.425$, consistent with existing studies \citep[e.g.][]{lelli_2023}. We note that using $\sigma_{inst}=0.33$ vs $\sigma_{inst}=0.43$ had a negligible impact on the derived kinematic properties.

        \subsection{Kinematic fits for HyLIRG-W}
            \label{sub:kinfits_17kms}

            In Sec.~\ref{sub:kinematic_modelling}, we discussed the differences between the kinematic models fit to the $\sim$17\,\kms and $\sim$35\,\kms cubes. In Fig.~\ref{fig:kinfit_3dbarolo_HyLIRGw_17kms}, we show the kinematic model fit to the 17\,\kms data cube. 

            \begin{figure*}
                \raggedright
                \includegraphics[width=\textwidth,trim={0cm 0cm 0cm 0cm},clip]{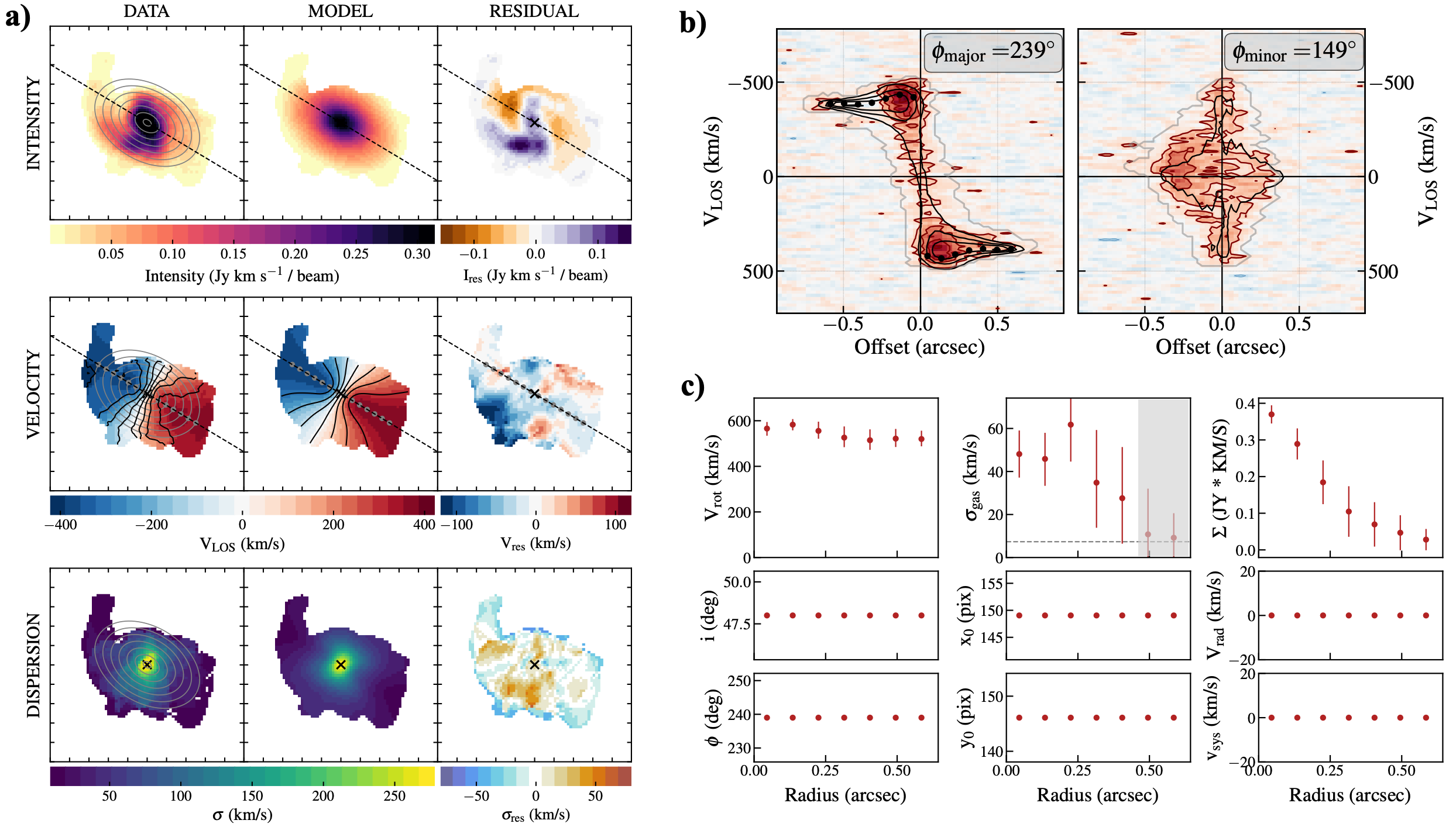}
                \\
                {\large\textbf{d)}}
                \includegraphics[width=\textwidth,trim={1.cm 0.7cm 1.1cm 0.2cm},clip]{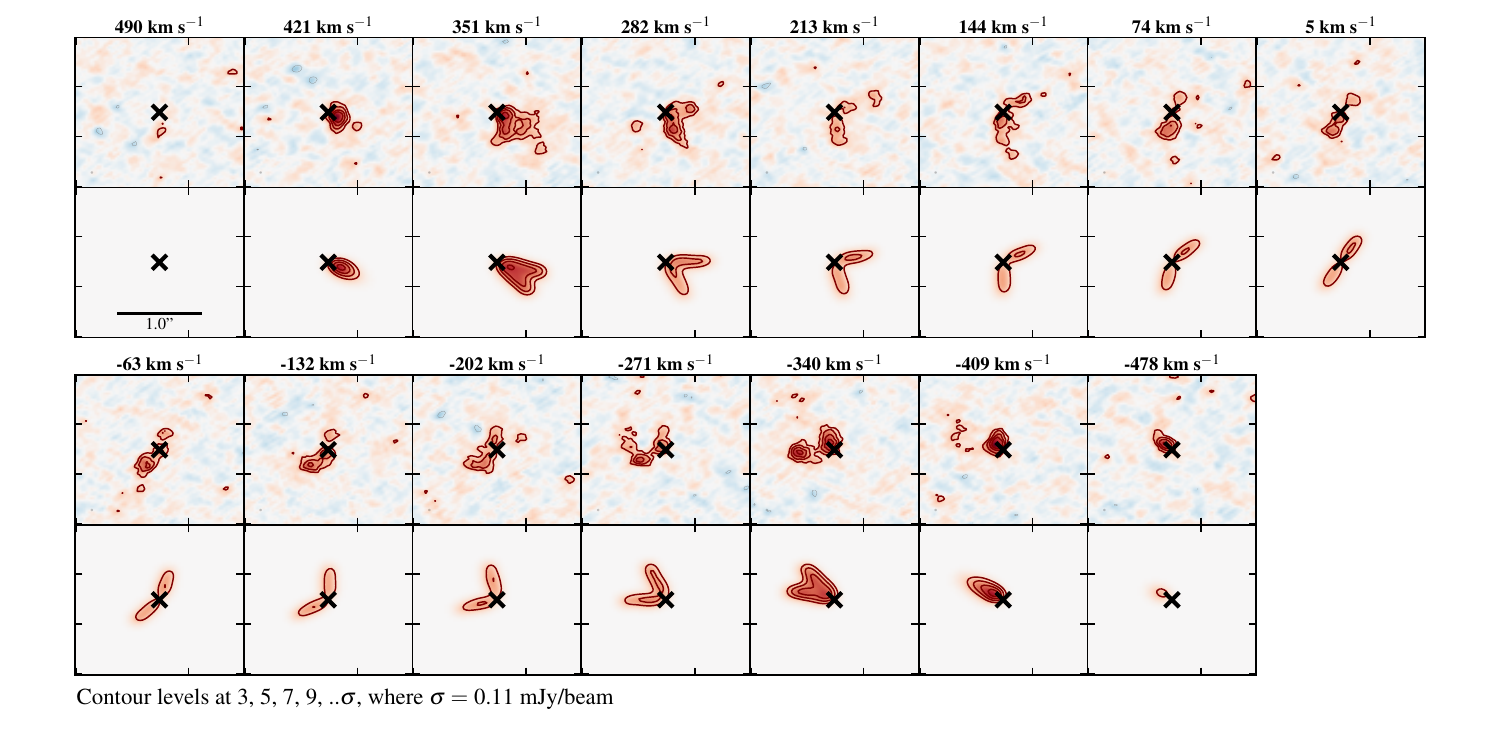}
                \caption{Same as Fig.~\ref{fig:kinfit_3dbarolo_HyLIRGw} but for the 17\,\kms data cube and showing every 4$^\mathrm{th}$ channel in panel d).
                }
                \label{fig:kinfit_3dbarolo_HyLIRGw_17kms}
            \end{figure*}

        \subsection{Literature comparison sample}
            \label{sub:lit_sample}

            In Sec.~\ref{sec:discussion}, we place our findings in the context of previous kinematic measurements. For this purpose, we use the following literature sample of cold gas measurements:

             \begin{itemize}[leftmargin=*, labelsep=0.6em, topsep=0.5pt, partopsep=0.5pt, parsep=0.5pt, itemsep=0.5pt]
                \item 42 local spiral galaxies from \cite{diteodoro_2021}, for which the kinematic properties are derived from \hi,
                \item 10 local spiral galaxies from PHANGS-ALMA for which the kinematics were consistently re-derived from the public CO(2--1) data by \cite{rizzo_2024},
                \item four local $M_\mathrm{\star}>10^{10}$\,\msun\ spiral galaxies from HERACLES for which the kinematics were derived from CO(2--1) in \cite{bacchini_2020a},
                \item nine highly star-forming $z<0.3$ galaxies from DYNAMO, for which the kinematics were derived from resolved CO(3--2) and CO(4--3) observations by \cite{girard_2021},
                \item the Cosmic Snake and A521, for which the kinematics were derived from resolved CO(4--3) observations \citep{girard_2019},
                \item the three extended main-sequence galaxies at $z=1.4, 1.5,$ and $2.7$ for which the resolved CO(2--1), CO(2--1), and CO(3--2) observations were modelled in \cite{kaasinen_2020},
                \item the archival sample from \cite{rizzo_2023} for which the kinematics were measured from ALMA observations of CO(2--1), \ci 2--1, CO(3--2), CO(5--4) and CO(6--5), 
                \item the two main-sequence galaxies at $z=1.46$ and $2.23$ in \cite{lelli_2023}, whose kinematic properties were inferred from CO(2--1) plus CO(3--2) and CO(3--2) plus CO(4--3), respectively,
                \item the rotating disc studied in CO(4--3) in \cite{pensabene_2025}
                \item seven lensed, dusty star-forming galaxies at $z=4-5$ observed in \cii\ and modelled in \cite{rizzo_2020,rizzo_2021} and \cite{lelli_2021},
                \item four discs studied in \cii\ in \cite{roman-oliveira_2023} (one previously studied in \cite{neeleman_2020}, and one previously studied in \cite{tsukui_2021}), and
                \item six disc galaxies in a $z=4.3$ protocluster, for which resolved \cii\ observations are modelled in \cite{venkateshwaran_2024}.
            \end{itemize}
            The table of literature measurements is available in the online supplementary materials.







\bsp	
\label{lastpage}
\end{document}